
\input harvmac
\input amssym.def
\input amssym.tex
\parskip=4pt \baselineskip=12pt
\hfuzz=20pt
\parindent 10pt

\def\goo{\cg_{\o}}
\def\trq{{\textstyle{3\over4}}}
\def\nt{\noindent}
\def\nl{\hfill\break}
\def\nli{\hfill\break\indent}
\def\np{\vfill\eject}

\font\iti=cmti9

\font\male=cmr9

\font\tfont=cmbx12 scaled\magstep1 

\global\secno=0

\def\wt{\widetilde{|\L\rg}}

\def\half{{\textstyle{1\over2}}}
\def\ha{{\textstyle{1\over2}}}
\def\trh{{\textstyle{3\over2}}}
\def\frh{{\textstyle{5\over2}}}
\def\srh{{\textstyle{7\over2}}}

\def\halfn{{\textstyle{2\over N}}}

\def\quarter{{\textstyle{1\over4}}}

\def\rg{\rangle} 

\font\fat=cmsy10 scaled\magstep5

\def\Bbullet{\raise-3pt\hbox{\fat\char"0F}}

\def\tV{\tilde V} \def\tW{\tilde W}
\def\tv{\tilde v}
\def\hV{\hat V} \def\hL{\hat L} \def\hLL{{\hat L}_{\rm long}}
\def\hI{\hat I}
\def\hs{\hat s} \def\hW{\hat W}

\font\verysmall=cmr5

\def\righa{{\longrightarrow\kern-14pt
\raise5pt\hbox{\verysmall 12}}}

\def\sigha{{\longrightarrow\kern-14pt
\raise5pt\hbox{\verysmall 34}}}

\def\upot{{\uparrow\kern-3pt
\hbox{\verysmall 12}}}

\def\upone{{\uparrow\kern-3pt
\hbox{\verysmall 1}}}

\def\uptwo{{\uparrow\kern-3pt
\hbox{\verysmall 2}}}

\def\dthree{{\downarrow\kern-3pt
\raise2pt\hbox{\verysmall 3}}}

\def\dfour{{\downarrow\kern-3pt
\raise2pt\hbox{\verysmall 4}}}

\def\dtf{{\downarrow\kern-3pt
\raise2pt\hbox{\verysmall 34}}}

\def\nwone{{\nwarrow\kern-3pt
\raise2pt\hbox{\verysmall 1}}}

\def\nwot{{\nwarrow\kern-5pt
\raise3pt\hbox{\verysmall 12}}}

\def\swthree{{\swarrow\kern-3pt
\raise2pt\hbox{\verysmall 3}}}

\def\swtf{{\swarrow\kern-5pt
\raise1pt\hbox{\verysmall 34}}}

\def\lone{{\leftarrow\kern-8pt
\raise5pt\hbox{\verysmall 1}}}

\def\ltwo{{\leftarrow\kern-8pt
\raise5pt\hbox{\verysmall 2}}}

\def\rone{{\rightarrow\kern-11pt
\raise5pt\hbox{\verysmall 1}}~~}

\def\rtwo{{\rightarrow\kern-11pt
\raise5pt\hbox{\verysmall 2}}~~}

\def\rthree{{\rightarrow\kern-11pt
\raise5pt\hbox{\verysmall 3}}~~}

\def\rfour{{\rightarrow\kern-11pt
\raise5pt\hbox{\verysmall 4}}~~}

\def\rtf{{\longrightarrow\kern-15pt
\raise5pt\hbox{\verysmall 34}}~~}

\def\lot{{\longleftarrow\kern-14pt
\raise5pt\hbox{\verysmall 12}}}

\def\dia{~~$\diamondsuit$}
\def\bsq{~~$\blacksquare$}
\def\bu{$\bullet$}

\def\cg{{\cal G}} \def\ch{{\cal H}}

 \def\cq{{\cal Q}} \def\car{{\cal R}}

\def\hcg{\hat{\cal G}} \def\hch{\hat{\cal H}}

\def\white#1{\mathop{\bigcirc}\limits_{#1}}
\def\gray#1{\mathop{\bigotimes}\limits_{#1}}
\def\riga{-\kern-4pt - \kern-4pt -}

\def\bbz{Z\!\!\!Z}
\def\bbc{C\kern-6pt I}
\def\bac{{C\kern-5.5pt I}}

\def\bbn{I\!\!N} \def\k{\kappa} \def\o{{\bar 0}} \def\I{{\bar 1}}

\def\a{\alpha} \def\b{\beta} \def\g{\gamma} \def\d{\delta}

\def\eps{{\epsilon}}
\def\l{\lambda} 
\def\r{\rho}

\def\G{\Gamma} 

\def\D{\Delta} \def\L{\Lambda}
\def\hD{\hat{\Delta}} \def\hpi{\hat{\pi}}

\def\ve{\varepsilon}

 \def\ra{\longrightarrow}
\def\mt{\mapsto}

\def\({\left(}
\def\){\right)}

\def\a{\alpha}
\def\b{\beta}
\def\d{\delta}
\def\g{\gamma}

\def\r{\rho}
\def\l{\lambda}
\def\D{\Delta}
\def\L{\Lambda}



\nref\LMRS{S. Lee, S. Minwalla, M. Rangamani and N. Seiberg, Adv.
Theor. Math. Phys. {\bf 2}, 697 (1998), hep-th/9806074.}

\nref\DEFP{E. D'Hoker, J. Erdmenger, D.Z. Freedman and M.
Perez-Victoria,
Nucl. Phys. {\bf B589}, 3-37 (2000) hep-th/0003218.}

\nref\FeFr{S. Ferrara and C. Fronsdal, Conformal fields in higher
 dimensions, in:  in {\it Rome 2000,  Recent developments in theoretical
 and experimental general relativity, gravitation and relativistic field theories, Pt.
 A},  pp. 508-527 and in: {\it Moscow 2000, Quantization, gauge theory, and strings, vol.
 1}, pp. 405-426; hep-th/0006009.}

\nref\DhP{E. D'Hoker and B. Pioline,
J. High En. Phys. {\bf 0007}, 021 (2000)   hep-th/0006103.}

\nref\NS{A. Nelson and M.J. Strassler, Phys. Rev. {\bf D60},
015004 (1999) hep-ph/9806346;\ J. High En. Phys. {\bf 0009}, 030
(2000) hep-ph/0006251.}

\nref\Osb{H. Osborn, Ann. Phys. (NY) {\bf 272}, 243 (1999),
hep-th/9808041.}

\nref\Park{J.-H. Park, Nucl. Phys. {\bf B559}, 455 (1999),
hep-th/9903230.}

\nref\FGPW{D.Z. Freedman, S.S. Gubser, K. Pilch and N.P. Warner,
Adv. Theor. Math. Phys. {\bf 3}, 363 (1999), hep-th/9904017.}

\nref\GrKa{M. Gremm and A. Kapustin,
J. High En. Phys. {\bf 9907}, 005 (1999)  hep-th/9904050.}

\nref\AGMOO{O. Aharony, S.S. Gubser, J. Maldacena, H. Ooguri and
Y. Oz, Phys. Rep. {\bf 323}, 184 (2000), hep-th/9905111.}

\nref\Reh{K.-H. Rehren,
Ann. Henri Poincare, {\bf 1},  607-623 (2000) hep-th/9905179;\
in: Proceedings of Zlatibor Summer School on Modern
Mathematical Physics, August 2004, eds. B. Dragovich et al, SFIN XVIII (A1) (Belgrade,
2005) pp. 95-118, hep-th/0411086.}

\nref\CDDF{A. Ceresole, G. Dall'Agata, R. D'Auria and S. Ferrara,
Phys. Rev. {\bf D61}, 066001 (2000), hep-th/9905226;
~Class. Quant. Grav. {\bf 17}, 1017 (2000) hep-th/9910066.}

\nref\BKRS{M. Bianchi, S. Kovacs, G. Rossi and Y.S. Stanev,
J. High En. Phys. {\bf 9909}, 020 (1999)   hep-th/9906188;\
Nucl. Phys. {\bf B584}, 216 (2000) hep-th/0003203.}

\nref\DFMMR{E. D'Hoker, D.Z. Freedman, S.D. Mathur, A. Matusis
and L. Rastelli, in: Shifman, M.A. (ed.) 'The Many Faces of the
Superworld', p. 332, hep-th/9908160.}

\nref\FeZa{S. Ferrara and A. Zaffaroni,
Conference Moshe Flato, Dijon, 1999, Proc.
Vol I - Quantization, Deformations and Symmetries,
Mathematical Physics Studies, 21
(Kluwer Academic Publ, Dordrecht, 2000)
pp. 177-188; hep-th/9908163.}

\nref\AFSZ{L. Andrianopoli, S. Ferrara, E. Sokatchev and B. Zupnik,
Adv. Theor. Math. Phys. {\bf 3}, 1149 (1999) hep-th/9912007.}

\nref\FS{S. Ferrara and E. Sokatchev,
Lett. Math. Phys. {\bf 52}, 247 (2000), 
hep-th/9912168;\
J. Math. Phys. {\bf 42}, 3015 (2001) hep-th/0010117.}

\nref\Pelc{O. Pelc, J. High En. Phys. {\bf 0003}, 012 (2000)
hep-th/0001054.}

\nref\Fa{S. Ferrara, TMR Conference, Paris, 1-7 Sep 1999,
hep-th/0002141.}

\nref\FSb{S. Ferrara and E. Sokatchev,
J. High En. Phys. {\bf 0005}, 038 (2000)  hep-th/0003051;\
Int. J. Mod. Phys. {\bf B14}, 2315 (2000) hep-th/0007058;\
New J. Phys. {\bf 4}, 2-22 (2002) hep-th/0110174.}

\nref\BaZu{F. Bastianelli and R. Zucchini,
J. High En. Phys. {\bf 0005}, 047 (2000)  hep-th/0003230.}

\nref\HeHo{P.J. Heslop and P.S. Howe, Class. Quant. Grav. {\bf 17},
3743 (2000) hep-th/0005135;\
TMR Conference, Paris, September
2000, hep-th/0009217.}

\nref\FSa{S. Ferrara and E. Sokatchev,
Int. J. Theor. Phys. {\bf 40}, 935 (2001) hep-th/0005151.}

\nref\DLS{M.J. Duff, J.T. Liu and K.S. Stelle,
J. Math. Phys. {\bf 42}, 3027 (2001) hep-th/0007120.}

\nref\LiSa{J.T. Liu and H. Sati,
Nucl. Phys. {\bf B605}, 116 (2001) hep-th/0009184.}

\nref\DFLV{R. D'Auria, S. Ferrara, M.A. Lledo and V.S.
Varadarajan, 
J. Geom. Phys. {\bf 40} 101-128, (2001) hep-th/0010124.}

\nref\KlSa{D. Klemm and W.A. Sabra, J. High En. Phys. {\bf 0102},
031 (2001)  hep-th/0011016.}

\nref\FrHL{D.Z. Freedman and P. Henry-Labordere, TMR Conference,
Paris, 7-13 Sep 2000, hep-th/0011086.}

\nref\MaWe{S.P. Martin and J.D. Wells,
Phys Rev {\bf D64}, 036010 (2001) hep-ph/0011382.}

\nref\Arg{P.C. Argyres, Nucl. Phys. Proc. Sup. {\bf B101}, 175
(2001) hep-th/0102006.}

\nref\AEPS{G. Arutyunov, B. Eden, A.C. Petkou and E. Sokatchev,
Nucl. Phys. {\bf B620}, 380 (2002) hep-th/0103230.}

\nref\BKRSa{M. Bianchi, S. Kovacs, G. Rossi and Y.S. Stanev,
J. High En. Phys. {\bf 0105}, 042 (2001)  hep-th/0104016.}

\nref\AES{G. Arutyunov, B. Eden and E. Sokatchev,
Nucl. Phys. {\bf B619}, 359 (2001) hep-th/0105254.}

\nref\HeHoa{P.J. Heslop and P.S. Howe,
Phys. Lett. {\bf 516B}, 367 (2001)
hep-th/0106238;\
Nucl. Phys. {\bf B626}, 265-286 (2002) hep-th/0107212;\
Aspects of N=4 SYM, hep-th/0307210.}

\nref\EdSo{B. Eden and E. Sokatchev, Nucl. Phys. {\bf B618}, 259 (2001)
hep-th/0106249.}

\nref\PeSa{S. Penati and A. Santambrogio,
Nucl. Phys. {\bf B614}, 367 (2001) hep-th/0107071.}

\nref\EFS{B. Eden, S. Ferrara and E. Sokatchev,
J. High En. Phys. {\bf 0111}, 020 (2001)  hep-th/0107084.}

\nref\KuSu{J. Kubo and D. Suematsu, Phys. Rev. {\bf D64}, 115014
(2001) hep-ph/0107133.}

\nref\West{P. West, Nucl. Phys. Proc. Sup. {\bf B101}, 112 (2001).}

\nref\Hes{P.J. Heslop, Class. Quant. Grav. {\bf 19}, 303 (2002)
hep-th/0108235;
J. High En. Phys. {\bf 0407}, 056 (2004) hep-th/0405245.}

\nref\Ryz{A.V. Ryzhov, J. High En. Phys. {\bf 0111}, 046 (2001)
hep-th/0109064;\
Operators in the D=4, N=4 SYM and the AdS/CFT
correspondence, hep-th/0307169, UCLA thesis, 169 pages.}

\nref\DHRy{E. D'Hoker and A.V. Ryzhov,
J. High En. Phys. {\bf 0202}, 047 (2002)  hep-th/0109065.}

\nref\HMMR{L. Hoffmann, L. Mesref, A. Meziane and W. Ruehl,
Nucl. Phys. {\bf B641}, 188-222 (2002) hep-th/0112191.}

\nref\ArSo{G. Arutyunov and E. Sokatchev,
Nucl. Phys. {\bf B635}, 3-32 (2002) hep-th/0201145;\
Class. Quant. Grav. {\bf 20}, L123-L131 (2003)
hep-th/0209103.}

\nref\HoFr{E. D'Hoker and D.Z. Freedman,
in: {\it Boulder 2001, Strings, branes and extra dimensions}, pp. 3-158,
hep-th/0201253.}

\nref\TaTe{T. Takayanagi and S. Terashima,
J. High En. Phys. {\bf 0206}, 036 (2002)  hep-th/0203093.}

\nref\SeSu{E. Sezgin and P. Sundell,
Nucl. Phys. {\bf B644}, 303-370 (2002) hep-th/0205131.}

\nref\BERS{M. Bianchi, B. Eden, G. Rossi and Y.S. Stanev,
Nucl. Phys. {\bf B646}, 69-101 (2002)
hep-th/0205321.}

\nref\APPSS{G. Arutyunov, S. Penati, A. C. Petkou, A.
Santambrogio and E. Sokatchev,
Nucl. Phys. {\bf B643}, 49-78 (2002) hep-th/0206020.}

\nref\FGP{S. Fernando, M. Gunaydin and O. Pavlyk,
J. High En. Phys. {\bf 0210}, 007 (2002)
hep-th/0207175.}

\nref\DoOs{F.A. Dolan and H. Osborn, Annals Phys.
{\bf 307} (2003) 41-89;
hep-th/0209056.}

\nref\Bei{N. Beisert, 
Nucl. Phys. {\bf B659}, 79-118 (2003) hep-th/0211032;\
Nucl. Phys. {\bf B676}, 3-42 (2004) hep-th/0307015;\
Phys. Rep. {\bf 405}, 1-202 (2004) hep-th/0407277.}

\nref\HHHR{E. D'Hoker, P. Heslop, P. Howe and A.V. Ryzhov,
J. High En. Phys. {\bf 0304}, 038 (2003)  hep-th/0301104.}

\nref\CLZ{Shun-Jen Cheng, Ngau Lam and R.B. Zhang,
J. Algebra, {\bf 273}, 780-805 (2004)  math.RT/0301183.}

\nref\BKS{N. Beisert, C. Kristjansen and M. Staudacher,
Nucl. Phys. {\bf B664}, 131-184 (2003)
hep-th/0303060.}

\nref\DMW{A. Dhar, G. Mandal and S.R. Wadia,
String bits in small radius AdS and weakly coupled N=4 Super
 Yang-Mills Theory: I, hep-th/0304062.}

\nref\BMS{M. Bianchi, J.F. Morales and H. Samtleben,
J. High En. Phys. {\bf 0307}, 062 (2003)
hep-th/0305052.}

\nref\DHHK{J.M. Drummond, P.J. Heslop, P.S. Howe and S.F. Kerstan,
J. High En. Phys. {\bf 0308}, 016 (2003)  hep-th/0305202.}

\nref\FMS{M. Fukuma, S. Matsuura and T. Sakai,
 Prog. Theor. Phys. {\bf 109}, 489-562 (2003).}

\nref\LNR{T. Leonhardt, R. Manvelyan and W. Ruehl,
Nucl. Phys. {\bf B667}, 413-434 (2003)
hep-th/0305235;\
The group approach to AdS space propagators: A fast algorithm,
J. Phys. {\bf A37}, 7051 (2004) hep-th/0310063.}

\nref\Ter{J. Terning,
in {\it Boulder 2002, Particle physics and cosmology}, pp. 343-443,
hep-th/0306119.}

\nref\BeSt{N. Beisert and M. Staudacher,
Nucl. Phys. {\bf B670}, 439-463 (2003)
The N=4 SYM Integrable Super Spin Chain,
hep-th/0307042.}

\nref\Kov{S. Kovacs,
Nucl. Phys. {\bf B684}, 3-74 (2004)
hep-th/0310193.}

\nref\BBMS{N. Beisert, M. Bianchi, J.F. Morales and H. Samtleben,
J. High En. Phys. {\bf 0402}, 001 (2004)  hep-th/0310292;~~
J. High En. Phys. {\bf 0407}, 058 (2004)
hep-th/0405057.}

\nref\Kuj{J. Kujawa, Crystal structures arising from
representations of $GL(m|n)$, math.RT/0311251.}

\nref\Met{R.R. Metsaev,
Phys. Lett. {\bf B590}, 95-104 (2004)
hep-th/0312297;
Phys. Rev. {\bf D71} (2005) 085017; hep-th/0410239;
Class. Quant. Grav. {\bf 22} 2777-2796 (2005) hep-th/0412311;
AdS friendly light-cone formulation of
conformal field theory, hep-th/0512330.}

\nref\Sch{H.J. Schnitzer,
Nucl. Phys. {\bf B695}, 283-300 (2004) hep-th/0310210;\
Nucl. Phys. {\bf B695}, 267-282 (2004) hep-th/0402219.}

\nref\CMT{C. Csaki, P. Meade and J. Terning,
J. High En. Phys. {\bf 0404}, 040 (2004) hep-th/0403062.}

\nref\Rue{W. Ruehl,
Nucl. Phys. {\bf B705}, 437-456 (2005) hep-th/0403114;\
Phys. Lett. {\bf B605}, 413-418 (2005) hep-th/0409252.}

\nref\Zou{K. Zoubos,
J. High En. Phys.
{\bf 0501}, 031 (2005) hep-th/0403292.}

\nref\LePa{S. Lee and J.J. Park,
J. High En. Phys. {\bf 0406},  038 (2004) hep-th/0404051.}

\nref\DGS{F.A. Dolan, L. Gallot and E. Sokatchev,
 J. High En. Phys. {\bf 0409},  056 (2004) hep-th/0405180.}

\nref\AACDFR{D. Arnaudon, J. Avan, N. Crampe, A. Doikou, L. Frappat  and
E. Ragoucy,
J. Stat. Mech.: Theor. Exp.
(JSTAT) 08 (2004) P005; math-ph/0406021.}

\nref\NiOs{M. Nirschl and H. Osborn,
Nucl. Phys. {\bf B711}, 409-479 (2005)
hep-th/0407060.}

\nref\Dem{D.A. Demir,
J. High En. Phys. {\bf 0511},     003  (2005) hep-ph/0408043.}

\nref\AJS{B. Eden, C. Jarczak and E. Sokatchev,
Nucl. Phys. {\bf B712}, 157-195 (2005) hep-th/0409009.}

\nref\Bia{M. Bianchi,
Comptes Rendus Phys. {\bf 5}, 1091-1099, (2004) hep-th/0409292;
Fortsch. Phys. {\bf 53}, 665-691 (2005) hep-th/0409304.}

\nref\MoSa{J.F. Morales and H. Samtleben,
Phys. Lett. {\bf B607}, 286-293 (2005) hep-th/0411246.}

\nref\BLS{B.A. Burrington, J.T. Liu and W.A. Sabra,
Phys. Rev. {\bf D71}, 105015 (2005) hep-th/0412155.}

\nref\MaRu{R. Manvelyan and W. Ruehl,
Phys. Lett. {\bf B613}, 197-207 (2005) hep-th/0412252;
Nucl. Phys. {\bf B717}, 3-18 (2005) hep-th/0502123.}

\nref\CCTV{C. Carmeli, G. Cassinelli, A. Toigo and V.S.
Varadarajan,
Comm. Math. Phys. {\bf 263}, 217-258 (2006) hep-th/0501061.}

\nref\BGHR{A. Barabanschikov, L. Grant, L.L. Huang and S. Raju,
J. High En. Phys. {\bf 0601}, 160 (2006) hep-th/0501063.}

\nref\KrLo{K. Krasnov and J. Louko,
J. Math. Phys. {\bf 47}, 033513 (2006) math-ph/0502017.}

\nref\BiDi{M. Bianchi and V. Didenko, "Massive" Higher Spin
Multiplets and Holography, hep-th/0502220.}

\nref\GeSt{L. Genovese and Y.S. Stanev,
Nucl. Phys. {\bf B721}, 212-228 (2005) hep-th/0503084.}

\nref\DaGe{M. D'Alessandro and L. Genovese,
Nucl. Phys. {\bf B732}, 64-88 (2006) hep-th/0504061.}

\nref\ACSW{P.C. Argyres, M. Crescimanno, A.D. Shapere and J.R.
Wittig, Classification of N=2 Superconformal Field Theories with
Two-Dimensional Coulomb Branches, hep-th/0504070.}

\nref\BHR{M. Bianchi, P.J. Heslop and F. Riccioni, More on La
Grande Bouffe, J. High En. Phys. {\bf 0508},  088 (2005)
hep-th/0504156.}

\nref\Arv{P. Arvidsson, 
J. Math. Phys. {\bf 47}, 042301 (2006) hep-th/0505197.}

\nref\IINSY{M. Ibe, K.-I. Izawa, Yu Nakayama, Y. Shinbara and T.
Yanagida,
Phys. Rev. {\bf D73}, 015004  (2006) hep-ph/0506023;
Phys. Rev. {\bf D73},  035012 (2006) hep-ph/0509229.}

\nref\MiOl{G. Milanesi  and M. O'Loughlin,
J. High En. Phys. {\bf 0509}, 008  (2005) hep-th/0507056.}

\nref\HJS{J. Henn, C. Jarczak and E. Sokatchev,
Nucl. Phys. {\bf B730}, 191-209 (2005) hep-th/0507241.}

\nref\ZaPa{V.N. Zaikin and M.Ya. Palchik, Secondary Fields in D>2
Conformal Theories, hep-th/0509103.}

\nref\KMMS{J. Kinney, J. Maldacena, S. Minwalla and S. Raju, An
Index for 4 dimensional Super Conformal Theories, hep-th/0510251.}

\nref\Nak{Yu Nakayama, Index for Orbifold Quiver Gauge Theories,
hep-th/0512280;  Index for supergravity on $AdS_5 \times T^{1,1}$
and conifold gauge theory, hep-th/0602284.}

\nref\BRS{M. Berkooz, D. Reichmann and J. Simon, A Fermi Surface
Model for Large Supersymmetric $AdS_5$ Black Holes,
  hep-th/0604023.}


\nref\Nahm{W.~Nahm,
Nucl. Phys. {\bf B135}, 149 (1978).}

\nref\HLS{R. Haag, J.T. Lopuszanski and M. Sohnius,
Nucl. Phys. {\bf B88}, 257 (1975).}

\nref\FF{M. Flato and C. Fronsdal, Lett. Math. Phys. {\bf 8}, 159
(1984).}

\nref\DPm{V.K. Dobrev and V.B. Petkova,
Lett. Math. Phys. {\bf 9}, 287 (1985).}

\nref\DPf{V.K. Dobrev and V.B. Petkova, 
Fortschr. d. Phys. {\bf 35}, 537-572 (1987); first as ICTP Trieste preprint
IC/85/29 (March 1985).}

\nref\DPu{V.K. Dobrev and V.B. Petkova, 
Phys. Lett. {\bf 162B}, 127-132 (1985).}

\nref\DPp{V.K. Dobrev and V.B. Petkova, 
Proceedings, eds. A.O. Barut and H.D. Doebner,
Lecture Notes in Physics, Vol. 261 (Springer-Verlag, Berlin, 1986)
p. 291 
and p. 300.} 

\nref\Min{S. Minwalla, Adv. Theor. Math. Phys. {\bf 2}, 781-846
(1998).}

\nref\Dosix{V.K. Dobrev,
J. Phys. {\bf A35}, 7079-7100 (2002) ; hep-th/0201076.}

\nref\Sie{W. Siegel, Nucl. Phys. {\bf B177}, 325 (1981).}

\nref\HST{P.S. Howe, K.S. Stelle and P.K. Townsend,
Nucl. Phys. {\bf B192}, 332 (1981).}

\nref\GuMa{M. Gunaydin and N. Marcus, Class. Quant. Grav. {\bf 2}
L11 (1985).}


\lref\Mack{G. Mack, Comm. Math. Phys. {\bf 55}, 1 (1977).}

\lref\Yamn{H. Yamane, "Universal R-matrices for Quantum Groups
Associated to Simple Lie Superalgebras", Proc. Japan Acad. {\bf 67} Ser. A (1991)
108-112.}

\lref\Dobn{V.K. Dobrev,
Phys. Lett. {\bf 186B}, 43-51 (1987)}

\lref\KL{D. Kazhdan and G. Lusztig,~ Inv. Math. {\bf 53}, 165
(1979).}

\lref\Di{J. Dixmier, {\it Enveloping Algebras}, (North Holland,
New York, 1977).}

\lref\Kac{V.G. Kac, {\it Infinite--Dimensional Lie Algebras},
(Birkh\"auser, Boston, 1983).}

\lref\Kacs{V.G. Kac, Adv. Math. {\bf 26}, 8-96 (1977);\
Comm. Math. Phys. {\bf 53}, 31-64 (1977);\
the second paper
is an adaptation for physicists of the first paper.}

\lref\Kacl{V.G. Kac, Lect. Notes in Math. {\bf 676}
(Springer-Verlag, Berlin, 1978) pp. 597-626.}

\lref\Kacc{V.G. Kac, ``Characters of typical representations of classical Lie superalgebras'',
Comm. Algebra {\bf 5}, 889-897 (1977).}

\lref\DPe{V.K. Dobrev and V.B. Petkova, `Odd Reflection',
contribution to ``Concise Encyclopedia of Supersymmetry'', eds. S.
Duplij, W. Siegel and J. Bagger (Kluwer Academic Publishers, 2003)
pp. 282-283.}

\lref\Doo{V.K. Dobrev, ``Multiplet classification of the reducible
elementary representations of real semi-simple Lie groups: the
$SO_e(p,q)$ example'', Lett. Math. Phys. {\bf 9},  205-211 (1985);
Talk at the I National Congress of Bulgarian Physicists (Sofia,
1983) and INRNE Sofia preprint (1983).}

\lref\Dor{V.K. Dobrev, Reports Math. Phys. {\bf 25}, 159-181 (1988).}

\lref\Sha{N.N. Shapovalov, Funkts. Anal. Prilozh. {\bf 6} (4) 65
(1972);\ English translation: Funct. Anal. Appl. {\bf 6}, 307
(1972).}

\lref\Docond{V.K. Dobrev,
J. Phys. A: Math. Gen. {\bf 28}, 7135-7155 (1995).}

\lref\Serg{V.V. Serganova, Appendix to the paper:
D.A. Leites, M.V. Saveliev and V.V. Serganova, in: Proc. of
Group Theoretical Methods in Physics, Yurmala, 1985
(Nauka, Moscow, 1985, in Russian) p. 377; English translation in VNU
(Sci. Press, Dordrecht, 1987).} 

\lref\PeSk{I. Penkov and I. Skornyakov, C.R. Acad. Sc. Paris,
{\bf 299}, Serie I, 1005 (1984).} 

\lref\Pen{I. Penkov, J. Sov. Math. {\bf 51}, 2108 (1990).}


\lref\BL{I.N. Bernstein and D.A. Leites,
C.R. Acad. Bulg. Sci. {\bf 33}, 1049 (1980).}

\lref\JHKT{J. Van der Jeugt, J.W.B. Hughes, R.C. King and
J. Thierry-Mieg, Comm. Algebra, {\bf 18}, 3453 (1990);
J. Math. Phys. {\bf 31}, 2278 (1990) .} 

\lref\Jeu{J. Van der Jeugt,
Comm. Algebra, {\bf 19}, 199 (1991).}

\lref\Serga{V. Serganova, 
       Selecta Math. {\bf 2}, 607 (1996).} 

\lref\VZ{J. van der Jeugt and R.B. Zhang, 
  Lett. Math. Phys. {\bf 47}, 49 (1999).}

\lref\Bru{J. Brundan, 
   J. Amer. Math. Soc. {\bf 16} (2002) 185; 
Adv. Math. {\bf 182}, 28 (2004).} 

\lref\SuZh{Yucai Su and R.B. Zhang, Character and dimension formulae for
general linear superalgebra, math.QA/0403315.}


\null

\vskip 2truecm

\centerline{{\tfont Characters of the Positive Energy UIRs of}}
\vskip 2truemm
\centerline{{\tfont D=4 Conformal Supersymmetry}}

\vskip 1.5cm

\centerline{{\bf V.K. Dobrev}}
\vskip 0.5cm

\centerline{School of Informatics, University of Northumbria,}
\centerline{Ellison Building, Newcastle upon Tyne, NE1 8ST, UK}
\centerline{and}
\centerline{Institute of Nuclear Research and Nuclear Energy}
\centerline{Bulgarian Academy of Sciences,}
\centerline{72 Tsarigradsko Chaussee, 1784 Sofia, Bulgaria}
\centerline{(permanent address)}
\vskip 1.5cm

\centerline{{\bf Abstract}}

\midinsert\narrower\narrower{\male
We give character formulae for the
positive energy unitary irreducible
representations of the N-extended D=4 conformal
superalgebras  su(2,2/N). Using these we also derive
decompositions of long superfields
as they descend to the unitarity threshold.
These results are also applicable to irreps of the
complex Lie superalgebras sl(4/N). Our derivations
use results from the representation theory of su(2,2/N)
developed already in the 80s.
}\endinsert

\vskip 1.5cm

\newsec{Introduction}

\nt Recently, superconformal field theories in various dimensions
are attracting more interest, cf. \refs{\LMRS{--}\BRS} and
references therein. Particularly important are those for ~$D\leq
6$~ since in these cases the relevant superconformal algebras
satisfy \Nahm{} the Haag-Lopuszanski-Sohnius theorem \HLS. This
makes the classification of the UIRs of these superalgebras very
important. Until recently such classification was known only for
the ~$D=4$~ superconformal algebras ~$su(2,2/1)$ \FF{} and
~$su(2,2/N)$ \refs{\DPm{--}\DPp} (for arbitrary $N$). Recently,
the classification for ~$D=3$ (for even $N$), $D=5$, and $D=6$
(for $N=1,2$)~ was given in \Min{} (some results being
conjectural), and then the $D=6$ case (for arbitrary $N$) was
finalized in \Dosix.

Once we know the UIRs of a (super-)algebra the next question is to find
their characters, since these give the spectrum which is
important for the applications. Some results on the spectrum
were given in the early papers \refs{\Sie{--}\GuMa,\DPu} but it is
necessary to have systematic results for which the character formulae
are needed. This is the question we address in this paper for the UIRs of
$D=4$ conformal superalgebras $su(2,2/N)$.  From the mathematical
point of view this question is clear only for representations
with conformal dimension above the unitarity threshold
viewed as irreps of the corresponding complex superalgebra
$sl(4/N)$.  But for $su(2,2/N)$ even the UIRs above the unitarity
threshold are truncated for small values of spin and isospin.
More than that, in the applications the most important role
is played by the representations with ``quantized" conformal
dimensions at the unitarity threshold and at discrete points
below.  In the quantum field or string theory framework some of
these correspond to operators with ``protected" scaling dimension
and therefore imply ``non-renormalization theorems" at the
quantum level, cf., e.g., \HeHo,\FSa.

Thus, we need detailed knowledge about the structure of the UIRs
from the representation-theoretical point of view.  Fortunately,
such information is contained in \refs{\DPm{--}\DPp}.  Following
these papers in Section 2 we recall the basic ingredients of the
representation theory of the D=4 superconformal algebras.  In
particular we recall the structure of Verma modules and UIRs.
Using this information we are able to derive character formulae,
some of which are very explicit, cf. Section 3. We also pin-point
the difference between character formulae for $sl(4,N)$ and
$su(2,2/N)$ since for the latter we need to introduce and use the
notion of counter-terms in the character formulae.  The general
formulae are valid for arbitrary $N$. For illustration we give
more explicit formulae for $N=1,2$, but we leave the example $N=4$
for a follow-up paper, since that would take too many pages, and
the present paper is long enough.   In Section 4 we summarize our
results on the decompositions of long superfields as they descend
to the unitarity threshold. These results may be applied to the
problem of operators with protected dimensions.

\np

\newsec{Representations of D=4 conformal supersymmetry}

\subsec{The setting}

\nt
The superconformal algebras in $D=4$ are ~$\cg ~=~ su(2,2/N)$.
The even subalgebra of ~$\cg$~ is the algebra ~$\cg_0 ~=~ su(2,2) \oplus u(1)
\oplus su(N)$. We label their physically relevant representations of
~$\cg$~ by the signature:
\eqn\sgn{\chi ~=~ [\,d\,;\, j_1\,,\,j_2\,;\,z\,;\,r_1\,,\ldots,r_{N-1}\,]}
where ~$d$~ is the conformal weight, ~$j_1,j_2$~ are
non-negative (half-)integers which are Dynkin labels of the finite-dimensional
irreps of the $D=4$ Lorentz subalgebra ~$so(3,1)$~ of dimension
~$(2j_1+1)(2j_2+1)$, ~$z$~ represents the ~$u(1)$~ subalgebra which
is central for ~$\cg_0$~ (and for $N=4$ is central for $\cg$ itself),
and ~$r_1,\ldots,r_{N-1}$~ are
non-negative integers which are Dynkin labels of the finite-dimensional
irreps of the internal (or $R$) symmetry algebra ~$su(N)$.

We recall that the algebraic approach to $D=4$ conformal
supersymmetry developed in \refs{\DPm{--}\DPp} involves two
related constructions - on function spaces and as Verma modules.
The first realization employs the explicit construction of induced
representations of ~$\cg$~ (and of the corresponding supergroup
~$G ~=~ SU(2,2/N)$) in spaces of functions (superfields) over
superspace which are called elementary representations (ER). The
UIRs of $\cg$ are realized as irreducible components of ERs, and
then they coincide with the usually used superfields in indexless
notation. The Verma module realization is also very useful as it
provides simpler and more intuitive picture for the relation
between reducible ERs, for the construction of the irreps, in
particular, of the UIRs. For the latter the main tool is an
adaptation of the Shapovalov form \Sha{} to the Verma modules
\DPu,\DPp. Here we shall need only the second - Verma module -
construction.

\vskip 5mm

\subsec{Verma modules}

\nt  To introduce Verma modules
one needs the standard triangular decomposition: \eqn\trig{
\cg^\bac ~=~ \cg^+ \oplus \ch \oplus \cg^-} where ~$\cg^\bac ~=~
sl(4/N)$~ is the complexification of $\cg$, ~~$\cg^+$, $\cg^-$,
resp., are the subalgebras corresponding to the positive,
negative, roots of ~$\cg^\bac$, resp., and $\ch$ denotes the
Cartan subalgebra of ~$\cg^\bac$.

We consider lowest weight Verma modules,
so that ~$V^\L ~ \cong U(\cg^+) \otimes v_0\,$,
 where ~$U(\cg^+)$~ is the universal enveloping algebra of $\cg^+$,
~$\L\in\ch^*$~ is the lowest weight,
and ~$v_0$~ is the lowest weight vector $v_0$ such that:
\eqn\low{\eqalign{
 X \ v_0\ =&\ 0 \ , \quad X\in \cg^- \ , \cr
 H \ v_0 \ =&\ \L(H)\ v_0 \ , \quad H\in \ch \ .}}
Further, for simplicity we omit the sign
~$\otimes \,$, i.e., we write $P\,v_0\in V^\L$ with $P\in U(\cg^+)$.

The lowest weight $\L$ is characterized by its values on the Cartan
subalgebra ~$\ch$, or, equivalently, by its products with the
simple roots (given explicitly below). In general, these would
be ~$3+N$~ complex numbers, however, in order to be useful for
the representations of the real form ~$\cg$~ these values would
be restricted to be real and furthermore to correspond to the
signatures ~$\chi\,$, and we shall write ~$\L = \L(\chi)$, or
~$\chi = \chi(\L)$.
Note, however, that there are Verma modules to which correspond
no ERs, cf. \DPf{} and below.

If a Verma module ~$V^\L$~ is irreducible then it gives the lowest
weight irrep ~$L_\L$~ with the same weight. If a Verma module
~$V^\L$~ is reducible then it contains a maximal invariant
submodule ~$I^\L$~ and the lowest weight irrep ~$L_\L$~ with the
same weight is given by factorization: ~$L_\L ~=~ V^\L\,/\,I^\L$
\Di,\Kac,\Kacl.

Thus, we need first to know which Verma modules are reducible. The
reducibility conditions for highest weight Verma modules over
basic classical Lie superalgebra were given by Kac \Kacl.
Translating his conditions to lowest weight Verma modules we have
\DPf: ~~~A lowest weight Verma module ~$V^\L$~ is reducible only
if at least one of the following conditions is true:\foot{Many
statements below are true for any basic classical Lie
superalgebra, and would require changes only for the superalgebras
$osp(1/2N)$.} \eqna\redg
$$\eqalignno{ (\r - \L, \b) ~=& ~m (\b,\b)/2 \ , \qquad \b \in \D^+ \ ,
~~(\b,\b) \neq 0\ ,\quad m\in\bbn \ , &\redg a\cr
(\r - \L, \b) ~=& ~0 \ , \qquad \b \in \D^+ \ , ~~(\b,\b) = 0\
 , &\redg b\cr} $$
where ~$\Delta^+$~ is the positive root system of ~$\cg^\bac$,
~$\r\in\ch^*$~ is the very important in representation
theory element given by ~$\r = \r_\o - \r_\I\,$,
where ~$\r_\o\,,\r_\I$~ are the half-sums of the even, odd, resp.,
positive roots, ~$(\cdot,\cdot)$~ is the standard bilinear product
in ~$\ch^*$.

If a condition from \redg{a} is fulfilled then ~$V^\L$~ contains a
submodule which is a Verma module ~$V^{\L'}$~ with shifted weight
given by the pair ~$m,\b$~: ~$\L' ~=~ \L + m\b$. The embedding of
~$V^{\L'}$~ in ~$V^\L$~ is provided by mapping the lowest weight vector
~$v'_0$~ of ~$V^{\L'}$~ to the singular vector ~$v_s^{m,\b}$~
in ~$V^\L$~ which is completely determined by the conditions:
\eqn\lowp{\eqalign{
 X \ v_s^{m,\b}\ =&\ 0 \ , \quad X\in \cg^- \ , \cr
 H \ v_s^{m,\b} \ =&\ \L'(H)\ v_0 \ , \quad H\in \ch \ ,
~~~\L' ~=~ \L + m\b\ .}} Explicitly, ~$v_s^{m,\b}$~ is given by an
even polynomial in the positive root generators: \eqn\sing{
v_s^{m,\b} ~=~ P^{m,\b} \,v_0 \ , \quad P^{m,\b}\in U(\cg^+)\ .}
Thus, the submodule of ~$V^\L$~ which is isomorphic to ~$V^{\L'}$~
is given by ~$U(\cg^+)\, P^{m,\b} \,v_0\,$.
[More on the even case following the same approach may be seen in, e.g.,
\Doo,\Dor.]

If a condition from \redg{b} is fulfilled then ~$V^\L$~ contains a
submodule ~$I^{\b}$~ obtained from the Verma module ~$V^{\L'}$~
with shifted weight ~$\L' ~=~ \L + \b$ as follows. In this situation
~$V^\L$~ contains a singular vector
\eqn\lows{\eqalign{
 X \ v_s^{\b}\ =&\ 0 \ , \quad X\in \cg^- \ , \cr
 H \ v_s^{\b} \ =&\ \L'(H)\ v_0 \ , \quad H\in \ch \ ,
~~ \L' ~=~ \L + \b }}
Explicitly, ~$v_s^{\b}$~ is given by an odd polynomial in the positive
root generators:
\eqn\sing{ v_s^{\b} ~=~ P^{\b} \,v_0 \ , \quad P^{\b}\in U(\cg^+) \ .}
Then we have:
\eqn\subm{ I^{\b} ~=~ U(\cg^+)\, P^{\b} \,v_0 }
which is smaller than ~$V^{\L'} ~=~ U(\cg^+)\, v'_0$~ since
this polynomial is Grassmannian:
\eqn\gras{\( P^{\b} \)^2 ~=~ 0 \ .}
To describe this situation we say that ~$V^{\L'}$~ is
~{\bf oddly embedded}~ in ~$V^{\L}$.

Note, however, that the above formulae describe also more general
situations when the difference ~$\L'-\L = \b $~ is not a root, as
used in \DPf, and below.

The weight shifts ~$\L' = \L + \b$, \DPf\ when ~$\b$~ is an odd root are
generalized ~{\bf odd~reflections}, and for future
reference will be denoted as: \eqn\oddr{ \hs_\b \cdot \L ~\equiv~ \L
+ \b \ , \quad (\b,\b)= 0 , \ (\L,\b) = (\r,\b) \ .} Each such odd
reflection generates an infinite discrete abelian group:
\eqn\oddwb{\eqalign{ \tW_\b ~\equiv&~ \{ (\hs_\b)^n ~|~ n\in\bbz \}
\ ,\cr &\ell ((\hs_\b)^n) ~=~ n \ , }} where the unit element is
obviously obtained for $n=0$, and ~$(\hs_\b)^{-n}$~ is the inverse
of ~$(\hs_\b)^n$, and for future use we have also defined the length
function ~$\ell(\cdot)$~ on the elements of ~$\tW_\b\,$. This group
acts on the weights $\L$ extending \oddr{}: \eqn\oddw{ (\hs_\b)^n\,
\cdot \L ~=~ \L + n\b \ , \quad n\in\bbz \ , \quad (\b,\b)= 0 , \
(\L,\b) = (\r,\b) \ .} This is related to the fact that there is a
doubly-infinite chain of oddly embedded Verma modules whenever a
Verma module is reducible w.r.t. an odd root. This is explained in
detail and used for the classification of the Verma modules in \DPm,
and shall be used below.

Further, to be more explicit we need to recall the root
system of ~$\cg^\bac$~ - for definiteness - as used in \DPf{}. The
positive root system ~$\D^+$~ is comprised from ~$\a_{ij}\,$,
~$1\leq i <j \leq 4+N$. The even positive root system ~$\D^+_\o$~
is comprised from ~$\a_{ij}\,$, with\ $i,j\leq 4$~ and ~$i,j\geq
5$; ~the odd positive root system ~$\D^+_\I$~
is comprised from ~$\a_{ij}\,$, with ~$i\leq 4, j \geq 5$.
The simple roots are chosen as in (2.4) of \DPf{}:
\eqn\smplr{ \g_1 = \a_{12}\ , ~\g_2 = \a_{34}\ , ~\g_3 = \a_{25}\ ,
~\g_4 = \a_{4,4+N}\ , ~~\g_k = \a_{k,k+1}\ , ~~ 5\leq k\leq 3+N}
Thus, the Dynkin diagram is:
\eqn\dynk{
\vbox{\offinterlineskip\baselineskip=10pt
\halign{\strut#
\hfil
& #\hfil
\cr &\cr
& $\white{{1}} \riga \gray{{3}} \riga \white{{5}} \riga
\cdots \riga \white{{3+N}} \riga
\gray{{4}} \riga \white{{2}}$
\cr }}}
This is a non-distinguished simple root system with two odd
simple roots (for the various root systems of the basic classical
superalgebras we refer to \Kacs).

Let ~$\L ~=~ \L(\chi)$. The products of ~$\L$~ with the simple
roots are \DPf:
\eqna\valu
$$\eqalignno{
(\L,\g_a) ~=&~ -2j_a \ , \qquad a=1,2 \ , &\valu a\cr
(\L,\g_3) ~=&~ \half (d+z') + j_1 - m/N + 1 \ ,&\valu b\cr
(\L,\g_4) ~=&~ \half (d-z') + j_2 - m_1 + m/N + 1 \ ,
&\valu c\cr
&z' \equiv z (1-\d_{N4}) \cr
(\L,\g_j) ~=&~ r_{N+4-j} \ , \qquad 5\leq j\leq 3+N &\valu d}$$
These formulae give the correspondence between signatures $\chi$
and lowest weights $\L(\chi)$.\foot{For ~$N=4$~ the factor $u(1)$
in $\cg_0$ becomes central in $\cg$ and $\cg^\bac$. Consequently,
the representation parameter $z$ can not come from the products
of $\L$ with the simple roots, as indicated in \valu{}. In that
case the lowest weight is actually given by the sum ~$\L +
{\tilde \L}$, where ${\tilde \L}$ carries the representation
parameter $z$. This is explained in detail in \DPf{} and further
we shall not comment more on it, but the peculiarities for $N=4$
will be evident in the formulae.}

In the case of even roots ~$\b\in\Delta^+_\o$~ there are six roots
~$\a_{ij}\,$, $j\leq 4$,
coming from the ~$sl(4)$~ factor (which is complexification of $su(2,2)$)
and ~$N(N-1)/2$ roots ~$\a_{ij}\,$, $5\leq i$,
coming form the ~$sl(N)$~ factor
(complexification of $su(N)$).

The reducibility conditions w.r.t. to the positive roots coming
from ~$sl(4)(su(2,2))$~ coming from \redg{} (denoting $m\to
n_{ij}$ for $\b\to \a_{ij}$) are:
\eqna\redd
$$\eqalignno{
n_{12} ~~=&~~ 1 + 2j_1 ~\equiv n_1
&\redd a\cr
n_{23} ~~=&~~ 1 - d - j_1 - j_2 ~\equiv n_2
&\redd b\cr
n_{34} ~~=&~~ 1 + 2j_2 ~\equiv n_3 &\redd c\cr
n_{13} ~~=&~~ 2 - d + j_1 - j_2
~~=~~ n_1 + n_2 &\redd d\cr
n_{24} ~~=&~~ 2 - d - j_1 + j_2
~~=~~ n_2 + n_3 &\redd e\cr
n_{14} ~~=&~~ 3 - d + j_1 + j_2
~~=~~ n_1 + n_2 + n_3 &\redd f\cr}$$

Thus, reducibility conditions \redd{a,c} are fulfilled automatically
for ~$\L(\chi)$~ with $\chi$ from \sgn{} since we always have:
~$n_1,n_3\in\bbn$.
 There are no such conditions for the ERs since they are induced
from the finite-dimensional irreps of the Lorentz subalgebra
(parametrized by $j_1,j_2\,$.) However, to these two conditions
correspond differential operators of order ~$1+2j_1$~ and ~$1+2j_2$~
(as we mentioned above) and these annihilate all functions of the
ERs with signature $\chi$.

The reducibility conditions w.r.t. to the positive roots coming
from ~$sl(N)(su(N))$~ are all fulfilled for ~$\L(\chi)$~
with $\chi$ from \sgn.
In particular, for the simple roots from those
condition \redg{} is fulfilled with ~$\b\to\g_j\,$, $m=1+r_{N+4-j}\,$.
for every $j=5,6,\ldots,N+3$.
There are no such conditions for the ERs since they are induced
from the finite-dimensional UIRs of $su(N)$.
However, to these $N-1$ conditions correspond $N-1$ differential
operators of orders $1+r_k$ (as we mentioned) and
the functions of our ERs are annihilated by all
these operators \DPf.\foot{Note that there are actually as many
operators as positive roots of $sl(N)$ but all are expressed in
terms of the $N-1$ above corresponding to the simple roots \DPf.}

For future use we note also the following decompositions:
\eqna\lamde
$$\eqalignno{ \L ~=&~ \sum_{j=1}^{N+3}\ \l_j\ \a_{j,j+1} ~=~ \L^s + \L^z + \L^u
&\lamde a\cr
\L^s \equiv& \sum_{j=1}^{3}\ \l_j\ \a_{j,j+1}\ , \quad \L^z \equiv \l_4\ \a_{45} \ , \quad
\L^u \equiv \sum_{j=5}^{N+3}\ \l_j\ \a_{j,j+1}\  &\lamde b\cr}$$
which actually employ the distinguished root system with one odd root $\a_{45}\,$.

The reducibility conditions for the
~$4N$~ odd positive roots of $\cg$ are \DPu,\DPf:
\eqna\redu
$$\eqalignno{
& d ~=~ d^1_{Nk} - z \d_{N4} &\redu {a.k}\cr
& d^1_{Nk} ~\equiv~ 4-2k +2j_2 +z+2m_k -2m/N \cr &&\cr
& d ~=~ d^2_{Nk} - z \d_{N4} &\redu {b.k}\cr
& d^2_{Nk} ~\equiv~ 2-2k -2j_2 +z+2m_k -2m/N \cr &&\cr
& d ~=~ d^3_{Nk} + z \d_{N4} &\redu {c.k}\cr
& d^3_{Nk} ~\equiv~ 2+2k-2N +2j_1 -z-2m_k +2m/N \cr &&\cr
& d ~=~ d^4_{Nk} + z \d_{N4} &\redu {d.k}\cr
& d^4_{Nk} ~\equiv~ 2k-2N -2j_1 -z-2m_k +2m/N \cr }$$
where in all four cases of \redu{} ~$k=1,\ldots,N$, ~$m_N\equiv 0$, and
\eqn\mkm{
m_k \equiv \sum_{i=k}^{N-1} r_i \ , \quad
m \equiv \sum_{k=1}^{N-1} m_k = \sum_{k=1}^{N-1} k r_k}
$m_k$ is the number of cells of the $k$-th row of the standard Young tableau,
$m$ is the total number of cells.
Condition \redu{a.k} corresponds to the root ~$\a_{3,N+5-k}\,$,
\redu{b.k} corresponds to the root ~$\a_{4,N+5-k}\,$,
\redu{c.k} corresponds to the root ~$\a_{1,N+5-k}\,$,
\redu{d.k} corresponds to the root ~$\a_{2,N+5-k}\,$.

Note that for a fixed module and fixed ~$i=1,2,3,4$~ only one of
the odd ~$N$~ conditions involving ~$d^i_{Nk}$~ may be satisfied. Thus, no
more than four (two, for $N=1$)  of the conditions \redu{} may hold for a given Verma module.

\nt
{\it Remark:}~~ {\iti Note that for ~$n_2\in\bbn$~ (cf. \redd{})
 the corresponding irreps of ~su(2,2)~ are finite-dimensional
 (the necessary and sufficient condition for this is:
~$n_1,n_2,n_3\in\bbn$).
Then the irreducible LWM ~L$_\L$~ of ~su(2,2/N)~ are also
finite-dimensional (and non-unitary) independently on whether
the corresponding Verma module ~V$^\L$~
is reducible w.r.t. any odd root.
If ~V$^\L$~ is not reducible w.r.t. any odd root, then these
finite-dimensional irreps are called 'typical' \Kacl, otherwise,
the irreps are called 'atypical' \Kacl. In our considerations
~$n_2\notin\bbn$~ in all cases, except the trivial 1-dimensional
UIR (for which $n_2=1$, cf. below).}\dia

We shall consider quotients of Verma modules
factoring out the ~{\it even}~ submodules for which the reducibility
conditions are always fulfilled. Before this we recall
the root vectors following \DPf. The positive (negative) root vectors
corresponding to ~$\a_{ij}$, ($-\a_{ij}$), are denoted by
~$X^+_{ij}$, ($X^-_{ij}$).
In the ~$su(2,2/N)$~ matrix notation the convention of \DPf,
(2.7), is:
\eqn\roov
{\eqalign{
X^+_{ij} ~=&~ \cases{ e_{ji} ~&~ for $(i,j) = (3,4),(3,j),(4,j),
~~5\leq j \leq N+4$ \cr
e_{ij} ~&~ otherwise } \cr
X^-_{ij} ~=&~ ^t(X^+_{ij}) }}
where $e_{ij}$ are $(N+4)\times(N+4)$ matrices with all elements zero
except the element equal to 1 on the intersection of the
i-th row and j-th column. The simple root vectors ~$X^+_i$~
follow the notation of the simple roots $\g_i\,$ \smplr{}:
\eqn\sroov{
X^+_1 \equiv X^+_{12}\ , ~X^+_2 \equiv X^+_{34}\ , ~X^+_3 \equiv X^+_{25}\ ,
~X^+_4 \equiv X^+_{4,4+N}\ , ~~X^+_k \equiv X^+_{k,k+1}\ , ~~ 5\leq k\leq 3+N}

The mentioned submodules are generated by the singular vectors
related to the even simple roots
~$\g_1,\g_2,\g_5,\ldots,\g_{N+3}$~\DPf{}:
\eqna\sings
$$\eqalignno{ v^1_s ~=&~ (X^+_1)^{1+2j_1}\, v_0 \ ,&\sings a\cr
 v^2_s ~=&~ (X^+_2)^{1+2j_2}\, v_0 \ , &\sings b\cr
v^j_s ~=&~ (X^+_j)^{1+r_{N+4-j}}\, v_0 \ ,
\quad j=5,\ldots,N+3 &\sings c\cr}$$
(for ~$N=1$~ \sings{c} being empty).
The corresponding submodules are ~$I^\L_k ~=~ U(\cg^+)\,v^k_s\,$,
and the invariant submodule to be factored out is:
\eqn\subb{ I^\L_c ~=~ \bigcup_k\, I^\L_k }
Thus, instead of ~$V^\L$~ we shall consider the factor-modules:
\eqn\fcc{ \tV^\L ~=~ V^\L\, /\, I^\L_c }
which are closer to the structure of the ERs.
In the factorized modules the singular vectors \sings{} become
null conditions, i.e., denoting by ~$\widetilde{|\L\rg}$~ the
lowest weight vector of ~$\tV^\L$, we have:
\eqna\nulm
$$\eqalignno{ &(X^+_1)^{1+2j_1}\, \widetilde{|\L\rg} ~=~ 0 \ , &\nulm a\cr
&(X^+_2)^{1+2j_2}\, \widetilde{|\L\rg} ~=~ 0 \ ,
&\nulm b\cr
&(X^+_j)^{1+r_{N+4-j}}\, \widetilde{|\L\rg} ~=~ 0 \ ,
\quad j=5,\ldots,N+3 &\nulm c\cr}$$

\vskip 5mm

\subsec{Singular vectors and invariant submodules
at the unitary reduction points}

\nt
We first recall the result of \DPu{} (cf. part (i) of the Theorem
there) that the following is the complete list of lowest weight
(positive energy) UIRs of $su(2,2/N)$~:
\eqna\unitt
$$\eqalignno{
&d ~\geq~ d_{\rm max} ~=~ \max (d^1_{N1}, d^3_{NN})\ , &\unitt a\cr
&d ~=~ d^4_{NN} \geq d^1_{N1}\ , ~~ j_1=0\ , &\unitt b\cr
&d ~=~ d^2_{N1} \geq d^3_{NN}\ , ~~ j_2=0\ ,&\unitt c\cr
&d ~=~ d^2_{N1} = d^4_{NN}\ , ~~ j_1 = j_2=0\ ,&\unitt d\cr }$$
where ~$d_{\rm max}$~ is the threshold of the continuous unitary
spectrum.\foot{Note that from \unitt{a} follows:
$$ d_{\rm max} ~\geq~ 2 + j_1 + j_2 + m_1 \ , $$
the equality being achieved only when ~$d^1_{N1}= d^3_{NN}\,$,
while from \unitt{b,c} follows:
$$ d ~\geq~ 1 + j_1 + j_2 + m_1\ , \qquad j_1j_2=0 \ , $$
the equality being achieved only when ~$d^4_{NN}= d^1_{N1}\,$, or
~$d^2_{N1}= d^3_{NN}\,$, for \unitt{b}, \unitt{c}, resp.
Recalling the unitarity conditions \Mack{} for the conformal algebra
su(2,2)~:
$$\eqalignno{
&d ~\geq~ 2 + j_1 + j_2 \ , \qquad j_1j_2> 0 \ ,\cr
&d ~\geq~ 1 + j_1 + j_2 \ , \qquad j_1j_2= 0 \ ,}$$
we see that the superconformal unitarity conditions are more
stringent that the conformal ones.}
 Note that in case (d) we have $d=m_1$, $z=2m/N -m_1\,$,
and that it is trivial for $N=1$ since then the internal symmetry
algebra $su(N)$ is trivial and by definition $m_1=m=0$
(the resulting irrep is 1-dimensional with
$d=z=j_1=j_2=0$). The UIRs for N=1 were first given in \FF.

Next we note that if ~$d ~>~ d_{\rm max}$~ the factorized Verma
modules are irreducible and coincide with the UIRs ~$L_\L\,$.
These UIRs are called ~${\bf long}$~ in the modern literature,
cf., e.g., \FGPW,\FS,\FSa,\AES,\BKRSa,\EdSo,\HeHoa.
Analogously, we shall use for the cases when ~$d= d_{\rm
max}\,$, i.e., \unitt{a}, the terminology of ~{\bf semi-short}~ UIRs,
introduced in \FGPW,\FSa, while the cases \unitt{b,c,d} are also called ~{\bf
short}~ UIRs, cf., e.g., \FS,\FSa,\AES,\BKRSa,\EdSo,\HeHoa.

Next consider in more detail the UIRs at the four distinguished
reduction points determining the list above:
\eqn\diss{\eqalign{
& d^1_{N1} ~=~ 2 +2j_2 +z+2m_1 -2m/N\ , \cr
& d^2_{N1} ~=~ z+2m_1 -2m/N \ , \quad (j_2=0)\ ,\cr
& d^3_{NN} ~=~ 2 +2j_1 -z +2m/N \ ,\cr
& d^4_{NN} ~=~ -z +2m/N \ , \quad (j_1=0)\ . }}

First we recall the singular vectors corresponding
to these points. The above reducibilities occur for the following
odd roots, resp.:
\eqn\disr{ \a_{3,4+N} \ , \quad \a_{4,4+N} \ , \quad
\a_{15} \ , \quad \a_{25} \ . }
The second and the fourth are the two odd simple roots:
\eqn\smpl{ \g_3 ~=~ \a_{25} \ , \quad \g_4 ~=~ \a_{4,4+N} }
and the other two are simply related to these:
\eqn\nsmpl{ \a_{15} ~=~ \a_{12} + \a_{25} ~=~ \g_1 + \g_3 \ , \qquad
\a_{3,4+N} ~=~ \a_{34} + \a_{4,4+N} ~=~ \g_2 + \g_4 \ .}

Thus, the corresponding singular vectors are:
\eqna\singg
$$\eqalignno{
v^1_{\rm odd} ~=&~ P_{3,4+N} \, v_0 ~=~ \(
 X^+_4 X^+_2 (h_2-1) - X^+_2 X^+_4 h_2
 \)\, v_0 ~= &\singg a\cr
=&~ \( 2j_2 X^+_2 X^+_4 - (2j_2+1) X^+_4 X^+_2 \)\, v_0 ~=
&\cr
=&~ \( 2j_2 X^+_{3,4+N} - X^+_4 X^+_2 \) \, v_0
\ ,\qquad d=d^1_{N1}
&\singg {a'}\cr
v^2_{\rm odd} ~=&~ X^+_4\, v_0 \ ,\qquad d=d^2_{N1}
&\singg b\cr
v^3_{\rm odd} ~=&~ P_{15} \, v_0 ~=~
\( X^+_3 X^+_1 (h_1-1) - X^+_1 X^+_3 h_1
 \)\, v_0 ~= &\singg c\cr
=&~ \( 2j_1 X^+_1 X^+_3 - (2j_1+1) X^+_3 X^+_1 \)\, v_0 ~=
&\cr
=&~ \( 2j_1 X^+_{15} - X^+_3 X^+_1 \)\, v_0
\ ,\qquad d=d^3_{NN}
&\singg {c'}\cr
v^4_{\rm odd} ~=&~ X^+_3\, v_0\ ,\qquad d=d^4_{NN} \ ,
&\singg d\cr }$$
where ~$X^+_{3,4+N} = [X^+_2,X^+_4]$~ is the odd root vector
corresponding to the root ~$\a_{3,4+N}\,$,
~$X^+_{15} = [X^+_1,X^+_3]$~ is the odd root vector
corresponding to the root ~$\a_{15}\,$, ~$h_1\,,h_2\,\in\ch$~ are
Cartan generators corresponding to the roots ~$\g_1\,,\g_2\,$,
(cf. \DPf), and passing from the
\singg{a}, (\singg{c}), to the next line we have used the fact
that ~$h_2\,v_0 = -2j_2\,v_0\,$, ($h_1\,v_0 = -2j_1\,v_0\,$),
consistently with \valu{b}, (\valu{a}).
These vectors are given in (8.9a),(8.7b),(8.8a),(8.7a), resp., of
\DPf{}.

These singular vectors carry over for the factorized Verma modules $\tV^\L$~:
\eqna\tsing
$$\eqalignno{
\tv^1_{\rm odd} ~=&~ P_{3,4+N} \, \widetilde{|\L\rg} ~=~
\( X^+_4 X^+_2 (h_2-1) - X^+_2 X^+_4 h_2\)
\, \widetilde{|\L\rg} ~=
&\tsing a\cr
=&~ \( 2j_2 X^+_{3,4+N} - X^+_4 X^+_2 \) \, \widetilde{|\L\rg}
\ ,\qquad d=d^1_{N1}
&\tsing {a'}\cr
\tv^2_{\rm odd} ~=&~ X^+_4\, \widetilde{|\L\rg} \ ,\qquad d=d^2_{N1}
&\tsing b\cr
\tv^3_{\rm odd} ~=&~ P_{15} \, \widetilde{|\L\rg} ~=~
\( X^+_3 X^+_1 (h_1-1) - X^+_1 X^+_3 h_1\)\, \widetilde{|\L\rg} ~=
&\tsing c\cr
=&~ \( 2j_1 X^+_{15} - X^+_3 X^+_1 \)\, \widetilde{|\L\rg}
\ ,\qquad d=d^3_{NN}
&\tsing {c'}\cr
\tv^4_{\rm odd} ~=&~ X^+_3\, \widetilde{|\L\rg}\ ,\qquad d=d^4_{NN} \ .
&\tsing d\cr }$$

For $j_1=0$, $j_2=0$, resp., the vector $v^3_{\rm odd}$,
$v^1_{\rm odd}$, resp., is
a descendant of the singular vector $v^1_s$, $v^2_s$, resp.,
cf. \sings{a}, \sings{b}, resp.
In the same situations the tilde counterparts
$\tv^1_s$, $\tv^2_s$ are just zero - cf.
\nulm{a}, \nulm{b}, resp. However, then there is another
independent singular vector of $\tV^\L$
in both cases. For $j_1=0$ it
corresponds to the sum of two roots: $\a_{15}+\a_{25}$
(which sum is not a root!) and is given by formula (D.1) of \DPf:
\eqn\nnra{ \tv^{34} ~=~ X^+_3\,X^+_1\,X^+_3\, \widetilde{|\L\rg} ~=~
X^+_3\,X^+_{15} \, \widetilde{|\L\rg}
\ ,\qquad d=d^3_{NN} \ ,\ j_1=0 }
Checking singularity we see at once that $X^-_k\, \tv^{34}=0$
for $k\neq 3$. It remains to calculate the action of $X^-_3$~:
$$\eqalignno{ X^-_3\, \tv^{34} ~=&~ h_3 \,X^+_1\,X^+_3\, \widetilde{|\L\rg}
- X^+_3\,X^+_1\,h_3\, \widetilde{|\L\rg} ~=~ \cr
 ~=&~ X^+_1\,X^+_3\, (h_3-1) \,\widetilde{|\L\rg}
- X^+_3\,X^+_1\,h_3\, \widetilde{|\L\rg} ~=~ 0\ , }$$
~$h_3\,,h_4\,\in\ch$~ are Cartan generators corresponding to the
roots ~$\g_3\,,\g_4\,$, (cf. \DPf),
 the first term is zero since $\L(h_3)-1 = \half
(d-d^3_{NN}) =0$, while the second term is
zero due to \nulm{a} for $j_1=0$.\nl
For $j_2=0$ there is a singular vector
corresponding to the sum of two roots: $\a_{3,4+N}+\a_{4,4+N}$
(which sum is not a root) and is given in \DPf{} (cf. the formula before
(D.4) there):
\eqn\nnrb{ \tv^{12} ~=~ X^+_4\,X^+_2\,X^+_4\, \widetilde{|\L\rg} ~=~
X^+_4\,X^+_{3,4+N} \, \widetilde{|\L\rg}
\ ,\qquad d=d^1_{N1} \ ,\ j_2=0 }
As above, one checks that $X^-_k\, v^{12}=0$
for $k\neq 4$, and then calculates:
$$\eqalignno{ X^-_4\, \tv^{12} ~=&~ h_4 \,X^+_2\,X^+_4\, \widetilde{|\L\rg}
- X^+_4\,X^+_2\,h_4\, \widetilde{|\L\rg} ~=~ \cr
 ~=&~ X^+_2\,X^+_4\, (h_4-1) \,\widetilde{|\L\rg}
- X^+_4\,X^+_2\,h_4\, \widetilde{|\L\rg} ~=~ 0 }$$
using $\L(h_4)-1 = \half (d-d^1_{N1}) =0$,
and \nulm{b} for $j_2=0$.

To the above two singular vectors in the ER picture correspond
second-order super-differential operators given explicitly in
formulae (11a,b) of \DPu, and in formulae (D3),(D5) of \DPf.\foot{
Note that w.r.t. $V^\L$ the analogues of the
vectors $\tv^{34}$ and $\tv^{12}$ are
not singular, but subsingular vectors. Indeed, consider the
vector in $V^\L$ given by the same $U(\cg^+)$ monomial as $\tv^{34}$~:
 ~$v^{34} ~=~ X^+_3\,X^+_1\,X^+_3\,$. Clearly, $X^-_k\, v^{34}=0$
for $k\neq 3$. It remains to calculate the action of $X^-_3$~:
$$\eqalignno{ X^-_3\, v^{34} ~=&~ h_3 \,X^+_1\,X^+_3\, v_0
- X^+_3\,X^+_1\,h_3\, v_0 ~=~ \cr
 ~=&~ X^+_1\,X^+_3\, (h_3-1) \, v_0
- X^+_3\,X^+_1\,h_3\, v_0 ~=~ - X^+_3\,X^+_1\, v_0
 }$$
where the first term is zero as above,
 while the second term is a descendant of the
singular vector $v^1_s = X^+_1\, v_0 $, (cf. \sings{a} for $j_1=0$),
which fulfills the definition of subsingular vector.
Analogously, for the vector
 ~$v^{12} ~=~ X^+_4\,X^+_2\,X^+_4$~ we have $X^-_k\, v^{12}=0$
for $k\neq 4$, and:
$$ X^-_4\,v^{12}= X^-_4\,X^+_4\,X^+_2\,X^+_4
 ~=~ - X^+_4\,X^+_2\, v_0 \ , $$
(using $\L(h_4)-1$), which is a descendant
of the singular vector $v^2_s = X^+_2\, v_0 $,
cf. \sings{b} for $j_2=0$.}

{}From the expressions of the singular vectors follow, using
\subm{}, the explicit formulae for the corresponding
invariant submodules ~$I^\b$~ of the modules ~$\tV^\L$~
as follows:
\eqna\isub
$$\eqalignno{
I^1 ~=&~ U(\cg^+)\, P_{3,4+N} \, \widetilde{|\L\rg} ~=~
U(\cg^+)\,\(X^+_4 X^+_2(h_2-1) - X^+_2 X^+_4 h_2 \) \,
\widetilde{|\L\rg} ~=~ \quad &\isub a\cr =&~ U(\cg^+)\, \( 2j_2
X^+_{3,4+N} - X^+_4 X^+_2 \) \, \widetilde{|\L\rg} \ ,\qquad
d=d^1_{N1} \ ,\ j_2 > 0 \ , &\isub {a'}\cr I^2 ~=&~ U(\cg^+)\,
X^+_4\, \widetilde{|\L\rg} \ ,\qquad d=d^2_{N1} \ , &\isub b\cr I^3
~=&~ U(\cg^+)\, P_{15} \, \widetilde{|\L\rg} ~=~ U(\cg^+)\,\(X^+_3
X^+_1(h_1-1) - X^+_1 X^+_3 h_1 \) \, \widetilde{|\L\rg} ~=~\quad
&\isub c\cr =&~ U(\cg^+)\, \( 2j_1 X^+_{15} - X^+_3 X^+_1 \)\,
\widetilde{|\L\rg} \ ,\qquad d=d^3_{NN}\ , \ j_1> 0\ , &\isub
{c'}\cr I^4 ~=&~ U(\cg^+)\, X^+_3\, \widetilde{|\L\rg}\ ,\qquad
d=d^4_{NN} \ , &\isub d\cr I^{12} ~=&~ U(\cg^+)\, \tv^{12} ~=~
X^+_4\,X^+_2\,X^+_4\, \widetilde{|\L\rg} \ ,\qquad d=d^1_{N1} \ ,\
j_2 = 0 \ , &\isub e\cr I^{34} ~=&~ U(\cg^+)\, \tv^{34} ~=~
X^+_3\,X^+_1\,X^+_3 \, \widetilde{|\L\rg} \ ,\qquad d=d^3_{NN}\ , \
j_1 = 0\ . &\isub f\cr }$$ Sometimes we shall indicate the signature
~$\chi(\L)$, writing, e.g., $I^1(\chi)$; sometimes we shall indicate
also the resulting signature, writing, e.g., $I^1(\chi,\chi')$ -
this is a redundancy since it is determined by what is displayed
already: $\chi' = \chi(\L+\b)$, but will be useful to see
immediately in the concrete situations without calculation.

The invariant submodules were used in \DPu{} in the construction
of the UIRs, as we shall recall below.

\vskip 5mm

\subsec{Structure of single-reducibility-condition Verma modules and
UIRs}

\nt We discuss now the reducibility of Verma modules at the four
distinguished points \diss{}. We note a partial ordering of these
four points: \eqn\parto{ d^1_{N1} ~>~ d^2_{N1} \ , \qquad d^3_{NN}
~>~ d^4_{NN} \ ,} or more precisely: \eqn\partoo{ d^1_{N1} ~=~
d^2_{N1} +2\ , ~~ (j_2=0) ; \qquad d^3_{NN} ~=~ d^4_{NN} +2\ , ~~
(j_1=0) \ .} Due to this ordering at most two of these four points
may coincide. Thus, we have two possible situations: of Verma
modules (or ERs) reducible at one and at two reduction points from
\diss{}.

In this Subsection we deal with the situations in which ~{\it no
two}~ of the points in \diss{} coincide. According to \DPu{}
(Theorem) there are four such situations involving UIRs:
\eqna\dist
$$\eqalignno{
&d ~=~ d_{\rm max} ~=~ d^1_{N1} > d^3_{NN}\ , &\dist a\cr
&d ~=~ d^2_{N1} > d^3_{NN}\ , ~~ j_2=0\ ,&\dist b\cr
&d ~=~ d_{\rm max} ~=~ d^3_{NN} > d^1_{N1}\ , &\dist c\cr
&d ~=~ d^4_{NN} > d^1_{N1}\ , ~~ j_1=0\ . &\dist d\cr}$$

We shall call these cases ~{\bf single-reducibility-condition
(SRC)}~ Verma modules or UIRs, depending on the context.
In addition, as already stated, we use for the cases when ~$d= d_{\rm
max}\,$, i.e., \dist{a,c}, the terminology of semi-short UIRs,
\FGPW,\FSa, while the cases \dist{b,d} are also called
short UIRs, \FS,\FSa,\AES,\BKRSa,\EdSo,\HeHoa.

As we see the SRC cases have supplementary conditions as specified.
And due to the inequalities there are the following
additional restrictions which are correspondingly given as:
$$\eqalignno{
z ~>&~ j_1-j_2 -m_1 +2m/N \ , &\dist {a'}\cr
z ~>&~ j_1 +1 -m_1 +2m/N \ , &\dist {b'}\cr
z ~<&~ j_1-j_2 -m_1 +2m/N \ , &\dist {c'}\cr
z ~<&~ -1-j_2 -m_1 +2m/N \ . &\dist {d'}\cr}$$
Using these inequalities the unitarity conditions \dist{} may be
rewritten more explicitly:
$$\eqalignno{
&d ~=~ d^1_{N1} ~=~ d^a ~\equiv~ 2 +2j_2 +z+2m_1 -2m/N ~>~
2+ j_1+j_2 + m_1 \ , &\dist {a''}\cr
&d ~=~ d^2_{N1} ~=~ z+2m_1 -2m/N
~>~ j_1 +1 + m_1 \ , \qquad j_2=0
\ , &\dist {b''}\cr
&d ~=~ d^3_{NN} ~=~ d^c ~\equiv~ 2 +2j_1 -z +2m/N ~>~
2 + j_1 + j_2 +m_1
\ , &\dist {c''}\cr
&d ~=~ d^4_{NN} ~=~ -z +2m/N ~>~ 1+j_2 +m_1
\ , \qquad j_2=0 \ , &\dist {d''}
}$$
where we have introduced notation $d^a,d^c$ to designate two
of the SRC cases.

To finalize the structure we should check the even reducibility
conditions \redd{b,d,e,f}. It is enough to note that the
conditions on ~$d$~ in \dist{a'',c''}:
$$d ~>~ 2+j_1+j_2+m_1 $$
and in \dist{b'',d''}:
$$d ~>~ 1+j_1+j_2 +m_1\ , ~~(j_1j_2=0) $$
are incompatible with \redd{b,d,e,f}, except in two cases. The
exceptions are in cases \dist{b'',d''} when $d=2+j_1+j_2 =z$ and
$j_1j_2=0$. In these cases we have $n_{14}=1$ in \redd{f} and
there exists a Verma submodule $V^{\L+\a_{14}}$. However, the
$su(2,2)$ signature $\chi_0(\L+\a_{14})$ is unphysical:
$[j_1-\ha,-\ha;3+j_1]$ for $j_2=0$, and
$[-\ha,j_2-\ha;3+j_1]$ for $j_1=0$. Thus, there is no such
submodule of $\tV^\L$.

Thus, the factorized Verma modules ~$\tV^\L$~ with the unitary signatures
from \dist{} have only one invariant (odd) submodule which has to be
factorized in order to obtain the UIRs.
These odd embeddings are given explicitly as:
\eqn\embs{\tV^\L ~\rightarrow~ \tV^{\L+\b} }
where we use the convention \DPm{} that arrows point to the
oddly embedded module, and there are the following cases for $\b$~:
\eqna\paorz
$$\eqalignno{
\b ~=&~ \a_{3,4+N}\ , \quad {\rm for}~ \dist{a}, \quad j_2> 0 ,
&\paorz{a} \cr
=&~ \a_{4,4+N}\ , \quad {\rm for}~ \dist{b},
&\paorz{b} \cr
=&~ \a_{15}\ ,\quad {\rm for}~ \dist{c}, \quad
j_1> 0, &\paorz{c}\cr
=&~ \a_{25}\ ,
\quad {\rm for}~ \dist{d}, &\paorz{d}\cr
=&~ \a_{3,4+N}+\a_{4,4+N}\ , \quad {\rm for}~ \dist{a}, \quad
j_2=0, &\paorz{e} \cr
=&~ \a_{15}+\a_{25}\ ,
\quad {\rm for}~ \dist{c},\quad j_1=0 \quad &\paorz{f}
 }$$

This diagram gives the UIR ~$L_\L$~ contained in ~$\tV^\L$~ as follows:
\eqn\genun{
L_\L ~=~ \tV^\L/I^\b \ , }
where ~$I^\b$~ is given by ~$I^1$, $I^2$, $I^3$, $I^4$,
$I^{12}$, $I^{34}$, resp., (cf. \isub{}),
in the cases \paorz{a,b,c,d,e,f}, resp.

It is useful to record the signatures of the shifted lowest weights,
i.e., ~$\chi' ~=~ \chi(\L+\b)$. In fact, for
future use we give the signature changes for arbitrary roots.
The explicit formulae are \DPm,\DPf:
\eqna\sgnn \eqna\sgnnn
$$\eqalignno{
\b=\a_{3,N+5-k}~: & ~\chi' ~=~ [d+\half;\, j_1,j_2-\half;\,
z + \eps_N\,;\,r_1,\ldots,r_{k-1}-1,r_k+1,\ldots,r_{N-1}], \qquad &\sgnn a\cr
& j_2> 0 \ , \quad r_{k-1}> 0 &\sgnn {a'}\cr
&&\cr
\b=\a_{4,N+5-k}~: & ~\chi' ~=~ [d+\half;\, j_1,j_2+\half;\,
z + \eps_N\,;\,r_1,\ldots,r_{k-1}-1,r_k+1,\ldots,r_{N-1}], \qquad &\sgnn b\cr
& r_{k-1}> 0 &\sgnn {b'}\cr
&&\cr
\b=\a_{1,N+5-k}~: & ~\chi' ~=~ [d+\half;\, j_1-\half,j_2\,;\,
z - \eps_N\,;\,r_1,\ldots,r_{k-1}+1,r_k-1,\ldots,r_{N-1}], \qquad &\sgnn c\cr
& j_1> 0 \ , \quad r_{k}> 0 &\sgnn {c'}\cr
&&\cr
\b=\a_{2,N+5-k}~: & ~\chi' ~=~ [d+\half;\, j_1+\half,j_2\,;\,
z - \eps_N\,;\,r_1,\ldots,r_{k-1}+1,r_k-1,\ldots,r_{N-1}],
\qquad\qquad &\sgnn d\cr
& r_{k}> 0 &\sgnn {d'}\cr
&&\cr
&k = 1,\ldots,N\ , \qquad
\eps_N ~\equiv ~ \halfn - \half &\sgnnn {}\ .
 }$$
For each fixed ~$\chi$~ the lowest weight ~$\L(\chi')$~ fulfills
the same odd reducibility condition as ~$\L(\chi')$.
We need also the special cases used in \paorz{e,f}:
$$\eqalignno{
\b_{12}=\a_{3,4+N} + \a_{4,4+N}~: & ~\chi'_{12} ~=~ [d+1;\, j_1,0;\,
z +2 \eps_N\,;\,r_1+2,r_2,\ldots,r_{N-1}], &\sgnn {e}\cr
& j_2= 0,\ d=d^1_{N1} \cr
&&\cr
\b_{34}=\a_{15}+\a_{25}~: & ~\chi'_{34} ~=~ [d+1;\, 0,j_2\,;\,
z -2 \eps_N\,;\,r_1,...,r_{N-2},r_{N-1}+2],\qquad &\sgnn {f}\cr
& j_1= 0,\ d=d^3_{NN} \cr }
$$
The lowest weight $\L(\chi'_{12})$ fulfils \dist{b},
while the lowest weight $\L(\chi'_{34})$ fulfils \dist{d}.

The embedding diagram \embs{} is a piece of a much richer
picture \DPm. Indeed, notice that if \redg{b} is fulfilled for some odd root
~$\b$, then it is fulfilled also for an infinite number of Verma
modules ~$V_\ell ~=~ V^{\L + \ell\b}$~ for all ~$\ell\in\bbz$.
These modules form an infinite chain complex of oddly embedded modules:
\eqn\embsi{\cdots \ra ~{V}_{-1}~ \ra ~{V}_0~
\ra ~{V}_1~ \ra \cdots}
Because of \gras{} this is an exact sequence with one nilpotent
operator involved in the whole chain.
Of course, once we restrict to the factorized modules ~$\tV^\L$~
the diagram will be shortened - this is evident from the
signature changes \sgnn{a,b,c,d}. In fact, there are only a
finite number of factorized nodules for $N>1$, while for $N=1$
the diagram continues to be infinite to the left. Furthermore,
when ~$\b=\b_{12}\,, \b_{34}$~ from the end of the restricted
chain one transmutes - via the embeddings \isub{e,f}, resp. - to
the chain with ~$\b=\a_{4,N+4}\,,\a_{25}\,$, resp. More explicitly, when
~$\b=\b_{12}\,, \b_{34}\,$, then the module ~$V_1$~
plays the role of ~$V_0$~ with ~$\b=\a_{4,N+4}\,,\a_{25}\,$.
All this is explained in detail in \DPm. Furthermore, when
a factorized Verma module ~$\tV^\L = \tV^\L_0$~ contains an UIR
then not all modules ~$\tV^\L_\ell$~ would contain an UIR,
\DPf,\DPu. From all this what is important from the view of
modern applications can be summarized as follows:\nl
\bu ~~The semi-short SRC UIRs (cf. \dist{a,c}) are obtained by
factorizing a Verma submodule ~$\tV^{\L+\b}$~ containing either
another semi-short SRC UIR of the same type
(cf. \paorz{a,c}) or containing a
short SRC UIR of a different type
(cf. \paorz{e,f}). In contrast, short SRC UIRs (cf.
\dist{b,d}) are obtained by factorizing a Verma submodule
~$\tV^{\L+\b}$~ whose irreducible factor-module is not unitary
(cf. \paorz{b,d}).

\vskip 5mm

\subsec{Structure of double-reducibility-condition
Verma modules and UIRs}

\nt
We consider now the situations in which ~{\it two}~ of the points in
\diss{} coincide. According to \DPu{} (Theorem) there are four
such situations involving UIRs:
\eqna\disd
$$\eqalignno{
&d ~=~ d_{\rm max} ~=~ d^{ac} ~\equiv~ d^1_{N1} = d^3_{NN}\ , &\disd a\cr
&d ~=~ d^1_{N1} = d^4_{NN}\ , ~~ j_1=0\ , &\disd b\cr
&d ~=~ d^2_{N1} = d^3_{NN}\ , ~~ j_2=0\ ,&\disd c\cr
&d ~=~ d^2_{N1} = d^4_{NN}\ , ~~ j_1=j_2=0\ .&\disd d\cr
}$$

We shall call these ~{\bf double-reducibility-condition (DRC)}
Verma modules or UIRs. As in the previous subsection we shall
use for the cases when ~$d= d_{\rm max}\,$, i.e., \disd{a}, also
the terminology of semi-short UIRs,
\FGPW,\FSa, while the cases \disd{b,c,d} shall also be called
short UIRs, \FS,\FSa,\AES,\BKRSa,\EdSo,\HeHoa.

For later use we list more explicitly the values of ~$d$~ and ~$z$~
$$\eqalignno{
d ~=&~   d^{ac} ~=~ d^1_{N1} = d^3_{NN} ~=~ 2 + j_1 + j_2 + m_1\ , &\cr
&z ~=~ j_1-j_2 +2m/N -m_1 \ ;&\disd {a'}\cr
d ~=&~ d^1_{N1} = d^4_{NN} ~=~ 1 + j_2 + m_1
\ , ~~ j_1=0\ , &\cr
&z ~=~ -1-j_2 + 2m/N -m_1\ ;
&\disd {b'}\cr
d ~=&~ d^2_{N1} = d^3_{NN} ~=~ 1 + j_1 + m_1
\ , ~~ j_2=0\ ,\cr
&z ~=~ 1+j_1 + 2m/N -m_1 \ ;
&\disd {c'}\cr
d ~=&~ d^2_{N1} = d^4_{NN} ~=~ m_1
\ , ~~ j_1=j_2=0\ , \cr
&z ~=~ 2m/N -m_1 \ . &\disd {d'}\cr
}$$
We noted already that for ~$N=1$~ the last case, \disd{d,d'}, is
trivial. Note also that for $N=2$ we have: ~$2m/N -m_1= m-m_1=0$.

To finalize the structure we should check the even reducibility
conditions \redd{b,d,e,f}. It is enough to note that the
values of ~$d$~ in \disd{}  are incompatible with \redd{b,d,e,f},
except in a few cases. The exceptions are:
\eqna\disda
$$\eqalignno{
d ~=&~ d^1_{N1} = d^3_{NN} ~=~ 2 + j_1 + j_2 \ , \qquad m_1=0
&\disda {a}\cr d ~=&~ d^1_{N1} = d^4_{NN} ~=~ 1 + j_2 + m_1 \ , ~~
j_1=0\ , \qquad m_1 =0,1 &\disda {b}\cr d ~=&~ d^2_{N1} = d^3_{NN}
~=~ 1 + j_1 + m_1 \ , ~~ j_2=0\ ,\qquad m_1 =0,1 &\disda {c}\cr d
~=&~ d^2_{N1} = d^4_{NN} ~=~ m_1 \ , \qquad j_1=j_2=0\ , \quad m_1
=0,1,2 &\disda {d}\cr }$$ \bu\ In case \disda{a} we have $n_{14}=1$
in \redd{f} and there exists a Verma submodule $V^{\L+\a_{14}}$ with
$su(2,2)$ signature $\chi_0(\L+\a_{14}) =
[j_1-\ha,j_2-\ha;3+j_1+j_2]$. As we can see this signature is
unphysical for $j_1j_2= 0$. Thus, there is the even submodule
$\tV^{\L+\a_{14}}$ of $\tV^\L$ only if $j_1j_2\neq 0$.\nl \bu\ In
case \disda{b} there are three subcases:\nl $m_1 =0$, $j_2=\ha$\ ;
then $d=\trh$, $n_{24}=1$, $n_{14}=2$. The signatures of the
embedded submodules of $V^\L$ are: $\chi_0(\L+\a_{24}) =
[\ha,0;\frh]$, $\chi_0(\L+2\a_{14}) = [-1,-\ha;\srh]$. Thus, there
is only the even submodule $\tV^{\L+\a_{24}}$ of $\tV$.\nl $m_1 =0$,
$j_2=0$\ ; then $d=1$, $n_{13}=1$, $n_{24}=1$, $n_{14}=2$. The
signatures of the embedded submodules of $V^\L$ are:
$\chi_0(\L+\a_{13}) = [-\ha,\ha;2]$, $\chi_0(\L+\a_{24}) =
[\ha,-\ha;2]$, $\chi_0(\L+2\a_{14}) = [-1,-1;3]$, and are all
unphysical. However, the Verma module $V^\L$ has a subsingular
vector of weight $\a_{23}+\a_{14}$, cf.  \Docond{}, and thus, the
factorized Verma module $\tV^\L$ has the submodule
$\tV^{\L+\a_{23}+\a_{14}}$.\nl $m_1 =1$\ ; then $n_{14}=1$, but as
above there is no nontrivial even submodule of $\tV^\L$.\nl \bu\  The case
\disda{c} is dual to \disda{b} so we list shortly the three
subcases:\nl $m_1 =0$, $j_1=\ha$\ ; then $d=\trh$, $n_{13}=1$,
$n_{14}=2$. There is only the even submodule $\tV^{\L+\a_{13}}$ of
$\tV$.\nl $m_1 =0$, $j_1=0$\ ; then $d=1$, $n_{13}=1$, $n_{24}=1$,
$n_{14}=2$. This subcase coincides with the second subcase of
\disda{b}.\nl $m_1 =1$\ ; then $n_{14}=1$ and as above there is no
nontrivial submodule of $\tV^\L$.\nl \bu\ In case \disda{d} there
are again three subcases:\nl $m_1=0$\ ; then all quantum numbers in
the signature are zero and the UIR is the one-dimensional trivial
irrep.\nl $m_1 =1$\ ; then $d=1$, $n_{13}=1$, $n_{24}=1$,
$n_{14}=2$. Though this subcase has nontrivial isospin from
$su(2,2)$ point of view it has the same structure as the second
subcase of \disda{b} and the factorized Verma module $\tV^\L$ has
the submodule $\tV^{\L+\a_{23}+\a_{14}}$.\nl $m_1 =2$\ ; then $d=2$,
$n_{14}=1$ and as above there is no nontrivial even submodule of
$\tV^\L$.

The embedding diagrams for the corresponding modules ~$\tV^\L$~
when there are no even embeddings are:
\eqn\embd{
\matrix{
\tV^{\L+\b'}&& \cr
&&\cr
\uparrow &&\cr
&&\cr
\tV^\L &\rightarrow & \tV^{\L+\b} \cr
}}
where
\eqna\paor
$$\eqalignno{
(\b,\b') ~=&~\cr ~=&~
(\a_{15},\a_{3,4+N}), \quad {\rm for}~ \disd{a}, \quad m_1j_1j_2> 0
&\paor{a} \cr
~=&~ (\a_{15},\a_{3,4+N}+\a_{3,4+N}),
\quad {\rm for}~ \disd{a},\quad j_1> 0,\ j_2=0 &\paor{b} \cr
~=&~ (\a_{15}+\a_{25},\a_{3,4+N}),\quad {\rm for}~ \disd{a}, \quad
j_1=0,\ j_2> 0 &\paor{c} \cr
~=&~
(\a_{15}+\a_{25},\a_{3,4+N}+\a_{3,4+N}),
\quad {\rm for}~ \disd{a},\quad j_1=j_2=0\quad &\paor{d}
\cr ~=&~
(\a_{25},\a_{3,4+N}), \quad {\rm for}~ \disd{b}, \quad j_2> 0  ,
\ 2j_2 +m_1 \geq 2
&\paor{e} \cr ~=&~
(\a_{25},\a_{3,4+N}+\a_{4,4+N}), \quad {\rm for}~ \disd{b}, \quad
j_2=0 , \ m_1 >0
&\paor{f} \cr ~=&~
(\a_{15},\a_{4,4+N}), \quad {\rm for}~ \disd{c}, \quad j_1> 0  ,
\ 2j_1 +m_1 \geq 2
&\paor{g} \cr ~=&~
(\a_{15}+\a_{25},\a_{4,4+N}), \quad {\rm for}~ \disd{c}, \quad
j_1= 0 , \ m_1 >0  &\paor{h} \cr ~=&~
(\a_{25},\a_{4,4+N}), \quad {\rm for}~ \disd{d},
\quad m_1 \neq 1
&\paor{i} }$$
This diagram gives the UIR ~$L_\L$~ contained in ~$\tV^\L$~ as follows:
\eqna\genuna
$$\eqalignno{
L_\L ~=&~ \tV^\L /I^{\b,\b'} \ , \quad
I^{\b,\b'} ~=~ I^\b \cup I^{\b'} &\genuna {}}$$
where ~$I^\b$, $I^{\b'}$ are given in \isub{},
accordingly to the cases in \paor{}.

The embedding diagrams for the corresponding modules ~$\tV^\L$~
when there are even embeddings are:
\eqn\embde{
\matrix{
&&\tV^{\L+\b'}&& \cr
&&&&\cr
&&\uparrow &&\cr
&&&&\cr
\tV^{\L+\b_e} & \leftarrow & \tV^\L & \rightarrow & \tV^{\L+\b} \cr
}}
where
\eqna\paory
$$\eqalignno{
(\b,\b',\b_e) ~=&~\cr ~=&~
(\a_{15},\a_{3,4+N},\a_{14}), \quad {\rm for}~ \disd{a}, \quad j_1j_2> 0,
~~m_1=0
&\paory{a} \cr
~=&~
(\a_{25},\a_{3,4+N},\a_{24}),
\quad {\rm for}~ \disd{b}, \quad
j_2=\ha\,,\ m_1=0\qquad
&\paory{b} \cr
~=&~
(\a_{25},\a_{3,4+N}+\a_{4,4+N},\a_{23}+\a_{14}),
\quad {\rm for}~ \disd{b}, \quad j_2=m_1 = 0
&\paory{c} \cr
~=&~
(\a_{15},\a_{4,4+N},\a_{13}), \quad {\rm for}~ \disd{c}, \quad
j_1=\ha\,,\ m_1=0
&\paory{d} \cr ~=&~
(\a_{15}+\a_{25},\a_{4,4+N},\a_{23}+\a_{14}),
 \quad {\rm for}~ \disd{c}, \quad j_1=m_1 = 0 \qquad
&\paory{e} \cr ~=&~
(\a_{25},\a_{4,4+N},\a_{23}+\a_{14}),  \quad {\rm for}~ \disd{d},
\quad  m_1 = 1
&\paory{f} }$$
This diagram gives the UIR ~$L_\L$~ contained in ~$\tV^\L$~ as follows:
\eqna\genunaz
$$\eqalignno{
L_\L ~=&~ \tV^\L /I^{\b,\b',\b_e} \ , \quad
I^{\b,\b'} ~=~ I^\b \cup I^{\b'}\cup \tV^{\L+\b_e} &\genunaz {}}$$

Naturally, the two odd embeddings in
\embd{} or \embde{} are the combination of the different cases of
\embs{}. Similarly, like \embs{} is a piece of the richer picture
\embsi{}, here we have the following analogues of \embsi{}
\DPm\foot{These diagrams are essential parts of much richer
diagrams
(which we do not need since we consider only UIRs-related modules)
which are explicitly described for any $N$ in \DPm, and
shown there in Fig. 1 (for N=1) and Fig. 2 (for N=2).}
\eqn\embdi{
\matrix{
&& \vdots && &&\cr
&&&&&&\cr
&& \uparrow && &&\cr
&&&&&&\cr
 & & V_{01} & & & &
 \cr
&&&&&&\cr
&& \uparrow && &&\cr
&&&&&&\cr
\cdots & \rightarrow & V_{00} & \rightarrow & V_{10} & \rightarrow &
\cdots \cr
&&&&&&\cr
&& \uparrow && &&\cr
&&&&&&\cr
&& \vdots && &&\cr
} \quad \qquad N=1  }
where ~$V_{k\ell} ~\equiv~ V^{\L+k\b+\ell\b'}$,
and ~$\b,\b'$~ are the roots appearing in \paor{a,e,g,i},
(or \paory{a,b,d,f})
\eqn\embdii{
\matrix{
&& \vdots && \vdots &&\cr
&&&&&&\cr
&& \uparrow && \uparrow &&\cr
&&&&&&\cr
\cdots & \rightarrow & V_{10} & \rightarrow & V_{11} & \rightarrow &
\cdots \cr
&&&&&&\cr
&& \uparrow && \uparrow &&\cr
&&&&&&\cr
\cdots & \rightarrow & V_{00} & \rightarrow & V_{01} & \rightarrow &
\cdots \cr
&&&&&&\cr
&& \uparrow && \uparrow &&\cr
&&&&&&\cr
&& \vdots && \vdots &&\cr
} \quad \qquad N>1  }

The difference between the cases ~$N=1$~ and ~$N>1$~ is due to
the fact that if \redg{b} is fulfilled for ~$V_{00}\,$~ w.r.t.
two odd roots ~$\b,\b'$~
then for ~$N>1$~ it is fulfilled also for all Verma modules
~$V_{k\ell}\,$~ again w.r.t. these odd roots ~$\b,\b'$,
while for $N=1$ it is fulfilled only for ~$V_{k0}\,$~ w.r.t.
the odd root ~$\b$~ and only for ~$V_{0\ell}\,$~ w.r.t.
the odd root ~$\b'$.

In the cases \paor{b,c,d,f,h} (or \paory{c,e}) we have the same
diagrams though their parametrization is more involved \DPm{} (cf.
also what we said about transmutation for the single chains after
\embsi{}). However, for the modules with ~$0\leq k,\ell\leq 1$~
(which we use) we have simply as before ~$V_{k\ell} ~=~
V^{\L+k\b+\ell\b'}$~ for the appropriate $\b,\b'$.

The richer structure for $N>1$ has practical consequences for the
calculation of the character formulae, cf. next Section.

\np

\newsec{Character formulae of positive energy UIRs}

\subsec{Character formulae: generalities}

\nt
In the beginning of this subsection we follow \Di.
Let ~$\hcg$~ be a simple Lie algebra of rank ~$\ell$~
with Cartan subalgebra
~$\hch$, root system ~$\hD$, simple root system ~$\hpi$.
Let ~$\G$, (resp. $\G_+$), be the set of all integral, (resp.
integral dominant), elements of $\hch^*$, i.e., $\l
\in \hch^*$ such that $(\l , \a_i^\vee) \in \bbz$, (resp. $\bbz_+$),
for all simple roots $\a_i\,$, ($\a_i^\vee \equiv 2\a_i/(\a_i,\a_i)$).
Let $V$ be a lowest weight module with lowest weight $\L$ and
lowest weight vector $v_0\,$. It has the following decomposition:
\eqn\wei{ V ~=~ \mathop{\oplus}\limits_{\mu\in\G_+} V_\mu ~~, \ \
~~~V_\mu ~=~ \{ u \in V ~\vert ~Hu = (\l + \mu)(H)u, \
\forall ~H\in\ch \} }
(Note that $V_0 = \bbc v_0\,$.) Let $E(\ch^*)$ be
the associative abelian algebra consisting of the series
$\sum_{\mu \in \ch^*} c_{\mu} e(\mu)$ , where $c_{\mu} \in \bbc ,
 ~c_{\mu} = 0$ for $\mu$ outside the union of a finite number of
sets of the form $D(\l) = \{ \mu \in \ch^* \vert \mu \geq
\l \}$~, ~using some ordering of $\ch^*$,
e.g., the lexicographic one; the formal
exponents $e(\mu)$ have the properties:~
$e(0) = 1, \ e(\mu) e(\nu) = e(\mu + \nu)$.

Then the (formal) character of $V$ is defined by:
\eqn\cha{ch_0~V ~~=~~ \sum_{\mu \in\G_+} (\dim \ V_\mu) ~e(\L+\mu)
 ~~=~~ e(\L) \sum_{\mu\in\G_+} (\dim \ V_\mu) ~e(\mu) }
(We shall use subscript '0' for the even case.)

For a Verma module, i.e.,
~$V = V^\L$~ one has $\dim \ V_\mu = P(\mu)$,
where ~$P(\mu)$ is a generalized
partition function, $P(\mu) = \#$ of ways $\mu$ can be presented
as a sum of positive roots $\beta$, each root taken with its
multiplicity $\dim \CG_{\beta}$ ($=1$ here),
$P(0)\equiv 1$. Thus, the character formula for Verma modules is:
\eqn\chv{ch_0~V^\L ~~=~~ e(\L)\sum_{\mu\in\G_+} P(\mu) e(\mu) ~~=~~ e(\L)
\prod_{\a \in\D^+}(1 - e(\a))^{-1} }

Further we recall the
standard reflections in $\hch^*$~:
\eqn\rfl{s_\a(\l) ~=~ \l - (\l, \a^\vee)\a \,,
\quad \l \in \hch^* \,, \quad \a\in\hD }
The Weyl group ~$W$~ is generated by the simple reflections
$s_i \equiv s_{\a_i}$, $\a_i\in\hpi\,$.
Thus every element $w\in W$ can
be written as the product of simple reflections. It is said that
$w$ is written in a reduced form if it is written with the
minimal possible number of simple reflections; the number of
reflections of a reduced form of $w$ is called the length of $w$,
denoted by $\ell (w)$.

The Weyl character
formula for the finite-dimensional irreducible LWM $L_\L$
over ~$\hcg$, i.e., when ~$\L\in -\G_+\,$,
has the form:\foot{A more general character formula involves
the Kazhdan--Lusztig polynomials ~$P_{y,w}(u)$, $y,w\in W$ \KL.}
\eqn\chm{
ch_0~L_\L ~~=~~ \sum_{w\in W}(-1)^{\ell(w)} ~ch_0~V^{w\cdot\L} \,, \quad
\L\in -\G_+}
where the dot ~$\cdot$~ action is defined by $w\cdot \l = w(\l -
\r) +\r$.
For future reference we note:
\eqn\rft{ s_\a ~\cdot ~\L ~~=~~ \L ~+~ n_\a \a }
where
\eqn\nchs{ n_\a ~~=~~ n_\a(\L) ~~\doteq~~ (\r-\L,\a^\vee)
~~=~~ (\rho-\L)(H_\a)
\,, \quad \a\in\D^+ }

In the case of basic classical Lie superalgebras the first
character formulae were given by Kac \Kacl,\Kacc.\foot{Kac considers
highest weight modules but his results are immediately
transferable to lowest weight modules.} For all such
superalgebras (except $osp(1/2N)$)
the character formula for Verma modules is \Kacl,\Kacc:
\eqn\chv{ch~V^\L
~=~ e(\L)
\( \prod_{\a \in\D^+_\o}(1 - e(\a))^{-1} \)
\( \prod_{\a \in\D^+_\I}(1 + e(\a)) \)
}
Note that the factor ~$\prod_{\a \in\D^+_\o}(1 - e(\a))^{-1}$~
represents the states of the even sector: ~$V_0^\L ~\equiv~
U((\cg^\bac_+)_{(0)})\,v_0\,$ (as above in the even case),
while ~$\prod_{\a \in\D^+_\I}(1 + e(\a))$
represents the states of the odd sector: ~$\hV^\L ~\equiv~
\(U(\cg^\bac_+)/U((\cg^\bac_+)_{(0)})\)\,v_0\,$. Thus, we
may introduce a character for ~$\hV^\L$~ as follows:
\eqn\chv{ch~\hV^\L ~\equiv~ \prod_{\a \in\D^+_\I}(1 + e(\a)) . }

In our case,
~$\hV^\L$~ may be viewed as the result of all possible applications
of the $4N$ odd generators $X^+_{a,4+k}$ on $v_0\,$, i.e., ~$\hV^\L$~
has ~$2^{4N}$~ states (including the vacuum). Explicitly, the
basis of ~$\hV^\L$~ may be chosen as in \DPp: \eqn\bas{ \eqalign{ \Psi_{\bar \ve}
~=&~ \( \ \prod_{k=N}^1\ (X^+_{1,4+k})^{\ve_{1,4+k}} \) \ \( \
\prod_{k=N}^1\ (X^+_{2,4+k})^{\ve_{2,4+k}} \) \ \times \cr
&\times\ \( \ \prod_{k=1}^N\ (X^+_{3,4+k})^{\ve_{3,4+k}} \) \ \( \
\prod_{k=1}^N\ (X^+_{4,4+k})^{\ve_{4,4+k}} \) \, v_0 \ , \cr &
\ve_{aj} = 0,1}} where ~${\bar \ve}$~ denotes the set of all
~$\ve_{ij}\,$.\foot{The order chosen in \bas{} was important in
the proof of unitarity in \DPu,\DPp{} and for that purposes one
may choose also an order in which the vectors on the first row are
exchanged with the vectors on the second row. For our purposes the
order is important as far as to avoid impossible states - this is
much of the analysis done in the next subsections.}
 Thus, the
character of ~$\hV^\L$~ may be written as: \eqna\chvv
$$\eqalignno{
ch~\hV^\L ~=&~ \sum_{\bar \ve}~ e(\Psi_{\bar \ve}) ~= &\chvv a\cr
 =&~ \sum_{\bar \ve}~
\( \ \prod_{k=1}^N\ e(\a_{1,4+k})^{\ve_{1,4+k}} \)
\ \( \ \prod_{k=1}^N\ e(\a_{2,4+k})^{\ve_{2,4+k}} \) \ \times \cr
&\times\ \( \ \prod_{k=1}^N\ e(\a_{3,4+k})^{\ve_{3,4+k}} \)
\ \( \ \prod_{k=1}^N\ e(\a_{4,4+k})^{\ve_{4,4+k}} \) ~= &\chvv
b\cr ~=&~ \sum_{\bar \ve}~ e \(\ \sum_{k=1}^N \sum_{a=1}^4\
\ve_{a,4+k}\,\a_{a,4+k} \)
&\chvv c\cr
}$$
(Note that in the above formula there is no actual dependence from
$\L$.)

We shall use the above to write for the character of $V^\L$~:
\eqn\chv{\eqalign{
ch~V^\L ~=&~ ch\ \hV^\L\ \cdot\  ch_0\ V_0^{\L}   ~= \cr
=&~ \sum_{\bar \ve}~ e \(\ \sum_{k=1}^N \sum_{a=1}^4\
\ve_{a,4+k}\,\a_{a,4+k} \)\ \cdot\ e(\L)\
\( \prod_{\a \in\D^+_\o}(1 - e(\a))^{-1} \) ~= \cr
=&~ \sum_{\bar \ve}~ e \(\ \L \ +\ \sum_{k=1}^N \sum_{a=1}^4\
\ve_{a,4+k}\,\a_{a,4+k} \)\
\( \prod_{\a \in\D^+_\o}(1 - e(\a))^{-1} \) ~=
\cr =&~ \sum_{\bar \ve}~ ch_0\ V_0^{\L \ +\ \sum_{k=1}^N \sum_{a=1}^4\
\ve_{a,4+k}\,\a_{a,4+k} }
}}
where $ch_0\ V_0^{\L}$ is the character   obtained by
restriction of $V^\L$ to $V_0^{\L}$:
\eqn\chevv{ch_0\ V_0^{\L} ~=~ e(\L^z)\ \cdot\ ch_0\ V^{\L^s}\ \cdot\ ch_0\ V^{\L^u} }
where we use the decomposition ~$\L = \L^s + \L^z + \L^u$ from \lamde{a},
and $V^{\L^s}$, $V^{\L^u}$, resp., are Verma modules over the complexifications of
$su(2,2)$, $su(N)$, resp., cf. Appendix C.

  Analogously, for the factorized
Verma modules ~$\tV^\L$~ the character formula is:
\eqn\chlr{\eqalign{
ch~\tV^\L ~=&~ ch\ \hV^\L\ \cdot\ ch_0\ \tV_0^\L ~= \cr
=&~ \sum_{\bar \ve}~  ch_0\ \tV_0^{\L \ +\ \sum_{k=1}^N \sum_{a=1}^4\
\ve_{a,4+k}\,\a_{a,4+k} }}}
where ~$ch_0\ \tV_0^\L$~  is the character   obtained by
restriction of ~$\tV^\L$~ to ~ $\tV_0^\L ~\equiv~ U((\cg^\bac_+)_{(0)})\,\wt\,$,
or more explicitly:
\eqn\cheve{ch_0\ \tV_0^\L ~=~ e(\L^z)\ \cdot\ ch_0\ L_{\L^s}\ \cdot\ ch_0\ L_{\L^u} }
where we use the decomposition ~$\L = \L^s + \L^z + \L^u$ from \lamde{a},
and character formulae (C.2),(C.3),(C.4) for the irreps of the even subalgebra (from Appendix C).

Formula \chlr{} represents  the expansion of
the corresponding superfield in components, and each component
has its own even character. We see that this expansion is given exactly
by the expansion of the odd character \chvv{}.

We have already displayed how the UIRs ~$L_\L$~ are obtained as
factor-modules of the (even-submodules-factorized) Verma modules
$\tV^\L$. Of course, this factorization means that the odd singular
vectors of $\tV^\L$ from \tsing{} are becoming null conditions in
$L_\L\,$. However, this is not enough to determine the character
formulae even when considering our UIRs as irreps of the
complexification $sl(4/N)$. The latter is a well known feature
even in the bosonic case. Here the situation is much more
complicated and much more refined analysis is necessary.

The most important aspect of this analysis is the determination
of the superfield content. (This analysis was used in \DPu,\DPp{} but
was not explicated enough.) This is given by the positive norm
states ~$\hL_\L$~ among all states in the odd sector $\hV^\L$.
Of course,  ~$\hL_\L$~ may have less than ~$2^{4N}$~ states.

For future use we introduce notation for the levels of the
different chiralities ~$\ve_i$~
and the overall level ~$\ve$~
\eqn\levc{ \ve_i ~=~ \sum_{k=1}^N\ \ve_{i,4+k} \ , \quad
i=1,2,3,4 \ , \qquad \ve ~=~ \ve_1 +\ve_2 +\ve_3 +\ve_4 \ .}

The odd null conditions entwine with the even null
conditions as we shall see. The even null conditions
follow from the even singular vectors in \sings{}
(alternatively, one may say that they carry over from the even
null conditions \nulm{} of $\tV^\L$).
We write down the even null conditions first since they hold for any
positive energy UIR:
\eqna\nulln
$$\eqalignno{
(X^+_1)^{1+2j_1}\, |\L\rg ~=&~ 0 \ , &\nulln a\cr
(X^+_2)^{1+2j_2}\, |\L\rg ~=&~ 0 \ , &\nulln b\cr
 (X^+_j)^{1+r_{N+4-j}}\, |\L\rg ~=&~ 0 \ ,
\quad j=5,...,N+3 &\nulln c\cr}$$
(\nulln{c} being empty for ~$N=1$), where
by ~$|\L\rg$~ we shall denote the lowest weight vector of the UIR
~$L_\L\,$.

\vskip 5mm

\subsec{Character formulae for the long UIRs}

\nt As we mentioned if ~$d ~>~ d_{\rm max}$~ there are no further
reducibilities, and the UIRs ~$L_\L ~=~ \tV^\L$~ are called ~{\it
long}~ since ~$\hL_\L$~ may have the maximally possible
number of states ~$2^{4N}$ (including the vacuum state).

However, the actual number of states may be less than ~$2^{4N}$~
states due to
the fact that - depending on the values of $j_a$ and $r_k$ -
not all actions of the odd generators on the vacuum would be allowed.
The latter is obvious from formulae \sgnn{}. Using the latter we
can give the resulting signature of the state ~$\Psi_{\bar \ve}$~:
\eqn\signs{\eqalign{
\chi\( \Psi_{\bar \ve} \) ~=~
[\,&d+\half\ve\,;\, j_1+\half(\ve_2-\ve_1),\,j_2 + \half(\ve_4-\ve_3)
\,;\, z + \eps_N(\ve_3+\ve_4-\ve_1-\ve_2)
\,;\cr & \ldots\,,
 r_i + \ve_{1,N+4-i} - \ve_{1,N+5-i}
+ \ve_{2,N+4-i} - \ve_{2,N+5-i} -\cr
 &- \ve_{3,N+4-i} + \ve_{3,N+5-i}
- \ve_{4,N+4-i} + \ve_{4,N+5-i}
\,,\ldots]
}}

Thus, only if ~$j_1,j_2 ~\geq~ N/2$~ and ~$r_i ~\geq 4$ (for all
$i$) the number of states is ~$2^{4N}$ \DPu, and the character
formula for the irreducible lowest weight module is \chlr{}:
\eqna\chll
$$\eqalignno{
ch~L_\L ~=&~ ch~\tV^\L ~=~ ch\ \hV^\L\ \cdot\ ch_0\ \tV_0^\L\  \ ,   \qquad d ~>~
d_{\rm max} \ , &\chll a\cr & j_1,j_2 \geq N/2, ~~r_i \geq 4
,~i=1,\ldots,N-1\ , &\chll b}$$

The general formula for ~$ch~L_\L$~ shall be written in a similar
 fashion:
\eqn\chgg{ ch~L_\L ~=~ ch\ \hL_\L\ \cdot\ ch_0\ \tV_0^\L \ \ .}

{\bf Moreover, from now on we shall write only the formulae for
~$ch~\hL_\L\,$.} ~Thus, formula \chll{} shall  be written equivalently
as:
\eqn\chggg{ ch~\hL_\L ~=~ ch~\hV^\L \ ,
\quad j_1,j_2 \geq N/2, ~~~r_i \geq 4, \forall\, i\ .}
As we have noted after \chlr{}  we do not lose information
using this factorized form which has the advantage of brevity.

If the auxiliary conditions \chll{b} are not fulfilled then
a careful analysis is necessary.
To simplify the exposition we classify the states by the
following quantities:
\eqn\indd{\eqalign{
\ve^c_j ~\equiv&~ \ve_{1} - \ve_{2} \ ,\cr
 \ve^a_j ~\equiv&~ \ve_{3} - \ve_{4} \ , \cr
 \ve_r^i ~\equiv&~
\ve_{1,5+i} + \ve_{2,5+i} + \ve_{3,4+i} + \ve_{4,4+i}
- \ve_{1,4+i} - \ve_{2,4+i} - \ve_{3,5+i} - \ve_{4,5+i}
\ ,\cr
 &i=1,\ldots,N-1 \ .}}

This gives the following necessary conditions on ~$\ve_{ij}$~
for a state to be allowed:
\eqna\conz
$$\eqalignno{
\ve^c_j ~\leq&~ 2j_1 \ ,&\conz a\cr
\ve^a_j ~\leq&~ 2j_2 \ ,&\conz b\cr
 \ve_r^i ~\leq&~ r_{N-i}
 \ , \quad i=1,\ldots,N-1 \ .
 &\conz c\cr }$$

These conditions are also sufficient only for $N=1$ (when
\conz{c} is absent). The exact conditions are:

\nt {\bf Criterion:}~~ The necessary and sufficient conditions
for the state ~$\Psi_{\bar \ve}$~ of level ~$\ve$~ to be allowed
are that conditions \conz{} are fulfilled and that
the state is a descendant of an allowed state of level
~$\ve-1$.\dia

The second part of the Criterion will take care first of all of
chiral (or anti-chiral) states when some ~$\ve_{aj}$~ contribute
to opposing sides of the inequalities in \conz{a} and \conz{c},
(or \conz{b} and \conz{c}). This phenomena happens for $j_1=r_i =0$,
(or $j_2=r_i =0$).

We shall give now the most important such occurrences.
Take first chiral states, i.e., all ~$\ve_{3,4+k} =
\ve_{4,4+k} =0$. Fix ~$i=1,\ldots,N-1$.
It is easy to see that the following states are not allowed:
\eqna\impz
$$\eqalignno{\psi_{ij} ~=~ \phi_{ij}\, |\L\rg ~=&~
X^+_{1,i+4}\, X^+_{2,i+5} \, X^+_{{a_{1},i+6}} \, \ldots\,
X^+_{a_{j-1},i+4+j}\, |\L\rg \ , \quad a_n=1,2 &\impz{} \ ,\cr &
j=1,\ldots,N-i,\quad j_1=r_{N-i} = \cdots = r_{N-i-j+1} = 0\ ,\cr
&{\rm in~addition,~for}~ N>2,\, i>1 ~{\rm holds}~ r_{N-i+1} \neq 0\
.}$$ {\it Demonstration:}~~ {\iti Naturally, this statement is
nontrivial only when these states are allowed by condition \conz{a}
(i.e., the number of $a_n$ being equal to 2 is not less than the
number of $a_n$ being equal to 1), thus we restrict to those. By
design these states fulfil also \conz{c}, (\conz{b} is not
relevant), however, they are not descendants of  allowed states.
First, all states ~$\hat\psi_{ij} = X^+_{2,i+5} \, X^+_{a_{1},i+6}
\, \ldots\, X^+_{a_{j-1},i+4+j}\, |\L\rg$~ violate \conz{c} with
$r_{N-i}=0$. Next, the state ~$\psi_{i1}$~ is not allowed since in
addition to ~$\hat\psi_{11}$~ also the state ~$X^+_{1,i+4}\,
|\L\rg$~ is not allowed (it violates \conz{a} with $j_1=0$). Due to
this the state ~$\psi_{i2}$~ is not descendant of any allowed
states, and so on, for all ~$\psi_{ij}\,$. Note that the last part
of the proof trivializes unless all $a_n=2$.}\bsq \nl {\it
Remark:}~~ {\iti The additional condition on the last line of
\impz{} is there, since if ~$r_{N-i+1}=0$,  the states ~$\psi_{ij}\,
|\L\rg$~ (for $i>1$) violate \conz{c} with $r_{N-i+1}=0$ and are
excluded without use of the
 Criterion.}\dia

Consider now anti-chiral states, i.e., such that ~$\ve_{1,4+k}
~=~ \ve_{2,4+k} ~=~ 0$, for all ~$k=1,\ldots,N$.
Fix ~$i=1,\ldots,N-1$. Then the following
anti-chiral states are not allowed:
\eqna\impy
$$\eqalignno{ \psi'_{ij} ~=~ \phi'_{ij}\, |\L\rg ~=&~
X^+_{3,i+5}\, X^+_{4,i+4} \, X^+_{b_{1},i+3} \,
\ldots\, X^+_{b_{j-1},i+5-j} \,|\L\rg \ ,
\quad b_n=3,4 \ ,&\impy{}
\cr & j=1,\ldots,i,
\quad j_2=r_{N-i} = \cdots = r_{N-i+j-1} =0\ ,\cr
&{\rm in~addition,~for}~ N>2,\, i>1 ~{\rm holds}~ r_{N-i-1} \neq 0\ .}
$$

Furthermore, any combinations of ~$\phi_{ij}$~
and ~$\phi'_{i'j'}$~ are not allowed.

Note that for ~$N\geq 4$~ the states in \impz{},\impy{} do not
exhaust the states forbidden by our Criterion. For example, for ~$N=4$~
there are the following forbidden states:
$$\eqalignno{
\psi_{4} ~=~ \phi_{4}\, |\L\rg ~=&~
X^+_{28}\, X^+_{17} \, X^+_{16} \, X^+_{25}\, |\L\rg \ ,  \quad
j_1=r_{1} = r_{2} = r_{3} = 0  \ , &\impz{'}\cr
\psi'_{4} ~=~ \phi'_{4}\, |\L\rg ~=&~
X^+_{45}\, X^+_{36} \, X^+_{37} \, X^+_{48}\, |\L\rg \ , \quad
j_2=r_{1} = r_{2} = r_{3} = 0\ . &\impy{'}
}$$

Summarizing the discussion so far, the general character formula
may be written as follows:
\eqna\chlll
$$\eqalignno{
ch~\hL_\L ~=&~
ch~\hV^\L ~-~ \car_{\rm long} \ , \qquad d ~>~ d_{\rm max} \ ,&\chlll{}\cr
\car ~=&~ e(\hV^\L_{\rm excl})
~=~ \sum_{\rm excluded\atop states}\ e(\Psi_{\bar \ve})\ ,
  }$$
where the counter-terms denoted by ~$\car_{\rm long}$~ are
determined by ~$\hV^\L_{\rm excl}$~ which is the collection of all
states (i.e., collection of $\ve_{jk}$) which violate the conditions
\conz{}, or are impossible in the sense of \impz{} and/or \impy{}.
Of course, each excluded state is accounted for only once even if it
is not allowed for several reasons.\foot{We should stress that the
necessity of the counter-terms above is related to the fact that our
representations of ~$su(2,2/N)$~ have physical meaning and the
states of ~$\hL_\L$~ represent components of a superfield. There are
no counter-terms when we consider these UIRs as irreps of $sl(4/N)$.
Thus, formula \chlll{} and almost all character formulae derived
further in this Section  are character formulae of $sl(4/N)$ by just
dropping the counter-term $\car$, cf. next Section.}

Finally, we consider two important conjugate special cases.

First, the chiral sector of $R$-symmetry scalars with $j_1=0$.
Taking into account \conz{a,c} (\conz{b} is trivially satisfied
for chiral states) and our Criterion it is easy to see that
the appearance of the generators $X^+_{1,4+k}$ is restricted as follows.
The generator $X^+_{15}$ may appear only in the state
\eqn\resa{ X^+_{15}\,X^+_{25}\,|\L\rg }
and its descendants.  The generator $X^+_{16}$ may only appear
either in states descendant to the state \resa{} or in the state
\eqn\rresa{ X^+_{16}\,X^+_{25}\,|\L\rg }
and its descendants including only generators ~$X^+_{a,5+\ell}\,$,
~$a=1,2$, ~$\ell>1$.  Further, the restrictions are described
recursively, namely, fix ~$\ell$~ such that ~$1 ~<~ \ell
~\leq~ N-1$. The generator ~$X^+_{1,5+\ell}$~ may only appear
either in states containing generators ~$X^+_{1,5+j}\,$, where
$0\leq j <\ell$, or in the state
\eqn\possi{  X^+_{1,5+\ell}\,X^+_{2,4+\ell}\,X^+_{2,3+\ell}\,
\cdots\,X^+_{2,5} \, |\L\rg }
and its descendants including only generators ~$X^+_{a,5+\ell'}\,$,
~$a=1,2$, ~$\ell'>\ell$.

The chiral part of the basis is further restricted. Namely,
there are only ~$N$~ chiral states that can be built from
the generators ~$X^+_{2,4+k}$~ alone, given as follows:
\eqn\possy{ X^+_{2,4+k}\,\cdots\,X^+_{25} \, |\L\rg \ , \quad
k=1,\ldots,N,  \ ,\quad j_1=r_{i} = 0, ~\forall\,i \ .}
This follows from \conz{c} which in this case is reduced to
~$\ve_{1i} ~\leq~ \ve_{1,i+1}$~ for ~$i=1,\ldots,N-1$.

Second, the anti-chiral sector of $R$-symmetry scalars with $j_2=0$.
Taking into account \conz{b,c} (\conz{a} is trivially satisfied
for anti-chiral states) and our Criterion it is easy to see that
the appearance of the generators $X^+_{3,4+k}$ is restricted as follows.
The generator $X^+_{3,4+N}$ may appear only in the state
\eqn\reso{ X^+_{3,4+N}\,X^+_{4,4+N}\,|\L\rg }
and its descendants.  The generator $X^+_{3,3+N}$ may only appear
either in states descendant to the state \reso{} or in the state
\eqn\rreso{ X^+_{3,3+N}\,X^+_{4,4+N}\,|\L\rg }
and its descendants including only generators ~$X^+_{a,4+N-\ell}\,$,
~$a=3,4$, ~$\ell>1$.  Further,  fix ~$\ell$~ such that ~$1 ~<~ \ell
~\leq~ N-1$. The generator ~$X^+_{3,4+N-\ell}$~ may only appear
either in states containing generators ~$X^+_{3,4+N-j}\,$, where
$0\leq j <\ell$, or in the state
\eqn\poss{ X^+_{3,4+N-\ell}\,X^+_{4,5+N-\ell}\,X^+_{4,6+N-\ell}\,
\cdots\,X^+_{4,4+N} \, |\L\rg }
and its descendants including only generators ~$X^+_{a,4+N-\ell'}\,$,
~$a=3,4$, $\ell'>\ell$.

The anti-chiral part of the basis is further restricted. Namely,
there are only ~$N$~ anti-chiral states that can be built from
the generators ~$X^+_{4,4+k}$~ alone, given as follows:
\eqn\posss{ X^+_{4,5+N-k}\,X^+_{4,6+N-k}\,
\cdots\,X^+_{4,4+N} \, |\L\rg \ , \quad k=1,\ldots,N,
\ ,\quad j_2=r_{i} = 0, ~\forall\,i \ .}
This follows from \conz{c} which for such states becomes
~$\ve_{4,4+N-i} ~\leq~ \ve_{4,5+N-i}$~ for ~$i=1,\ldots,N-1\,$.

\vskip 5mm

\subsec{Character formulae of SRC UIRs}

\nt Here we consider the four SRC cases.

\vskip 3mm

\nt\bu{\bf a}~~~~~
$d   ~=~ d^1_{N1} ~=~ d^a ~\equiv~ 2 +2j_2 +z+2m_1 -2m/N ~>~
 d^3_{NN}$ \ .

\nt \bu  ~~~~~Let first  $j_2>0$.\nl
In these semi-short SRC cases holds the odd null condition
(following from the singular vector (8.9a) of \DPf{}, cf. also
\singg{a'}, \tsing{a'}, \isub{a}):
\eqn\nulla{ \eqalign{
P_{3,4+N}\, |\L\rg ~=&~ \(X^+_4 X^+_2
(h_2-1) - X^+_2 X^+_4 h_2 \) \, |\L\rg ~=\cr
=&~ \( 2j_2 X^+_{3,4+N} - X^+_4 X^+_2\)\, |\L\rg ~=~ 0 }}
where ~$X^+_{3,4+N} = [X^+_2,X^+_4]$. Clearly, condition
\nulla{} means that the generator ~$X^+_{3,4+N}$~ is eliminated
from the basis that is built on the lowest weight vector
~$|\L\rg\,$. Thus, for $N=1$ and if $r_1> 0$ for $N>1$
the character formula is:
\eqn\chs{\eqalign{
ch~\hL_\L ~=&~
 \prod_{\a \in\D^+_\I \atop \a\neq \a_{3,4+N}}
\ (1 +e(\a)) ~-~ \car ~, \cr
&~ d ~=~ d_{\rm max} ~=~ d^1_{N1} > d^3_{NN} \ ,
\quad j_2r_1> 0 \ .}}
There are no counter-terms when
~$j_1~\geq~ N/2$,\ $j_2 ~\geq~ (N-1)/2$~ and ~$r_i ~\geq 4$ (for all
$i$), and then the number of states is ~$2^{4N-1}$.
The change of statement (as compared to the long superfields)
w.r.t. $j_2$ comes because of the elimination of the generator
$X^+_{3,4+N}\,$.

\nt{\it Remark:}~~{\iti
For the finite-dimensional irreps of ~sl(4/N)~ (in fact, of all
basic classical Lie superalgebras) such situations are called
'singly atypical' and the character formulae look exactly as
\chs{} with $\car=0$, cf. \BL,\JHKT,\Jeu.\foot{For
character formulae of finite-dimensional irreps beyond the singly
atypical case cf. \Serga, \VZ, \Bru, \SuZh, and references therein.}}\dia

When there are no counter-terms (also for the complex
$sl(4/N)$ case) this formula follows easily from \genun{}.
Indeed, in the case at
hand ~$I^\b=I^1$, (cf. \isub{a}); then from
~$L_\L ~=~ \tV^\L/I^1$~ follows:
\eqn\chard{ \eqalign{
ch~L_\L ~=&~ ch~\tV^\L ~-~ ch~I^1 \ , \qquad {\rm or~ equivalently,}\cr
ch~\hL_\L ~=&~ ch~\hV^\L ~-~ ch~\hI^1 \ ,}}
where ~$\hI^1$~ is the projection of ~$I^1$~ to the odd sector.
Naively, the character of ~$\hI^1$~ should be given by the
character of ~$\hV^{\L+\a_{3,4+N}}$, however,
as discussed in general - cf. \subm{}, ~$I^1$~ is smaller than
~$\hV^{\L+\a_{3,4+N}}$~ and its character is
given with a prefactor\foot{This technique was applied
first when deriving the characters of the ~$N=2$~ super-Virasoro
algebras, cf. \Dobn.}:
\eqn\chii{ ch~\hI^1 ~=~ {1 \over 1+e(\a_{3,4+N})}~
ch~\hV^{\L+\a_{3,4+N}} ~=~
{e(\a_{3,4+N}) \over 1+e(\a_{3,4+N})}~
ch~\hV^{\L} \ .}
Now \chs{} (with $\car=0$) follows from the combination of
\chard{} and \chii.

Formula \chs{} may also be described by using the
odd reflection \oddr{} with ~$\b = \a_{3,4+N}$~:
\eqna\chodr
$$\eqalignno{
ch~\hL_\L ~=&~ ch~\hV^\L ~-~ {1 \over 1+e(\a_{3,4+N})}~
ch~\hV^{\hs_{\a_{3,4+N}}\cdot\L} ~-~ \car ~= &\chodr{a}\cr
=&~ ch~\hV^\L ~-~ \hs_{\a_{3,4+N}}\, \cdot ch~\hV^\L ~-~ \car ~= &\chodr{b}\cr
=&~ \sum_{\hs\in\hW_{\a_{3,4+N}}}\ (-1)^{\ell(\hs)}\ \hs\cdot ch~\hV^\L
~-~ \car \ ,&\chodr{c}\cr
}$$
where ~$\hW_{\b} ~\equiv~ \{1,\hs_\b\}$~
is a two-element semi-group restriction of ~$\tW_{\b}\,$, and
we have formalized further by introducing notation for the
action of an odd reflection on characters:
\eqn\oddre{
\hs_\b\, \cdot ch~V^\L ~=~ {1\over 1+e(\b)}\ ch~V^{\hs_\b\,\cdot\L}
~=~ {1\over 1+e(\b)}\ ch~V^{\L+\b}
~=~ {e(\b)\over 1+e(\b)}\ ch~V^{\L} \ .}

In particular, we shall show that in many cases
 character formulae \chs{},\chodr{} may be written as follows:
\eqn\chct{ch~\hL_\L ~=~
\sum_{\hs\in\hW_\b}\ (-1)^{\ell(\hs)}\ \hs\cdot \( ch~\hV^\L
~-~ \car_{\rm long} \) \ ,}
where ~$\car_{\rm long}$~ represents the counter-terms for the
long superfields for the same values of ~$j_1$~ and ~$r_i$~ as
$\L$, while the value of $j_2$ is zero when $j_2$ from $\L$ is zero,
otherwise it has to be the generic value $j_2\geq N/2$.
(As we know, restriction \conz{b} trivializes for $j_2\geq N/2$
and thus the structure of the irrep is the same for any such
generic value.)

Writing \chs{} as \chodr{} (or \chct{}) may look as a complicated
way to describe the cancellation of a factor from the character
formula for $\hV^\L$, however, first of all it is related to the
structure of $\tV^\L$ given by \genun{}, and furthermore may be
interpreted - when there are no counter-terms - as the following
decomposition: \eqn\dec{ \hV^\L ~=~ \hL_\L ~\oplus \hL_{\L+\b} \ ,}
for ~$\b = \a_{3,4+N}\,$. Indeed, for generic signatures
~$\hL_{\L+\b}$~ is isomorphic to ~$\hL_{\L}$~ as a vector space
(this is due to the fact that ~$V^{\L+\b}$~ has the same
reducibilities as ~$V^{\L}$, cf. Section 2), they differ only by the
vacuum state. Thus, when there are no counter-terms, both
~$\hL_{\L}$~ and ~$\hL_{\L+\b}$~ have the same ~$2^{4N-1}$~ states.
If we describe them for shortness as: \eqn\basz{ \Phi_i\,|\L\rg \ ,
\qquad \Phi_i\,|\L+\b\rg \ ,} where none of ~$\Phi_i$~ contains
~$X^+_{3,4+N}$~ and recall that the embedding of ~$\hV^{\L+\b}$~
into ~$\hV^{\L}$~ is given essentially by the generator
~$X^+_{3,4+N}$~ (cf. \isub{a})
 then we see that after the embedding
the states in \basz{} restore all ~$2^{4N}$~ states in ~$\hV^\L$~:
\eqn\basy{ \Phi_i\,\wt\ , \qquad X^+_{3,4+N}\,\Phi_i\,\wt \ .}

It is more important that there is a similar decomposition valid
for many cases beyond the generic, i.e., we have:
\eqn\deca{ \(\hL_{\rm long}\)_{\vert_{d=d^a}} ~=~ \hL_\L ~\oplus \hL_{\L+\a_{3,4+N}} \ ,
\qquad N=1 ~{\rm or}~ r_1 >0~{\rm for}~ N>1\ ,}
where ~$\hL_{\rm long}$~ is a long superfield with the same
values of ~$j_1$~ and ~$r_i$~ as $\L$,
while the value of $j_2$ has to be specified as above for
 ~$\car_{\rm long}\,$, and equality is as vector spaces.

For ~$N>1$~ there are possible additional truncations of the
basis. To make the exposition easier we need additional notation.
Let ~$i_0$~ be an integer such that
~$0\leq i_0\leq N-1\,$, and ~$r_i=0$~ for ~$i\leq i_0\,$,
and if ~$i_0<N-1$~ then ~$r_{i_0+1}> 0$.\foot{This is
formally valid for ~$N=1$~ with ~$i_0=0$~ since ~$r_{0}\equiv 0$~ by
convention. This shall be used to make certain statements
valid for general $N$.}

Let now ~$N>1$~ and ~$i_0>0$; then the generators ~$X^+_{3,4+N-i}\,$,
~$i=1,\ldots,i_0$, are eliminated from the basis.\nl
{\it Demonstration:}~~{\iti
First we consider the vector:
\eqn\adsin{ \eqalign{
P_{3,3+N}\, v_0 ~=&~ \( 2j_2 X^+_{3,3+N}
- X^+_{4,3+N}\, X^+_{2}\)\, v_0 ~= \cr
~=&~ 2j_2 \( X^+_{3,4+N} X^+_{3+N} - X^+_{3+N} X^+_{3,4+N} \)\, v_0
~- \cr &-
~ \( X^+_{4} X^+_{3+N} - X^+_{3+N} X^+_{4} \)\,X^+_2\, v_0 ~=\cr
~=&~ \( P_{3,4+N}\, X^+_{3+N} - X^+_{3+N}\, P_{3,4+N} \)\, v_0\ . }}
For ~$r_1=0$~ it is descendant of \sings{c} and \singg{} and
leads to the null condition:
\eqn\adnua{ P_{3,3+N}\, |\L\rg ~=~ \( 2j_2 X^+_{3,3+N}
- X^+_{4,3+N}\, X^+_{2} \)\, |\L\rg ~=~ 0 }
which naturally follows from \nulln{c} and \nulla{}, and
which means that the generator ~$X^+_{3,3+N}$~ is eliminated
from the basis.
Analogously, we define the vectors:
\eqn\addsi{P_{3,4+N-i}\, v_0 ~=~ \( 2j_2 X^+_{3,4+N-i}
- X^+_{4,4+N-i} X^+_{2} \)\, v_0 }
which are recursively related:
\eqn\adsin{ \eqalign{
P_{3,4+N-i}\, v_0 ~=&~ 2j_2 \( X^+_{3,5+N-i} X^+_{4+N-i} -
X^+_{4+N-i} X^+_{3,5+N-i} \)\, v_0 ~- \cr &-
~ \( X^+_{4,5+N-i} X^+_{4+N-i} - X^+_{4+N-i} X^+_{4,5+N-i}
\)\,X^+_2\, v_0 ~=\cr
~=&~ \( P_{3,5+N-i}\, X^+_{4+N-i} - X^+_{4+N-i}\, P_{3,5+N-i}
\)\, v_0\ . }}
Thus, in the situation: ~$r_i ~=~ 0$, $i=1,\ldots,i_0\,$,
there are the following null conditions:
\eqn\adnu{ P_{3,4+N-i}\, |\L\rg ~=~ \( 2j_2 X^+_{3,4+N-i}
- X^+_{4,4+N-i}\,X^+_{2}\)\, |\L\rg ~=~ 0 \ ,
\quad r_j=0, ~~1\leq j\leq  i\leq i_0 \ .}
These are recursively descendant null conditions which means that
a condition for fixed $i$ is a descendant of the one for $i-1$
(since $X^+_{4+N-i}\, |\L\rg = 0$ due to \nulln{c}).
Conditions \adnu{} mean that the generators ~$X^+_{3,4+N-i}\,$,
~$i=1,\ldots,i_0$, are eliminated from the basis.}\bsq

{}From the above follows that for $i_0>0$ the decomposition \deca{}
can not hold. Indeed, the generators ~$X^+_{3,4+N-i}\,$,
~$i=1,\ldots,i_0\,$, are eliminated from the irrep $\hL_\L$ due to
the fact that we are at a reducibility point, but there is no reason
for them to be eliminated from the long superfield. Certainly, some
of these generators are present in the second term
$\hL_{\L+\a_{3,4+N}}$ in \deca{}, but that would be only those which
in the long superfield were in states of the kind:
~$\Phi\,X^+_{3,4+N}\,|\L\rg$, and, certainly, such states do not
exhaust the occurrence of the discussed generators in the long
superfield. Symbolically, instead of the decomposition \deca{} we
shall write: \eqn\dectaz{ \(\hL_{\rm long}\)_{\vert_{d=d^a}} ~=~
\hL_\L\ \oplus\ \hL_{\L+\a_{3,4+N}} \oplus\ \hL'_{\L}\ ,\qquad N>1\
,\ i_0>0\ ,} where we have represented the excess states by the last
term with prime. With the prime we stress that this is not a genuine
irrep, but just a book-keeping device. Formulae as \dectaz{} in
which not all terms are genuine irreps shall be called ~{\it
quasi-decompositions}.

The corresponding character formula is:
\eqna\chsa
$$\eqalignno{
ch~\hL_\L ~=&~
 \prod_{\a \in\D^+_\I \atop {\a\neq \a_{3,5+N-k}
\atop k=1,\ldots,1+i_0}}
\ (1 +e(\a)) ~-~ \car ~= &\chsa {a}\cr &&\cr
=&~ \sum_{\hs\in\hW^a_{i_0}}\ (-1)^{\ell(\hs)}\ \hs\cdot ch~\hV^\L
~-~ \car ~= &\chsa {b}\cr
=&~ \sum_{\hs\in\hW^a_{i_0}}\ (-1)^{\ell(\hs)}\ \hs\cdot \( ch~\hV^\L
~-~ \car_{\rm long} \) \ , &\chsa {c}\cr
&\hW^a_{i_0} ~\equiv~
 \hW_{\a_{3,N+4}}\times \hW_{\a_{3,N+3}} \times \cdots \times
 \hW_{\a_{3,N+4-i_0}}
 \ ,&\chsa {d} \cr
&~ d ~=~ d_{\rm max} ~=~ d^1_{N1} > d^3_{NN} \ ,
\quad j_2> 0, ~r_i =0, ~i\leq i_0\  .
}$$
The restrictions \conz{} used to determine the counter-terms are,
of course, with ~$\ve_{3,5+N-k}=0$, ~$k=1,\ldots,1+i_0\,$.
Formulae \chs{},\chodr{},\chct{} are special cases of \chsa{a,b,c}, resp., for
$i_0=0$. The maximal number of states in ~$\hL_\L$~
is ~$2^{4N-1-i_0}$. This is the number of states that is obtained
from the action of the Weyl group ~$\hW^a_{i_0}$~ on
~$ch~\hV^\L$, while the actual counter-term is obtained from
the action of the Weyl group on ~$\car_{\rm long}\,$.

In the extreme case of $R$-symmetry scalars: ~$i_0=N-1$, i.e.,
~$r_i=0$, $i=1,\ldots,N-1$, or, equivalently, ~$m_1 ~=~ 0 ~=~ m$,
all the ~$N$~ generators ~$X^+_{3,4+k}$~ are eliminated.
The character formula is again \chsa{} taken with ~$i_0 ~=~ N-1$.

\vskip 3mm

\nt\bu ~~~~~Let now $j_2=0$.\nl
Then all null conditions above (valid for $j_2>0$) follow from \nulm{b},
so these conditions do not mean elimination of the mentioned
vectors.
As we know in this situation we have the singular vector \nnrb{}
which leads to the following null condition:
\eqn\nulnraz{ X^+_{3,4+N}\,X^+_{4,4+N} \, |\L\rg ~=~
X^+_4\,X^+_2\,X^+_4 \, |\L\rg ~=~ 0 \ .}
The state in \nulnraz{} and all of its ~$2^{4N-2}$~
descendants are zero for any $N$.
Thus, the character formula is similar to \chodr{}, but with
~$\a_{3,4+N}$~
replaced by ~$\b_{12} = \a_{3,4+N} + \a_{4,4+N}$, (cf. \sgnn{e}):
\eqna\chodra
$$\eqalignno{
ch~\hL_\L ~=&~ \sum_{\hs\in\hW_{\b_{12}}}\
(-1)^{\ell(\hs)}\ \hs\cdot ch~\hV^\L
~-~ \car ~= &\chodra{a}\cr
=&~ \sum_{\hs\in\hW_{\b_{12}}}\
(-1)^{\ell(\hs)}\ \hs\cdot \( ch~\hV^\L
~-~ \car_{\rm long} \) \ ,\quad N=1\ {\rm or}\ r_1>0\ ,
&\chodra{b}}$$
where ~$\hW_{\b_{12}} ~\equiv~ \{ 1,\b_{12}\}$.

Note that for $N=1$ formula \chodra{} is equivalent to \chs{}
since due to \conz{b} the generator ~$X^+_{3,4+N}$~
could appear only together with ~$X^+_{4,4+N}$~
but the resulting state \nulnraz{} is zero.

Here holds a decomposition similar to \deca{}:
\eqn\decaa{ \(\hL_{\rm long}\)_{\vert_{d=d^a}}
 ~=~ \hL_\L ~\oplus \hL_{\L+\b_{12}} \ ,
\qquad N=1 ~{\rm or}~ r_1 >0~{\rm for}~ N>1\ ,}
where $\hL_{\rm long}$~ is with the same values of
~$j_1,j_2(=0),r_i$~ as ~$\L$. Note, however, that the UIR
~$\hL_{\L+\b_{12}}$~ belongs to type {\bf b} below.

There are more eliminations for ~$N>1$~ when ~$i_0>0$.
For instance we can show that all states as in \poss{} considered
for $\ell = 1,\ldots,i_0\,$ are not allowed.\nl
{\it Demonstration:}~~{\iti
We show this by induction. Consider first the case $\ell=1$~:
\eqna\nulnrb
$$\eqalignno{
& X^+_{3,3+N}\,X^+_{4,4+N} \, |\L\rg ~=~
\( X^+_{3+N} X^+_{3,4+N} - X^+_{3,4+N} X^+_{3+N} \)
\,X^+_{4,4+N} \, |\L\rg ~=&\cr
&=~ - X^+_{3,4+N}\, X^+_{3+N} \,X^+_{4,4+N} \, |\L\rg ~=~
-X^+_{3,4+N}\, X^+_{4,3+N} \, |\L\rg\ , &\nulnrb {} }$$
where the first term is zero due to \nulnraz{},
 and the second term is transformed by pulling ~$X^+_{3+N}$~
 to the right, where it annihilates the vacuum (due to
 \nulln{c} with $j=N+3$ for $r_1=0$), and the resulting state is
 the forbidden ~$\psi'_{N-1,1}$~ from \impy{}. Thus, the above
 state is not allowed.\nl
Now fix ~$k$~ such that ~$1 ~<~ k ~\leq~ i_0$~ and suppose
that we have already shown that all states in \poss{} for $\ell< k$
are not allowed, and we shall show this for $\ell = k$.
Indeed, this state is not allowed:
\eqna\nulnrc
$$\eqalignno{
&X^+_{3,4+N-k}\,X^+_{4,5+N-k}\,X^+_{4,6+N-k}\,
\cdots\,X^+_{4,4+N} \, |\L\rg ~= &\nulnrc {}\cr
&=~ -X^+_{4,5+N-k}\,X^+_{3,4+N-k}\,X^+_{4,6+N-k}\,
\cdots\,X^+_{4,4+N} \, |\L\rg ~=\cr
&=~ -X^+_{4,5+N-k}\,
\( X^+_{4+N-k} X^+_{3,5+N-k} - X^+_{3,5+N-k} X^+_{4+N-k} \)
\,X^+_{4,6+N-k}\,\cdots\,X^+_{4,4+N} \, |\L\rg
 \ ,  }$$
where the first term on the last line is a state descendant of
\poss{} with ~$k\to k-1$, which is not allowed by the
induction hypothesis and the second term is zero due to pulling
~$X^+_{4+N-k}$~ to the right, where it annihilates the vacuum (due to
 \nulln{c} with $j=N+4-k$ for $r_k=0$).}\bsq

{}From the above follows that if $i_0>0$ the decomposition \decaa{}
does not hold. Instead, there is a quasi-decomposition similar to
\dectaz{}.

We can be more explicit in the case when all $r_i=0$. In that
case all the vectors ~$X^+_{3,5+N-k}\,$~ are eliminated from
all anti-chiral states.\nl
{\it Demonstration:}~~{\iti
We show this by induction in ~$k$~ starting with $k=1,2$.
Take first the generator ~$X^+_{3,4+N}\,$. As we know
 when ~$j_2=r_{i} = 0$, $\forall\,i\,$, the only anti-chiral state
containing it in a long superfield is the state \reso{} and its descendants.
However, here all these possible states are zero due to \nulnraz{}.
Thus, there are no anti-chiral states containing ~$X^+_{3,4+N}\,$.\nl
Take next the vector ~$X^+_{3,3+N}\,$. As we know the only anti-chiral states
containing it in a long superfield are the states \reso{},\rreso{},
and their descendants. The first is zero, while the second is
not allowed as we showed above. Thus, the vector ~$X^+_{3,3+N}$~
is eliminated from all anti-chiral states.\nl
Now fix ~$\ell$~ such that ~$1 ~<~ \ell ~\leq~ N-1$~ and suppose
that we have already shown elimination of ~$X^+_{3,5+N-k}$~ for
~$k=1,\ldots,\ell$, from all anti-chiral states. We want to show
elimination for $k=\ell+1$, i.e., of the generator
~$X^+_{3,4+N-\ell}\,$. As we know from the similar consideration
of long superfields all anti-chiral states including
~$X^+_{3,4+N-\ell}$~ and which are not yet excluded may be
written as the state \poss{} and its descendants
including only generators ~$X^+_{a,4+N-\ell'}\,$,
~$a=3,4$, $\ell'>\ell$. However, above we have shown that this
state is not allowed.
Thus, all generators ~$X^+_{3,4+k}$~ for ~$k=1,\ldots,N$~ are
eliminated from the anti-chiral part of the basis.}\bsq

The anti-chiral part of the basis is further restricted.
As we know, when ~$j_2=r_{i} = 0$, $\forall\,i\,$, there
are only ~$N$~ anti-chiral states that can be built from
the generators ~$X^+_{4,4+k}$~ alone, given in \posss{}.
Thus the corresponding character formula is:
\eqna\chsaaz
$$\eqalignno{
ch~\hL_\L ~=&~
 \sum_{k=1}^N \prod_{i=1}^k
e(\a_{4,5+N-i}) ~+~
\prod_{\a \in\D^+_\I \atop {\ve_1+\ve_2> 0}}
\ (1 +e(\a))
~-~ \car ~, &\chsaaz{}\cr &&\cr
&~ d ~=~ d_{\rm max} ~=~ d^1_{N1} > d^3_{NN} \ ,
\quad j_2=0, ~r_i =0, \forall\,i \ .
}$$

 \vskip 3mm

\nt{\bu{\bf b}}~~~~~
$d ~=~ d^2_{N1} ~=~ z+2m_1 -2m/N ~>~ d^3_{NN}\ , ~~ j_2=0$\ .

\nt
In these short single-reducibility-condition cases
holds the odd null condition (following from the singular
vector \singg{b} or \tsing{b})
\eqn\nullb{ X^+_4 \, |\L\rg ~=~ X^+_{4,4+N} \, |\L\rg ~=~ 0 \ .}
Since $j_2=0$ from \nulm{b} and \nullb{} follows the additional
null condition:
\eqn\nullbb{ X^+_{3,4+N} \, |\L\rg ~=~ [ X^+_2 , X^+_4] \, |\L\rg ~=~ 0 }

For ~$N>1$~ and ~$r_1> 2$~ each of these UIRs enters as the second
term in decomposition \decaa{}, when the first term is an UIR of
type {\bf a} with ~$j_2=0$, as explained above.

Further, for ~$N>1$~ there are additional null conditions
if ~$r_i=0$, $i\leq i_0\,$. Indeed, let ~$r_1=0$; then
from \nulm{c} and \nullbb{} follow the additional
null conditions:
\eqna\nullbbb
$$\eqalignno{
X^+_{4,3+N} \, |\L\rg ~=&~ [ X^+_{4,4+N} ,
X^+_{3+N}] \, |\L\rg ~=~ 0 \ , \quad r_1=0 \ ,
&\nullbbb a\cr
X^+_{3,3+N} \, |\L\rg ~=&~ [ X^+_{3,4+N} ,
X^+_{3+N}] \, |\L\rg ~=~ 0 \ , \quad r_1=0
&\nullbbb b\cr }$$
Analogously, in the situation: ~$r_i ~=~ 0$, $i=1,\ldots,i_0\,$,
there are recursive null conditions:
\eqna\adreu
$$\eqalignno{
X^+_{3,4+N-i} \, |\L\rg ~=&~ [ X^+_{3,5+N-i} ,
X^+_{4+N-i}] \, |\L\rg ~=~ 0 \ , \quad r_j =0, ~~1\leq j\leq i \leq i_0
&\adreu a\cr
X^+_{4,4+N-i} \, |\L\rg ~=&~ [ X^+_{4,5+N-i} ,
X^+_{4+N-i}] \, |\L\rg ~=~ 0 \ , \quad r_j =0, ~~1\leq j\leq i\leq i_0 \ ,
&\adreu b\cr }$$
Thus, ~$2(1+i_0)$~ generators ~$X^+_{3,5+N-k}\,$, $X^+_{4,5+N-k}\,$,
~$k=1,\ldots,1+i_0\,$, are eliminated.
The maximal number of states in ~$\hL_\L$~
is ~$2^{4N-2- 2i_0}$.

The corresponding character formula is:
\eqna\chsb
$$\eqalignno{
ch~\hL_\L ~=&~
\prod_{\a \in\D^+_\I \atop {\a\neq \a_{j,5+N-k}
\atop j=3,4\,, \ k=1,\ldots,1+i_0}}
\ (1 +e(\a)) ~-~ \car ~
= &\chsb {a}\cr &&\cr
=&~ \sum_{\hs\in\hW^b_{i_0}}\ (-1)^{\ell(\hs)}\ \hs\cdot ch~\hV^\L
~-~ \car \ , &\chsb {b}\cr
\hW^b_{i_0} ~\equiv&~
 \hW_{\a_{3,N+4}}\times \hW_{\a_{3,N+3}} \times \cdots \times
 \hW_{\a_{3,N+4-i_0}}\ \times \cr &\times
 \hW_{\a_{4,N+4}}\times \hW_{\a_{4,N+3}} \times \cdots \times
 \hW_{\a_{4,N+4-i_0}}
 \ ,&\chsb {c} \cr
 &~ d ~=~ d^2_{N1} > d^3_{NN}\ ,\quad j_2=0\ ,
~r_i =0, ~i\leq i_0\  ,
}$$
where determining the counter-terms we use
 ~$\ve_{j,5+N-k}=0$, ~$j=3,4$, ~$k=1,\ldots,1+i_0\,$.

In the case of ~$R$-symmetry scalars ($i_0=N-1$) we have:
\eqn\aree{ X^+_{3,4+k} \, |\L\rg ~=~ 0 \ , \quad
X^+_{4,4+k} \, |\L\rg ~=~ 0 \ , \qquad k=1,\ldots,N \ ,
\quad r_i=0, \forall\, i\ . }
The character formula is \chsb{} taken with ~$1+i_0 ~=~ N$.
These UIRs should be called chiral since all anti-chiral
generators are eliminated.

\vskip 3mm

The next two cases are conjugates of the first two and the
exposition will be compact.

\vskip 3mm

\nt{\bu{\bf c}}~~~~~
$d   ~=~ d^3_{NN} ~=~ d^c ~\equiv~ 2 +2j_1 -z +2m/N ~>~ d^1_{N1}$ \ .

\nt\bu~~~~~Let first $j_1> 0$.\nl
In these semi-short SRC cases
holds the odd null condition (following from the singular
vector (8.8a) of \DPf), here cf. \singg{c'} or \tsing{c'}):
\eqn\nullc{ P_{15} \, |\L\rg ~=~
\( 2j_1 X^+_{15} - X^+_3 X^+_1 \)\, |\L\rg ~=~ 0}
where ~$X^+_{15} = [X^+_1,X^+_3]$. Clearly, condition
\nullc{} means that the generator ~$X^+_{15}$~ is eliminated
from the basis.

Let now ~$i'_0$~ be an integer such that
~$0\leq i'_0\leq N-1\,$, and ~$r_{N-i}=0$~ for ~$i\leq i'_0\,$,
and if ~$i'_0<N-1$~ then ~$r_{N-1-i'_0}> 0$.\foot{This is
formally valid for ~$N=1$~ with ~$i'_0=0$~ since ~$r_{N}=0$~ by
convention.}
For ~$N>1$~ and ~$i'_0>0$~ there are additional truncations due
to the vectors (cf. (C.7) of \DPf):
\eqn\addsz{\eqalign{
P_{1,5+i}\, v_0 ~=&~ \( 2j_1 X^+_{1,5+i}
- X^+_{2,5+i}\) X^+_{1}\, v_0 ~=\cr
=&~ 2j_1 \( X^+_{1,4+i} X^+_{4+i} -
X^+_{4+i} X^+_{1,4+i} \)\, v_0 ~- \cr &-
~ \( X^+_{2,4+i} X^+_{4+i} - X^+_{4+i} X^+_{2,4+i} \)
X^+_1 \, v_0 ~=\cr
~=&~ \( P_{1,4+i}\, X^+_{4+i} - X^+_{4+i}\, P_{1,4+i} \)\, v_0 }}
which produced recursive null conditions:
\eqn\adnuu{ P_{1,5+i}\, |\L\rg ~=~ \( 2j_1 X^+_{1,5+i}
- X^+_{2,5+i} X^+_{1} \)\, |\L\rg ~=~ 0 \ ,
\quad r_{N-j} =0, ~~1\leq j\leq i\leq i'_0 \ ,}
which means that the generators ~$X^+_{1,5+i}$~ are eliminated
from the basis.

The corresponding character formula is:
\eqna\chsc
$$\eqalignno{
ch~\hL_\L ~=&~
 \prod_{\a \in\D^+_\I \atop {\a\neq \a_{1,4+k}
\atop k=1,\ldots,1+i'_0}}
\ (1 +e(\a)) ~-~ \car ~= &\chsc {a}\cr &&\cr
=&~ \sum_{\hs\in\hW^c_{i'_0}}\ (-1)^{\ell(\hs)}\ \hs\cdot ch~\hV^\L
~-~ \car ~= &\chsc {b}\cr
=&~ \sum_{\hs\in\hW^c_{i'_0}}\ (-1)^{\ell(\hs)}\ \hs\cdot
\( ch~\hV^\L ~-~ \car_{\rm long}\) \ , &\chsc {c}\cr
&\hW^c_{i'_0} ~\equiv~
 \hW_{\a_{15}}\times \hW_{\a_{16}} \times \cdots \times \hW_{\a_{1,5+i'_0}}
 \ ,&\chsc {d} \cr
&~ d ~=~ d_{\rm max} ~=~ d^3_{NN} > d^1_{N1} \ ,
\quad j_1> 0, ~r_{N-i} =0, ~i\leq i'_0\leq N-1\ .
}$$
This formula is valid also for ~$N=1$~ or when ~$r_{N-1}> 0$~
by setting ~$i'_0 =0$.
The maximal number of states in ~$\hL_\L$~ is ~$2^{4N-1-i'_0}$.
The restrictions \conz{} used for the counter-terms are
with ~$\ve_{1,N-i}=0$, ~$i=0,1,\ldots,i'_0\,$.

When ~$i'_0=0$~ holds decomposition similar to \deca{}:
\eqn\decc{ \(\hL_{\rm long}\)_{\vert_{d=d^c}}
 ~=~ \hL_\L ~\oplus \hL_{\L+\a_{15}} \ ,
\qquad N=1 ~{\rm or}~ r_{N-1} >0~{\rm for}~ N>1\ ,}
where ~$\hL_{\rm long}$~ is a long superfield with the same
values of ~$j_2$~ and ~$r_i$~ as $\L$,
while the value of $j_1$ is zero when $j_1$ from $\L$ is zero,
otherwise it has to be the generic value $j_1\geq N/2$.
{}From the above follows also that when  $i'_0>0$ the decomposition
\decc{} does not hold.

In the case of $R$-symmetry scalars ($i'_0=N-1$)
all the ~$N$~ generators ~$X^+_{1,4+k}$,
($k=1,\ldots,N)$ are eliminated.
The maximal number of states in ~$\hL_\L$~ is ~$2^{3N}$.

\vskip 3mm

\nt\bu ~~~Let now $j_1=0$.\nl
Then the null conditions above all follow from \nulm{a}
so these conditions do not mean elimination of the mentioned
vectors.
In this situation we have the singular vector \nnra{}
which leads to the following null condition:
\eqn\nulnray{ X^+_{15}\,X^+_{25} \, |\L\rg ~=~
X^+_3\,X^+_1\,X^+_3 \, |\L\rg ~=~ 0\ . }
The state in \nulnray{} and all of its ~$2^{4N-2}$~
descendants are zero for any $N$.
Thus, for ~$N=1$~ or if ~$r_{N-1}> 0$, the
character formula is as \chsc{} for ~$i'_0=0$, but with ~$\a_{15}$~
replaced by ~$\b_{34} = \a_{25} + \a_{25}$, (cf. \sgnn{f}):
\eqna\chodrc
$$\eqalignno{
ch~\hL_\L ~=&~ \sum_{\hs\in\hW_{\b_{34}}}\
(-1)^{\ell(\hs)}\ \hs\cdot ch~\hV^\L
~-~ \car ~= &\chodrc{a}\cr
=&~ \sum_{\hs\in\hW_{\b_{34}}}\
(-1)^{\ell(\hs)}\ \hs\cdot \( ch~\hV^\L
~-~ \car_{\rm long} \) \ ,\quad N=1\ {\rm or}\ r_{N-1}>0\ ,
&\chodrc{b}\cr
}$$
where ~$\hW_{\b_{34}} ~\equiv~ \{1,\b_{34}\}$.

For $N=1$ formula \chodrc{} is equivalent to \chsc{}
for ~$i'_0=0$~ since due to \conz{a} the generator ~$X^+_{15}$~
could appear only together with ~$X^+_{25}$~
but the resulting state \nulnray{} is zero.

For ~$i'_0= 0$~ holds the decomposition:
\eqn\deccc{ \(\hL_{\rm long}\)_{\vert_{d=d^c}}
 ~=~ \hL_\L ~\oplus \hL_{\L+\b_{34}} \ ,
\qquad N=1 ~{\rm or}~ r_{N-1} >0~{\rm for}~ N>1\ ,}
where $\hL_{\rm long}$~ is with the same values of
~$j_1(=0),j_2,r_i$~ as ~$\L$.
Note, however, that the UIR ~$\hL_{\L+\b_{34}}$~
belongs to type {\bf d} below.

There are more eliminations for ~$N>1$~ when ~$i'_0>0$.
For instance we can show that all states as in \possi{} considered
for $\ell = 1,\ldots,i'_0\,$ are not allowed.\nl
{\it Demonstration:}~~{\iti
We show this by induction. Consider first the case $\ell=1$~:
\eqna\nulnrbb
$$\eqalignno{
& X^+_{16}\,X^+_{25} \, |\L\rg ~=~
\( X^+_{15} X^+_{5} - X^+_{5} X^+_{15} \)
\,X^+_{25} \, |\L\rg ~=&\cr
&=~ X^+_{15}\, X^+_{5} \,X^+_{25} \, |\L\rg ~=~
X^+_{15}\, X^+_{26} \, |\L\rg\ , &\nulnrbb {} }$$
where the second term is zero due to \nulnray{},
 and the first term is transformed by pulling ~$X^+_{5}$~
 to the right, where it annihilates the vacuum (due to
 \nulln{c} with $j=1$ for $r_{N-1}=0$), and the resulting state is
 the forbidden $\psi_{11}\,$. Thus, the above
 state is not allowed. Further, we proceed by induction similarly
to the conjugate case, cf. \nulnrc{}.}\bsq

{}From the above follows that when $i'_0>0$ the
decomposition \deccc{} does not hold.

We can be more explicit in the case when all $r_i=0$. In that
case all the generators ~$X^+_{1,4+k}\,$, (for $k=1,\ldots,N$),
are eliminated from all chiral states.\nl
{\it Demonstration:}~~{\iti
Take first the vector ~$X^+_{15}\,$.
As we know  when ~$j_1=r_{i} = 0$, $\forall\,i\,$, the only chiral state
containing it in a long superfield is the state \resa{} and its descendants.
However, here all these possible states are zero due to \nulnray{}.
Thus, there are no chiral states  containing ~$X^+_{15}\,$.\nl
 Take next the vector ~$X^+_{16}\,$. As we know the only chiral states
containing it in a long superfield are the states \resa{},\rresa{},
and their descendants. The first is zero, while the second is
not allowed as we showed above.  Thus, the vector ~$X^+_{16}$~ is eliminated
from all chiral states.\nl
Now fix ~$\ell$~ such that ~$1 ~<~ \ell ~\leq~ N-1$~ and suppose
that we have already shown elimination of ~$X^+_{1,4+k}$~ for
~$k=1,\ldots,\ell$, from all anti-chiral states. We want to show
elimination of ~$X^+_{1,4+k}$~ for ~$k=\ell+1$.
As we know from the similar consideration
of long superfields all chiral states including
~$X^+_{1,5+\ell}$~ and which are not yet excluded may be
written as the state \possi{} and its descendants
including only generators ~$X^+_{a,5+\ell'}\,$,
~$a=1,2$, ~$\ell'>\ell$.
Then it is shown (analogously to \poss) that this state is also
not allowed.
Thus, all generators ~$X^+_{1,4+k}$~ for ~$k=1,\ldots,N$~ are
eliminated from the chiral part of the basis.}\bsq

The chiral part of the basis is further restricted. As we know,
when ~$j_1=r_{i} = 0$, $\forall\,i\,$, there
are only ~$N$~ chiral states that can be built from
the generators ~$X^+_{2,4+k}$~ alone, given in \possy{}.
Thus, the corresponding character formula is:
\eqna\chsccz
$$\eqalignno{
ch~\hL_\L ~=&~
\sum_{k=1}^{N}\ \prod_{i=1}^{k} \
e(\a_{2,4+i}) ~+~
\prod_{\a \in\D^+_\I \atop {\ve_3+\ve_4> 0}}
\ (1 +e(\a)) ~-~ \car ~, &\chsccz{}\cr &&\cr
d ~&=~ d_{\rm max} ~=~ d^3_{NN} > d^1_{N1} \ ,
\quad j_1=0, ~r_{i} =0, ~\forall\, i\ .}$$

\vskip 3mm

\nt{\bu{\bf d}}~~~~~
$d ~=~ d^4_{NN} ~=~ -z +2m/N ~>~ d^1_{N1}\ , ~~ j_1=0$\ .

\nt
In these short single-reducibility-condition cases
holds the odd null condition (following from the singular
vector \singg{d} or \tsing{d}):
\eqn\nulld{ X^+_3 \, |\L\rg ~=~ X^+_{25} \, |\L\rg ~=~
0 \ .}
Since $j_1=0$ from \nulm{a} and \nulld{} follows the additional
null condition:
\eqn\nulldd{ X^+_{15} \, |\L\rg ~=~ [ X^+_1 , X^+_3] \, |\L\rg ~=~ 0 }

For ~$N>1$~ and ~$r_{N-1} > 2$~ each of these UIRs enters as the second
term in decomposition \deccc{}, when the first term is an UIR of
type {\bf c} with ~$j_1=0$, as explained above.

Further, for ~$N>1$~ there are additional null conditions
if ~$r_{N-i}=0$, $i\leq i'_0\,$. These are recursive null conditions:
\eqna\nullddd
$$\eqalignno{
X^+_{1,5+i} \, |\L\rg ~=&~ [ X^+_{1,4+i} ,
X^+_{4+i}] \, |\L\rg ~=~ 0 \ , \quad r_{N-j} =0, ~~1\leq j\leq i\leq i'_0 \ ,
&\nullddd a\cr
X^+_{2,5+i} \, |\L\rg ~=&~ [ X^+_{2,4+i} ,
X^+_{4+i}] \, |\L\rg ~=~ 0 \ , \quad r_{N-j} =0, ~~1\leq j\leq i \leq i'_0
&\nullddd b\cr }$$
Thus, ~$2(1+i'_0)$~ generators ~$X^+_{1,4+k}\,$, $X^+_{2,4+k}\,$,
~$k=1,\ldots,1+i'_0\,$, are eliminated.
The maximal number of states in ~$\hL_\L$~
is ~$2^{4N-2- 2i'_0}$.

The corresponding character formula is:
\eqna\chsd
$$\eqalignno{
ch~\hL_\L ~=&~
\prod_{\a \in\D^+_\I \atop {\a\neq \a_{j,4+k}
\atop j=1,2\,, \ k=1,\ldots,1+i'_0}}
\ (1 +e(\a)) ~-~ \car ~
= &\chsd {a}\cr &&\cr
=&~ \sum_{\hs\in\hW^d_{i'_0}}\ (-1)^{\ell(\hs)}\ \hs\cdot ch~\hV^\L
~-~ \car \ , &\chsd {b}\cr
\hW^d_{i'_0} ~\equiv&~
 \hW_{\a_{15}}\times \hW_{\a_{16}} \times \cdots \times \hW_{\a_{1,5+i_0}}
 \ \times
 \hW_{\a_{25}}\times \hW_{\a_{26}} \times \cdots \times \hW_{\a_{2,5+i_0}}
 \ ,&\chsd {c} \cr
&~ d ~=~ d^4_{NN} > d^1_{N1}\ ,\quad j_1=0\ ,
~r_{N-i} =0, ~i\leq i'_0\leq N-1\,,\ r_{N-1-i'_0}> 0,
}$$
where $\car$ designates the counter-terms due to our Criterion,
in particular, due to \conz{} taken with ~$\ve_{j,4+k}=0$,
~$j=1,2$, ~$k=1,\ldots,1+i'_0\,$.

In the case of ~$R$-symmetry scalars we have:
\eqn\areee{ X^+_{1,4+k} \, |\L\rg ~=~ 0 \ , \quad
X^+_{2,4+k} \, |\L\rg ~=~ 0 \ , \qquad k=1,...,N \ ,
\quad r_i=0, \forall\, i }
The character formula is \chsd{} taken with ~$1+i'_0 ~=~ N$.
These are chiral UIRs
conjugate to the anti-chiral ones in \aree.

\vskip 5mm

\subsec{Character formulae of DRC UIRs}

\nt  Each of the DRC cases
is the obvious combination of two SRC cases and some results
follow from this. In fact, in the generic cases, we can give a
general character formula which follows directly from embedding diagram
\embdii{}.

So let first ~$N>1$~ and ~$r_1r_N-1>0$, (i.e., $i_0=i'_0=0$).
Then  holds the following character formula:
\eqna\chodrg
$$\eqalignno{
ch~\hL_\L ~=&~ \sum_{\hs\in\hW_{\b,\b'}}
(-1)^{\ell(\hs)}\ \hs\cdot ch~\hV^\L
~-~ \car ~= &\chodrg{a}\cr &&\cr
=&~ ch~\hV^\L ~-~ {1\over 1 +e(\b)}\ ch~\hV^{\L+\b}
~-~ {1\over 1 +e(\b')}\ ch~\hV^{\L+\b'}
~+\cr
&+~ {1\over (1 +e(\b))\ (1 +e(\b'))}\ ch~\hV^{\L+\b+\b'}
~-~ \car\ ,&\chodrg{b}\cr &&\cr
&\hW_{\b,\b'} ~\equiv~
\hW_{\b} \ \times \hW_{\b'}
 &\chodrg{c}
}$$
The above formula is proved similarly to what we had in the SRC cases.
It reflects the contribution of the modules on embedding diagram
\embdii{}.
In fact, the two terms with  minus sign
on the first line of \chodrg{b} take into
account the factorization of the  oddly embedded
submodules ~$I^{\b},I^{\b'}$, cf. \genuna{}, coming from the
modules ~$V_{10}\,$, $V_{01}\,$, resp. There can be no
contribution of the modules along the same lines of embeddings
~$V_{k0}\,$, $V_{0\ell}\,$, ~$k,\ell >1$, due to the Grassmannian nature
of the odd embeddings involved. Consequently, all modules ~$V_{k\ell}$~
for ~$k,\ell >1$ cannot contribute to the character formula of UIR
in ~$V_{00}\,$. Only the module ~$V_{11}$~ can contribute
since it is also a non-zero submodule of ~$V_{00}\,$.
However, since it is oddly embedded
in ~$V_{00}$~ via both submodules ~$V_{10}\,$, $V_{01}\,$,
its contribution is taken out {\it two
times} - once with ~$I^{\b}$, and a second time with ~$I^{\b'}$.
Thus, we need the term with  plus sign
on the second line of \chodrg{b} to restore its contribution once.\foot{
For more complicated application of similar arguments we refer to
\Dobn.}
We can not apply the same kind of arguments for ~$N=1$,
nevertheless, formula \chodrg{} holds also then for the
case \paor{a}, cf. Appendix A.1.

In accord with \chodrg{} for $N>1$  and ~$d=d^{ac}$~ holds the following
decomposition:
\eqn\decgg{ \(\hL_{\rm long} \)_{\vert_{d=d^{ac}}}
~=~ \hL_{\L} ~\oplus~ \hL_{\L +\b} ~\oplus~
\hL_{\L + \b'} ~\oplus~ \hL_{\L +\b+ \b'}\ ,
\qquad r_1r_{N-1} >0\ ,}
where ~$\hL_{\rm long}$~ is a long superfield with the same values of
~$r_i$~ as ~$\L$,
while the value of $j_1$, (resp. $j_2$), is zero when $j_1$,
(resp. $j_2$), from $\L$ is zero,
otherwise it has to be the generic value $j_1\geq N/2$,
(resp. $j_2\geq N/2$).

Next we consider the four DRC cases separately.

\vskip 3mm

\nt{\bu{\bf ac}}~~~
$d ~=~ d_{\rm max}~=~ d^1_{N1} = d^3_{NN} ~=~ d^{ac} ~\equiv~
2 + j_1 + j_2 + m_1$\ , ~~~$z ~=~ j_1-j_2 +2m/N -m_1$\ .

\nt
In these semi-short DRC cases hold the two null conditions
\nulla{} and \nullc{}.
In addition, for ~$N>1$~ if ~$r_i ~=~ 0$, $i=1,\ldots,i_0\,$,
hold \adnu{} and if ~$r_{N-i} ~=~ 0$, $i=1,\ldots,i'_0\,$,
hold   \adnuu{}.

There are two basic situations. The first is when ~$i_0+i'_0 \leq N-2$.
(This situation is not applicable for $N=1$.) This
means that not all $r_i$ are zero and all eliminations are as
described separately for cases ~\bu{\bf a}~ and ~\bu{\bf c}.
These semi-short UIRs may be called Grassmann-analytic
following \FSa, since odd generators from different chiralities
are eliminated.
The maximal number of states in ~$\hL_\L$~
is ~$2^{4N-2- i_0-i'_0}$.

The second is when ~$i_0+i'_0 \leq N-2$~ does not hold which means that all
~$r_i$~ are zero, ($R$-symmetry scalars, $m_1=0=m$), and
in fact we have ~$i_0=i'_0=N-1$~ and all generators
~$X^+_{1,4+k}$~ and ~$X^+_{3,4+k}$~ are eliminated.
The maximal number of states in ~$\hL_\L$~ is ~$2^{2N}$.

Note that below only one case is applicable for $N=1$.

\nt\bu ~~For ~$j_1j_2> 0$~ the corresponding character formulae are
combinations of \chsa{} and \chsc{}:
\eqna\chdac
$$\eqalignno{
ch~\hL_\L ~=&~
\prod_{\a \in\D^+_\I
\atop {\a\neq \a_{1,4+k}\,,
\atop {k=1,\ldots,1+i'_0
\atop {\a\neq \a_{3,5+N-j}\,,
\atop j=1,\ldots,1+i_0}}}}
\ (1 +e(\a)) ~-~ \car ~= &\chdac{a}\cr &&\cr
=&~ \sum_{\hs\in\hW^{ac}_{i_0,i'_0}}\ (-1)^{\ell(\hs)}\ \hs\cdot ch~\hV^\L
~-~ \car \ , &\chdac {b}\cr
=&~ \sum_{\hs\in\hW^{ac}_{i_0,i'_0}}\ (-1)^{\ell(\hs)}\ \hs\cdot
\( ch~\hV^\L ~-~ \car_{\rm long}\) \ , &\chdac {c}\cr
&\hW^{ac}_{i_0,i'_0} ~\equiv~
\hW^a_{i_0} \ \times \hW^c_{i'_0}
 \ ,&\chdac {d} \cr
 &d \ =\ d_{\rm max} \ =\ d^1_{N1} = d^3_{NN} = 2+j_1+j_2+m_1\,,
 \quad j_1j_2> 0,\cr
{\rm either} ~&~ i_0+i'_0 \leq N-2, \cr
&~ r_i =0,\ i=1,2,\ldots,i_0,N-i'_0,N-i'_0+1,\ldots,N-1, \cr
&~r_i> 0,\ i = i_0+1, N-i'_0-1\ ,\cr
{\rm or} ~&~ i_0=i'_0=N-1\ , \quad r_i=0, ~\forall\, i \ .}$$
The last subcase is of $R$-symmetry scalars. It is also the
only formula in the case under consideration - {\bf ac} -
valid for $N=1$ (where there are no counterterms since
\conz{a,b} bring no restrictions, cf. also Appendix A.1).

For ~$N>1$~ and ~$i_0=i'_0= 0$~ formula \chdac{} is equivalent to
\chodrg{} with ~$\b=\a_{15}\,$, ~$\b'=\a_{3,4+N}\,$. Also
\decgg{} holds  with these $\b,\b'$~:
\eqn\decac{ \(\hL_{\rm long}\)_{\vert_{d=d^{ac}}}
~=~ \hL_{\L} ~\oplus~ \hL_{\L +\a_{15}} ~\oplus~
\hL_{\L + \a_{3,4+N}} ~\oplus~ \hL_{\L +\a_{15}+ \a_{3,4+N}}\ ,
\qquad r_1r_{N-1} >0 \ , }
and with ~$\hL_{\rm long}$~ being a long superfield
with the same values of  ~$r_i$~ as $\L$ and with ~$j_1,j_2\geq N/2$.

All formulae below to the end of case {\bf ac} are for $N>1$.

\nt\bu ~~
For ~$j_1> 0,j_2= 0$~ the corresponding character formulae are
combinations of \chodra{} and \chsc{}~:
\eqna\chodrac
$$\eqalignno{
ch~\hL_\L ~=&~ \sum_{\hs\in\hW^{a'c}_{i'_0}}
(-1)^{\ell(\hs)}\ \hs\cdot ch~\hV^\L
~-~ \car ~=&\chodrac{a}\cr &&\cr
=&~ \sum_{\hs\in\hW^{a'c}_{i'_0}}
(-1)^{\ell(\hs)}\ \hs\cdot \( ch~\hV^\L
~-~ \car_{\rm long} \) \ , &\chodrac{b}\cr &&\cr
\hW^{a'c}_{i'_0} ~\equiv&~ \hW_{\b_{12}}\ \times\ \hW^c_{i'_0}
\ ,\qquad \b_{12} = \a_{3,4+N} + \a_{4,4+N}\ ,
&\chodrac{c}\cr
&~ d ~=~ d_{\rm max} ~=~ d^1_{N1} = d^3_{NN} = 2+j_1+m_1\  , \cr
&~ j_1> 0,\ j_2= 0 \ , \quad r_1>0 \ .
}$$

For ~$i_0=i'_0= 0$~ holds  decomposition \decgg{}:
\eqn\decacp{ \(\hL_{\rm long}\)_{\vert_{d=d^{ac}}}
~=~ \hL_{\L} ~\oplus~ \hL_{\L +\a_{15}} ~\oplus~
\hL_{\L + \b_{12}} ~\oplus~ \hL_{\L +\a_{15}+ \b_{12}}\ ,
\qquad r_1r_{N-1} >0 \ , }
where ~$\hL_{\rm long}$~ is a long superfield with the same values of
~$j_2(=0),r_i$~ as $\L$ and with ~$j_1\geq N/2$.
Note that the UIR ~$\hL_{\L +\a_{15}}$~ is also of the type {\bf
ac} under consideration, while the last two UIRs are short
from type {\bf bc} considered below.

For $R$-symmetry scalars we combine \chsaaz{} and \chsc{a}:
\eqn\chldacyy{\eqalign{
ch~\hL_\L ~=&~
\ \sum_{k=1}^N \prod_{i=1}^k
e(\a_{4,5+N-i}) ~+~
\prod_{\a \in\D^+_\I
\atop {\a\neq \a_{1,4+k}\,,
\atop {k=1,\ldots,N
\atop \ve_2> 0}}}
\ (1 +e(\a))
~-~ \car ~, \cr&\cr
&~ d ~=~ d_{\rm max} ~=~ d^1_{N1} = d^3_{NN} = 2+j_1\  , \cr
&~ j_1> 0, j_2= 0 \ , \quad r_i=0,\ \forall\, i \ .}}

\nt\bu ~~
For ~$j_1=0,j_2> 0$~ the corresponding character formulae are
combinations of \chodrc{} and \chsa{}~:
\eqna\chodracz
$$\eqalignno{
ch~\hL_\L ~=&~ \sum_{\hs\in\hW^{ac'}_{i_0}}
(-1)^{\ell(\hs)}\ \hs\cdot ch~\hV^\L
~-~ \car ~=&\chodracz{a}\cr &&\cr
=&~ \sum_{\hs\in\hW^{ac'}_{i_0}}
(-1)^{\ell(\hs)}\ \hs\cdot \( ch~\hV^\L
~-~ \car_{\rm long} \) \ , &\chodracz{b}\cr &&\cr
\hW^{ac'}_{i_0} ~\equiv&~ \hW_{\b_{34}}\ \times\ \hW^a_{i_0}
\ ,\qquad \b_{34} = \a_{15} + \a_{25}\ ,
&\chodracz{c}\cr
&~ d ~=~ d_{\rm max} ~=~ d^1_{N1} = d^3_{NN} = 2+j_2+m_1\  , \cr
&~ j_1= 0,\ j_2> 0 \ , \quad r_{N-1}>0 \ .
}$$

For ~$i_0=i'_0= 0$~ holds  decomposition \decgg{}:
\eqn\decacq{ \(\hL_{\rm long} \)_{\vert_{d=d^{ac}}}
~=~ \hL_{\L} ~\oplus~ \hL_{\L +\a_{3,4+N}} ~\oplus~
\hL_{\L + \b_{34}} ~\oplus~ \hL_{\L +\a_{3,4+N}+ \b_{34}}\ ,
\qquad r_1r_{N-1} >0\ , }
where ~$\hL_{\rm long}$~ is a long superfield with the same values of
~$j_1(=0),r_i$~ as $\L$ and with ~$j_2\geq N/2$.
Note that the UIR ~$\hL_{\L +\a_{3,4+N}}$~ is again of the type {\bf
ac} under consideration, while the last two UIRs are actually
from type {\bf ad} considered below.

For $R$-symmetry scalars we combine \chsa{a} and \chsccz{}:
\eqn\chldacxx{\eqalign{
ch~\hL_\L ~=&~
 \sum_{k=1}^{N}\ \prod_{i=1}^{k} \
e(\a_{2,4+i}) ~+~
\prod_{\a \in\D^+_\I
\atop {\a\neq \a_{3,4+k}
\atop { k=1,\ldots,N
\atop \ve_4> 0}}}
\ (1 +e(\a))
~-~ \car ~, \cr &\cr
d ~&=~ d_{\rm max} ~=~ d^1_{N1} = d^3_{NN} = 2+j_2 \  ,\cr
& \quad j_1=0, j_2> 0 \ , \quad r_i=0\,, \forall\, i\ . }}

\nt\bu ~~
For ~$j_1=j_2= 0$~ the corresponding character formulae are
combinations of \chodra{} and \chodrc{}~:
\eqna\chodraczz
$$\eqalignno{
ch~\hL_\L ~=&~ \sum_{\hs\in\hW^{a'c'}_{i'_0}}
(-1)^{\ell(\hs)}\ \hs\cdot ch~\hV^\L
~-~ \car ~=&\chodraczz{a}\cr &&\cr
=&~ \sum_{\hs\in\hW^{a'c'}_{i'_0}}
(-1)^{\ell(\hs)}\ \hs\cdot \( ch~\hV^\L
~-~ \car_{\rm long} \) \ , &\chodraczz{b}\cr &&\cr
\hW^{a'c'}_{i'_0} ~\equiv&~ \hW_{\b_{12}}\ \times\ \hW_{\b_{34}}
\ ,&\chodraczz{c}\cr
&~ d ~=~ d_{\rm max} ~=~ d^1_{N1} = d^3_{NN} = 2+m_1\  , \cr
&~ j_1=j_2= 0 \ , \quad r_1r_{N-1}>0 \ .
}$$

For ~$i_0=i'_0= 0$~ holds  decomposition \decgg{}:
\eqn\decacr{ \(\hL_{\rm long} \)_{\vert_{d=d^{ac}}}
~=~ \hL_{\L} ~\oplus~ \hL_{\L +\b_{12}} ~\oplus~
\hL_{\L + \b_{34}} ~\oplus~ \hL_{\L +\b_{12}+ \b_{34}}\ ,
\qquad r_1r_{N-1} >0\ , }
where ~$\hL_{\rm long}$~ is a long superfield with the same values of
~$j_1(=0),j_2(=0),r_i$~ as $\L$.
Note that the UIR ~$\hL_{\L +\b_{12}}$~ is of the type {\bf
bc}, ~$\hL_{\L +\b_{34}}$~ is of the type {\bf
ad}, ~$\hL_{\L +\b_{12} +\b_{34}}$~ is of the type {\bf
bd}, these three being  considered below.

For $R$-symmetry scalars we combine \chsaaz{} and \chsccz{}:
\eqn\chldaczz{\eqalign{
ch~\hL_\L ~=&~
 \sum_{k=1}^{N}\ \prod_{i=1}^{k} \
e(\a_{2,4+i}) ~+~
 \sum_{k=1}^N \prod_{i=1}^k
e(\a_{4,5+N-i}) ~+ \cr &\cr
&+~ \prod_{\a \in\D^+_\I
\atop {\ve_2> 0
\atop \ve_4> 0}}
\ (1 +e(\a))
~-~ \car ~, \cr &\cr
&~ d ~=~ d_{\rm max} ~=~ d^1_{N1} = d^3_{NN} = 2\ , ~~ z=0\ , \cr
&~ j_1=j_2= 0 \ , \quad r_i=0,\ \forall\, i \ .}}

\vskip 3mm

\nt{\bu{\bf ad}}~~~
$d ~=~ d^1_{N1} = d^4_{NN} ~=~ 1 + j_2 + m_1\ , ~~ j_1=0$,
~~~$z ~=~ 2m/N -m_1-1-j_2$\ .

\nt
In these short DRC cases hold the three null conditions \nulla{},
\nulld{} and \nulldd.
In addition, for $N>1$ if ~$r_i ~=~ 0$, $i=1,\ldots,i_0\,$,
hold \adnu{} and if ~$r_{N-i} ~=~ 0$, $i=1,\ldots,i'_0\,$,
hold \nullddd{}.

If ~$i_0+i'_0 \leq N-2$~ all eliminations are as
described separately for cases ~\bu{\bf a}~ and ~\bu{\bf d}.
All these are Grassmann-analytic UIRs.
The maximal number of states in ~$\hL_\L$~
is ~$2^{4N-3- i_0-2i'_0}$.
Interesting subcases are the so-called BPS states, cf.,
\AFSZ, \FSb, \FSa, \EdSo, \Ryz, \DHRy, \ArSo, \HHHR.
They are characterized by the number ~$\k$~ of odd generators which
annihilate them - then the corresponding case is called
~${\k\over 4N}$-BPS state.
For example consider ~$N=4$~ and $\quarter$-BPS cases with $z=0
~\Rightarrow ~d ~=~ 2m/N\,$. One such case is obtained for
$i_0=1,i'_0=0,j_2> 0$, then ~$d ~=~ \half(2r_2 + 3 r_3)$, $r_1=0$,
$r_2> 0$, $r_3 = 2(1+j_2)$.

For ~$j_2m_1> 0$~ the corresponding character formula is a
combination of \chsa{} and \chsd{}:
\eqna\chdad
$$\eqalignno{
ch~\hL_\L ~=&~ \prod_{\a \in\D^+_\I \atop {\a\neq \a_{3,5+N-k} \ ,
\atop {k=1,\ldots,1+i_0 \atop {\a\neq \a_{a,4+j} \ , \atop a=1,2\,,
\ j=1,\ldots,1+i'_0}}}} \ (1 +e(\a)) ~-~ \car ~= &\chdad{a}\cr &&\cr
=&~ \sum_{\hs\in\hW^{ad}_{i_0,i'_0}}\ (-1)^{\ell(\hs)}\ \hs\cdot
ch~\hV^\L ~-~ \car \ , &\chdad {b}\cr &\hW^{ad}_{i_0,i'_0} ~\equiv~
\hW^a_{i_0}\ \times \hW^d_{i'_0}
 \ ,&\chdad {c} \cr
&d ~=~ d^1_{N1} = d^4_{NN} ~=~ 1 + j_2 + m_1\ , \quad j_1=0,j_2> 0 \ ,
\quad i_0+i'_0 \leq N-2, \cr
&~ r_i =0,\ i=1,2,\ldots,i_0,N-i'_0,N-i'_0+1,\ldots,N-1, \cr
&~r_i> 0,\ i = i_0+1, N-i'_0-1\, .
}$$

For ~$i_0=i'_0= 0$~ some of these UIRs appear (up to two times)
in the decomposition \decacq{}.
More precisely, those with ~$r_i>2\d_{i,N-1}\,$, $i=1,N-1$,
appear as the term
~$\hL_{\L + \b_{34}}$, while those with
~$r_i>\d_{i1}+2\d_{i,N-1}\,$, $i=1,N-1$, appear also as the term
~$\hL_{\L +\a_{3,4+N}+ \b_{34}}\,$.

For ~$j_2= 0,m_1> 0$~ the corresponding character formula is a
combination of \chodra{} and \chsd{b}:
\eqna\chodrad
$$\eqalignno{
ch~\hL_\L ~=&~ \sum_{\hs\in\hW^{a'd}_{i'_0}}
(-1)^{\ell(\hs)}\ \hs\cdot ch~\hV^\L
~-~ \car \ ,&\chodrad{a}\cr &&\cr
\hW^{a'd}_{i'_0} ~\equiv&~ \hW_{\b_{12}}\ \times\ \hW^d_{i'_0}
\ ,&\chodrad{b}}$$
where ~$\b_{12} = \a_{3,4+N} + \a_{4,4+N}\,$.
For ~$i_0=i'_0= 0$~ some of these UIRs appear in the
decomposition \decacr{} or \decacq{}. More precisely, those with
~$r_i>2\d_{i,N-1}\,$, $i=1,N-1$,  appear as the term
~$\hL_{\L + \b_{34}}$~ of \decacr{}, while those with
~$r_i>\d_{i1}+2\d_{i,N-1}\,$, $i=1,N-1$,  appear as the term
~$\hL_{\L + \a_{3,4+N} +
\b_{34}}$~ of \decacq{} but only when
~$j_2=\half$~ in ~$\L$~ there.

In the case of $R$-symmetry scalars we have ~$i_0=i'_0=N-1$,
~$\k ~=~ 3N$~ and all generators
~$X^+_{1,4+k}\,$, ~$X^+_{2,4+k}\,$,
~$X^+_{3,4+k}\,$ are eliminated. Here holds ~$d = -z = 1+j_2\,$.
These anti-chiral irreps form one of the three series of ~{\bf
massless} UIRs; they
are denoted ~$\chi^+_s\,$, $s=j_2=0,\half,1,\ldots$, in Section 3 of \DPu.
Besides the vacuum they contain only ~$N$~ states in ~$\hL_\L$~
given by \posss{} for ~$k=1,\ldots,N$.
These should be called ultrashort UIRs.
The character formula can be written
in the most explicit way:
\eqna\chdadz
$$\eqalignno{
ch~\hL_\L ~=&~
  1 + \sum_{k=1}^N \prod_{i=1}^k
e(\a_{4,5+N-i}) \ , &\chdadz {}\cr &&\cr
&~d ~=~ d^1_{N1} = d^4_{NN} = 1 + j_2 = -z \ , \cr
&~ j_1=0 \ , \quad r_i=0, \ \forall\, i \ ,}$$
and it is valid for any $j_2\,$.
In the case under consideration - {\bf ad} - only the last
character formula is valid for $N=1$ (cf. Appendix A.1).

\vskip 3mm

The next case is conjugate to the previous one.

\vskip 3mm

\nt{\bu{\bf bc}}~~~
$d ~=~ d^2_{N1} = d^3_{NN} ~=~ 1 + j_1 + m_1\ , ~~ j_2=0$\ ,
~~~$z ~=~ 2m/N -m_1 +1+j_1$ \ .

\nt
In these short DRC cases hold the three null conditions \nullb{},
\nullbb{} and \nullc{}.
In addition, for $N>1$ if ~$r_i ~=~ 0$, $i=1,\ldots,i_0\,$,
hold \adreu{} and if ~$r_{N-i} ~=~ 0$, $i=1,\ldots,i'_0\,$,
hold \adnuu{}.

If ~$i_0+i'_0 \leq N-2$~ all eliminations are as
described separately for cases ~\bu{\bf b}~ and ~\bu{\bf c}.
These are also Grassmann-analytic UIRs.
The maximal number of states in ~$\hL_\L$~
is ~$2^{4N-3- 2i_0-i'_0}$.
Here for ~$N=4$~ one $\quarter$-BPS case is obtained for
$i_0=0,i'_0=1,j_1> 0$, then ~$d ~=~ \half(2r_2 + 3 r_1)$,
$r_1 = 2(1+j_1)$, $r_2> 0$, $r_3=0$.

For ~$j_1m_1> 0$~ the corresponding character formula is a
combination of \chsb{} and \chsc{}:
\eqna\chdbc
$$\eqalignno{
ch~\hL_\L ~=&~
\prod_{\a \in\D^+_\I
\atop {\a\neq \a_{1,4+k} \ ,
\atop {k=1,\ldots,1+i'_0
\atop {\a\neq \a_{a,5+N-j} \ ,
\atop a=3,4\,, \ j=1,\ldots,1+i_0}}}}
\ (1 +e(\a)) ~-~ \car ~= &\chdbc a\cr &&\cr
=&~ \sum_{\hs\in\hW^{bc}_{i_0,i'_0}}\ (-1)^{\ell(\hs)}\ \hs\cdot
ch~\hV^\L ~-~ \car \ , &\chdbc {b}\cr
&\hW^{bc}_{i_0,i'_0} ~\equiv~
\hW^b_{i_0}\ \times \hW^c_{i'_0}
 \ ,&\chdbc {c} \cr
&d ~=~ d^2_{N1} = d^3_{NN} ~=~ 1 + j_1 + m_1\ , \quad j_1> 0,j_2= 0 \ ,
\quad i_0+i'_0 \leq N-2, \cr
&~ r_i =0,\ i=1,2,\ldots,i_0,N-i'_0,N-i'_0+1,\ldots,N-1, \cr
&~r_i> 0,\ i = i_0+1,N-i'_0-1\, .
}$$
For ~$i_0=i'_0= 0$~ some of these UIRs appear in the
decomposition \decacp{}. More precisely, those with
~$r_i>2\d_{i1}$, $i=1,N-1$,  appear as the term
~$\hL_{\L + \b_{12}}$, while those with
~$r_i>2\d_{i1}+\d_{i,N-1}\,$, $i=1,N-1$,  appear as the term
~$\hL_{\L +\a_{15}+ \b_{12}}\,$.

For ~$j_1= 0,m_1> 0$~ the corresponding character formula is a
combination of \chodrc{} and \chsb{b}:
\eqna\chodrbc
$$\eqalignno{
ch~\hL_\L ~=&~ \sum_{\hs\in\hW^{bc'}_{i_0}}
(-1)^{\ell(\hs)}\ \hs\cdot ch~\hV^\L
~-~ \car \ ,&\chodrbc{a}\cr &&\cr
\hW^{bc'}_{i_0} ~\equiv&~ \hW_{\b_{34}}\ \times\ \hW^b_{i_0}
\ ,&\chodrbc{b}}$$
where ~$\b_{34} = \a_{15} + \a_{25}\,$.
For ~$i_0=i'_0= 0$~ some of these UIRs appear in the
decomposition \decacr{} or \decacp{}. More precisely, those with
~$r_i>2\d_{i1}\,$, $i=1,N-1$, appear as the term
~$\hL_{\L + \b_{12}}$~ of \decacr{}, while those with
~$r_i>2\d_{i1}+\d_{i,N-1}\,$, $i=1,N-1$, appear as the term
~$\hL_{\L +\a_{15}+ \b_{12}}$~ of \decacp{} but only when
~$j_1=\half$~ in ~$\L$~ there.

In the case of $R$-symmetry scalars we have ~$i_0=i'_0=N-1$,
~$\k ~=~ 3N$~ and all generators
 ~$X^+_{1,4+k}\,$, ~$X^+_{3,4+k}\,$, ~$X^+_{4,4+k}\,$,
 are eliminated.
These chiral irreps form another series of ~{\bf massless}
UIRs, conjugate to the
first above; they are denoted ~$\chi_s\,$,
$s=j_1=0,\half,1,\ldots$, in Section 3
of \DPu. Besides the vacuum they contain only ~$N$~ states in
~$\hL_\L$~ given
by \possy{} for ~$k=1,\ldots,N$.
These should also be called ultrashort UIRs.
The character formula is:
\eqna\chdbcz
$$\eqalignno{
ch~\hL_\L ~=&~
  1 + \sum_{k=1}^N \prod_{i=1}^k
e(\a_{2,4+i}) \ , &\chdbcz b\cr &&\cr
&~d ~=~ d^2_{N1} = d^3_{NN} = 1 + j_1 = z \ , \cr
&~ j_2=0 \ , \quad r_i=0, \ \forall\, i \ ,}$$
and it is valid for any $j_1\,$.
In the case under consideration - {\bf bc} - only the last
character formula is valid for $N=1$ (cf. Appendix A.1).

\vskip 3mm\nt{\bu{\bf bd}}~~~
$d ~=~ d^2_{N1} = d^4_{NN} ~=~ m_1\ , ~~ j_1=j_2=0$,
~~~$z ~=~ 2m/N -m_1$ \ .

In these short DRC cases hold the four null conditions \nullb{},
\nullbb{}, \nulld{} and \nulldd.

For $N=1$ this is the trivial irrep with $d=z=0$.
This follows from the fact that since ~$d=j_1=j_2=0$~ also holds the
even reducibility condition \redd{b} (and consequently
\redd{d,e,f}). Thus, we have the null conditions:
~$X^+_k \, |\L\rg ~=~ 0 $~ for all simple root generators (and
consequently for all generators) and the irrep consists only of
the vacuum ~$|\L\rg\,$.

For $N>1$ the situation is non-trivial.
In addition to the mentioned conditions, and if ~$r_i ~=~ 0$,
$i=1,\ldots,i_0\,$, hold
\adreu{} and if ~$r_{N-i} ~=~ 0$, $i=1,\ldots,i'_0\,$, hold \nullddd{}.

If ~$i_0+i'_0 \leq N-2$~ all eliminations are as
described separately for cases ~\bu{\bf b}~ and ~\bu{\bf d}.
These are also Grassmann-analytic UIRs.
The maximal number of states in ~$\hL_\L$~
is ~$2^{4N-4- 2i_0-2i'_0}$.
For ~$N=4$~ for the BPS cases we take ~$z=\half(r_3-r_1) =0
~\Rightarrow~ d ~=~ 2r_1+r_2\,$.
In the $\quarter$-BPS case we have $i_0=i'_0=0$,
$r_1=r_3> 0$.

For ~$i_0=i'_0= 0$~ some of these UIRs appear in the
decomposition \decacr{}. More precisely, those with
~$r_i>2\d_{i1}+2\d_{i,N-1}\,$, $i=1,N-1$, ~ appear as the term
~$\hL_{\L + \b_{12}+ \b_{34}}\,$.

Most interesting is the case ~$i_0+i'_0 ~=~ N-2$, then there
is only one non-zero $r_i\,$,
namely, ~$r_{1+i_0}=r_{N-1-i'_0}>
0$, while the rest ~$r_i$~ are zero.
Thus, the Young tableau
parameters are: ~$m_1 =r_{1+i_0}$, ~$m=(1+i_0)r_{1+i_0}\,$.\nl
An important subcase is when ~$d=m_1=1$, then ~$m=i_0+1=
N-1-i'_0\,$, ~$r_i=\d_{mi}$, and these irreps form the
third series of ~{\bf massless} UIRs. In Section 3 of \DPu{}
they are parametrized by $n\in\bbn$, $\half N \leq n < N$, and
denoted by ~$\chi'_n\,$, $n=m$, ($z= 2n/N-1$),
~$\chi'^+_n\,$, $n=N-m$, ($z= 1-2n/N$).
Note that for ~even $N$ there is the coincidence:
~$\chi'_{n} ~=~\chi'^+_{n}\,$, where ~$n ~=~ m ~=~ N-m ~=~ N/2$.
Here we shall parametrize these UIRs by the parameter ~$i_0 ~=~
0,1,\ldots,N-2$.

Another interesting subcase here is for even $N$ with
~$z=0 ~\Rightarrow~ d=m_1 = 2m/N$ ~$\Rightarrow ~i_0=i'_0=N/2 -1$
~$\Rightarrow ~m_1 =r_{N/2}$, ~$m={N\over 2}r_{N/2 }\,$. These are $\half$-BPS states
for ~$m_1>1$ and ~$\trq$-BPS for ~$m_1=1$.
The latter are the massless self-conjugate case: $\chi'_{n}$, $n =N/2$ mentioned above.
For ~$N=4$ we have: $i_0=i'_0=1$,
$r_1=r_3= 0$, $r_2> 0$, which is  massless for $r_2=1$.

Finally, in the case of $R$-symmetry scalars we have
~$i_0=i'_0=N-1$~ and all ~$4N$~ odd
generators ~$X^+_{1,4+k}\,$, ~$X^+_{2,4+k}\,$, ~$X^+_{3,4+k}\,$,
~$X^+_{4,4+k}\,$, are eliminated. More than this, all quantum
numbers are zero, (cf. \disd{d,d'}), and this is
the trivial irrep. The latter follows exactly as explained above
for the case $N=1$.

For ~$m_1> 0$~ the corresponding character formula is a
combination of \chsb{} and \chsd{}:
\eqna\chdbd
$$\eqalignno{
ch~\hL_\L ~=&~
\prod_{\a \in\D^+_\I
\atop {\a\neq \a_{j,5+N-k}
\atop {j=3,4\,, \ k=1,\ldots,1+i_0
\atop {\a\neq \a_{j',4+k'}
\atop j'=1,2\,, \ k'=1,\ldots,1+i'_0
}}}}
\ (1 +e(\a)) ~~-~~ \car ~= &\chdbd{a} \cr &&\cr
=&~ \sum_{\hs\in\hW^{bd}_{i_0,i'_0}}\ (-1)^{\ell(\hs)}\ \hs\cdot
ch~\hV^\L ~-~ \car \ , &\chdbd {b}\cr
&\hW^{bd}_{i_0,i'_0} ~\equiv~
\hW^b_{i_0}\ \times \hW^d_{i'_0}
 \ ,&\chdbd {c} \cr
&~ d ~=~ d^2_{N1} = d^4_{NN} = m_1 \ ,\quad j_1=j_2=0\ ,
 \quad i_0+i'_0 \leq N-2, \cr
&~ r_i =0,\ i=1,2,\ldots,i_0,N-i'_0,N-i'_0+1,\ldots,N-1, \cr
&~r_i> 0,\ i = i_0+1, N-i'_0-1\, ,
}$$
where $\car$ designates the counter-terms due to our Criterion,
in particular, due to \conz{}
taken with ~$\ve_{j,5+N-k}=0$, ~$j=3,4$, ~$k=1,\ldots,1+i_0\,$,
~$\ve_{j',4+k'}=0$, ~$j'=1,2$, ~$k'=1,\ldots,1+i'_0\,$.

Also for the third series of massless UIRs we can give a much more
explicit character formula without counter-terms. Fix
the parameter ~$i_0 ~=~ 0,1,\ldots,N-2$. Then there are only the
following states in ~$\hL_\L$~:
\eqna\posm
$$\eqalignno{
&X^+_{2,N+4-j}\,\cdots\,X^+_{2,N+4-i_0} \, |\L\rg \ ,
\quad j=0,1,\ldots,i_0 \ , &\posm a\cr
&X^+_{4,4+k}\,\cdots\,X^+_{4,N+3-i_0} \, |\L\rg \ ,
\quad k=1,\ldots,N-1-i_0 \ , &\posm b\cr}$$
altogether ~$N$~ states besides the vacuum.\nl
{\it Demonstration:}~~ {\iti Indeed, besides \posm{}
no other states involving generators ~$X^+_{a,4+k}$~ for $a=2,4$
are possible due to the restrictions \conz{}. Note that the
generators of the latter kind which do not appear in \posm{} are
eliminated due to \nullb, \adreu{b} and \nulld, \nullddd{b}.
We have to discuss the generators ~$X^+_{a,4+k}$~ for $a=1,3$.
Part of them are eliminated due to \adreu{a} and \nullddd{a}.
The rest are: ~$X^+_{1,N+4-j}\,$, ~$j=0,1,\ldots,i_0$~
and ~$X^+_{4,4+k}\,$, ~$k=1,\ldots,N-1-i_0\,$. They can not act
on the vacuum, so they can only act on some of the states
in \posm{a} or \posm{b}, resp. For two of these:
~$X^+_{1,N+4-i_0}$~ and ~$X^+_{3,N+3-i_0}$~ it is easy to see that
they can not act on any state. For the rest:
 ~$X^+_{1,N+4-j}\,$, ~$j=0,1,\ldots,i_0-1$~
and ~$X^+_{4,4+k}\,$, ~$k=1,\ldots,N-2-i_0\,$,
the only possibility for action
which can not be excluded in an obvious way, is:
\eqna\posmm
$$\eqalignno{
&X^+_{1,N+4-j}\,X^+_{2,N+3-j}\,\cdots\,X^+_{2,N+4-i_0} \, |\L\rg \ ,
\quad j=0,1,\ldots,i_0-1\ , &\posmm a\cr
&X^+_{3,4+k}\,X^+_{4,5+k}\,\cdots\,X^+_{4,N+3-i_0} \, |\L\rg \ ,
\quad k=1,\ldots,N-2-i_0 \ . &\posmm b\cr}$$
However, all these states are not allowed. This is shown as
for the states \poss{} and \possi. Thus, besides the vacuum,
~$\hL_\L$~ contains only the $N$ states given in \posm{}.}\bsq

The corresponding character formula for the massless UIRs of
this series is therefore:
\eqn\chdbdz{\eqalign{
ch~\hL_\L ~=&~
1 ~+~
\sum_{j=0}^{i_0}\ \prod_{i=j}^{i_0} \
e(\a_{2,N+4-i}) ~+~
 \sum_{k=1}^{N-1-i_0}\ \prod_{i=k}^{N-1-i_0} \
e(\a_{4,4+i})
 \ , \cr
&~d ~=~ d^2_{N1} = d^4_{NN} = m_1 = 1,\quad
i_0 ~=~ 0,1,\ldots,N-2 \ , \cr
& z = 2(i_0+1)/N - 1 \ , \quad
 j_1=j_2=0 \ , \quad r_i ~=~ \d_{i,i_0+1}\ .
}}

\vskip 5mm

\nt{\it Remark: ~~
In this paper we use the Verma (factor-)module realization of the
UIRs. We give here a short remark on what happens with the ER
realization of the
UIRs. As we know, cf. \DPf, the ERs are superfields depending on Minkowski
space-time and on $4N$ Grassmann coordinates ~$\theta_a^i$,
~$\bar{\theta}_b^k$, $a,b=1,2$, $i,k=1,\ldots,N$.
\foot{A mathematically precise formulation is given in \DPf, while for
the even case we refer to \Doo,\Dor.} \
There is 1-to-1 correspondence in these dependencies
and the odd null conditions. Namely, if the condition
~$X^+_{a,4+k}\ |\L\rg ~=~ 0$, $a=1,2$, holds, then the superfields of the
corresponding ER do not depend on the variable ~$\theta_a^k\,$,
while if the condition
~$X^+_{a,4+k}\ |\L\rg ~=~ 0$, $a=3,4$, holds, then the superfields of the
corresponding ER do not depend on the variable
~$\bar{\theta}_{a-2}^k\,$.
These statements were used in the proof of unitarity for the ERs
picture, cf. \DPp, but were not explicated. They were analyzed in
detail in the papers \AFSZ,\FS,\FSb,\FSa, using the notions of
'harmonic superspace analyticity' and Grassmann analyticity.}\dia

\np

\newsec{Discussion and Outlook}

\nt
First we summarize the results on decompositions of long
irreps as they descend to the unitarity threshold.

In the SRC cases we have embedding formula \embs{} and UIRs are
given by formula \genun{}. Starting from this in subsection (3.3)
we have established that for ~$d=d_{\rm max}$~
there hold the following decompositions:
\eqn\decog{
\(\hL_{\rm long}\)_{\vert_{d=d_{\rm max}}}
 ~=~ \hL_\L ~\oplus \hL_{\L+\b} \ , }
where there are two possibilities for $\L$ and four possibilities
for $\b$ as given in \paorz{a,c,e,f}, however, for $N>1$ there are
additional conditions on $r_i\,$. In more detail, $\L$ and $\b$
are specified as follows:
\eqna\dista
$$\eqalignno{
d ~=&~ d_{\rm max} ~=~ d^a ~=~ d^1_{N1}   > d^3_{NN}\ , \quad  ~r_1>0\ ,
&\dista a\cr
&\b ~=~ \a_{3,4+N}\ , \quad j_2> 0\ , &\dista {a'}\cr
&\b ~=~  \a_{3,4+N}+\a_{4,4+N}\ , \quad j_2=0
\ , &\dista {a''}\cr
d ~=&~ d_{\rm max} ~=~ d^c ~=~ d^3_{NN} > d^1_{N1}\ , \quad r_{N-1}>0\ ,
&\dista c\cr
&\b ~=~ \a_{15}\ ,\quad j_1> 0  \ , &\dista {c'}\cr
&\b ~=~ \a_{15}+\a_{25}\ ,\quad j_1= 0 \ . &\dista {c''}}$$
The corresponding four decompositions are given in formulae \deca{}, \decaa{},
\decc{}, \deccc{}, resp., and in each case it is explained how
$\hL_{\rm long}$ is specified. It is also noted that in cases
\dista{a'',c''} the UIRs $\hL_{\L+\b}$ are short from types given
in \dist{b,d}, resp., and with $r_1>2$, $r_{N-1}>2$, resp.

In the DRC cases we have embedding formulae \embd{}, \embdi{},
\embdii{}, and UIRs are given by formula \genuna{}. Starting from
this in subsection (3.4) we have established that for $N>1$ and
~$d=d_{\rm max}=d^{ac} $~ hold the following decompositions:
\eqn\decggg{ \(\hL_{\rm long}\)_{\vert_{d=d^{ac}}} ~=~ \hL_{\L}
~\oplus~ \hL_{\L +\b} ~\oplus~ \hL_{\L + \b'} ~\oplus~ \hL_{\L +\b+
\b'}\ , \qquad r_1r_{N-1} >0\ ,} where $\L$ is the semi-short DRC
designated as type {\bf ac} and there are  four possibilities for
$\b,\b'$ as given in \paor{a,b,c,d}. The corresponding four
decompositions are given in formulae \decac{}, \decacp{}, \decacq{},
\decacr{},  resp., and in each case it is explained how  $\hL_{\rm
long}$ is specified. Note that in \decac{} all UIRs are semi-short.
In \decacp{} the first two UIRs are  semi-short, the last two UIRs
are short of type {\bf bc}. From the latter two the first is with
$r_1>2,r_{N-1}>0$, ($r_1>2$ if $N=2$), the second is with
$r_1>2,r_{N-1}>1$, ($r_1>3$ if $N=2$). In \decacq{} the first two
UIRs are  semi-short, the last two UIRs are short of type {\bf ad}.
From the latter two the first is with $r_1>0,r_{N-1}>2$, ($r_1>2$ if
$N=2$), the second is with $r_1>1,r_{N-1}>2$, ($r_1>3$ if $N=2$). In
\decacr{} the first UIR is the semi-short, the other three UIRs are
short of types {\bf bc}, {\bf ad}, {\bf bd}, resp. From the latter
three the first is with $r_1>2,r_{N-1}>0$, the second is with
$r_1>0,r_{N-1}>2$, the third is with $r_1,r_{N-1}>2$, ($r_1>4$ if
$N=2$).

Summarizing the above, we note first that for $N=1$ all SRC cases
enter some decomposition \decog{}, while no DRC cases enter any
decomposition \decggg{}. For $N>1$ the situation is more diverse and
so we give the list of UIRs that do {\bf not} enter decompositions
\decog{} and \decggg{}:

\nt\bu ~~~{\bf SRC cases:}

\nt\bu{\bf a} ~~~
$d ~=~ d_{\rm max} ~=~ d^a ~=~ d^1_{N1}
~=~ 2 +2j_2 +z+2m_1 -2m/N ~>~ d^3_{NN}$ \ ,\nli
 $j_1,j_2$ arbitrary, ~$r_1=0$\ .

\vskip 3mm

\nt{\bu{\bf b}}~~~
$d ~=~ d^2_{N1} ~=~ z+2m_1 -2m/N ~>~ d^3_{NN}\ , ~~ j_2=0$\ ,\nli
$j_1$ arbitrary, ~$r_1\leq 2$\ .

\vskip 3mm

\nt\bu{\bf c} ~~~
$d ~=~ d_{\rm max} ~=~ d^c ~=~ d^3_{NN}
~=~ 2 +2j_1 -z+ 2m/N ~>~ d^1_{N1}$ \ ,\nli
$j_1,j_2$ arbitrary, ~$r_{N-1}=0$\ .

\vskip 3mm

\nt{\bu{\bf d}}~~~
$d ~=~ d^4_{NN} ~=~ -z+2m/N ~>~ d^1_{N1}\ , ~~ j_1=0$\ ,\nli
$j_1$ arbitrary, ~$r_{N-1}\leq 2$\ .

\vskip 3mm

\nt\bu ~~~{\bf DRC cases:}\nl
all non-trivial cases for $N=1$, while for $N>1$ the list is:

\nt\bu{\bf ac} ~~~
$d ~=~ d_{\rm max} ~=~ d^{ac} ~=~
d^1_{N1} = d^3_{NN} ~=~ 2 + j_1 + j_2 +
m_1$\ , ~~~$z ~=~ j_1-j_2 +2m/N -m_1\,$,\nli
$j_1,j_2$ arbitrary, ~$r_1r_{N-1}=0$\ .

\vskip 3mm

\nt{\bu{\bf ad}}~~~
$d ~=~ d^1_{N1} = d^4_{NN} ~=~ 1 + j_2 + m_1\ , ~~ j_1=0$,
~~~$z ~=~ -1-j_2+ 2m/N -m_1$\ ,\nli
$j_2$ arbitrary, ~$r_{N-1}\leq 2$\ ,
~~~$r_1=0\ {\rm for}\ N>2$.

\vskip 3mm

\nt{\bu{\bf bc}}~~~
$d ~=~ d^2_{N1} = d^3_{NN} ~=~ 1 + j_1 + m_1\ , ~~ j_2=0$,
~~~$z ~=~ 1+j_1+ 2m/N -m_1$\ ,\nli
$j_1$ arbitrary, ~$r_{1}\leq 2$\ ,
~~~$r_{N-1}=0\  {\rm for}\ N>2$.

\vskip 3mm

\nt{\bu{\bf bd}}~~~
$d ~=~ d^2_{N1} = d^4_{NN} ~=~ m_1\ , ~~ j_1=j_2=0$,
~~~$z ~=~ 2m/N -m_1$\ ,\nli
$r_1,r_{N-1}\leq 2$\ {\rm for}\ $N>2$,
~~~$r_1 \leq 4$\ {\rm for}\ $N=2$.

\vskip 3mm

We would like to point out possible application of our results to
current developments in conformal field theory.
Recently, there is interest in superfields with conformal
dimensions which are protected from renormalisation in the sense
that they cannot develop anomalous dimensions
\FSa,\AEPS,\BKRSa,\AES,\HeHoa,\DoOs.
Initially, the idea was that this happens because the representations
under which they transform determine these dimensions uniquely.
Later, it was argued that one can tell which operators will be
protected in the quantum theory simply by looking at the
representations they transform under and whether they can be
written in terms of single trace 1/2 BPS operators (chiral
primaries or CPOs) on analytic superspace \HeHoa{}.
In \DoOs{} it was shown how, at the unitarity threshold, a long
multiplet can be decomposed into four semi-short multiplets, and
decompositions similar to \decgg{}, i.e., involving the modules
in \embdii{} (as given in \DPm), were considered  for $N=2,4$.
However, the decompositions of \DoOs{}
are justified on the dimensions of the
finite-dimensional irreps of the Lorentz  and  $su(N)$
subalgebras involved in the superfields involved in the
decompositions, and in particular,
the latter hold also when $r_1r_{N-1}=0$.

Independently of the above, we would like to make a ~{\it
mathematical}~ remark.  As a by-product of our analysis we have
obtained character formulae for the complex Lie superalgebras
$sl(4/N)$. The point is that our character formulae have as
starting point character formulae of Verma modules and
factor-modules over $sl(4/N)$.  Thus, almost all character
formulae in Section 3, more precisely, formulae \chlll{}, \chs{},
\chodr{}, \chct{}, \chsa{}, \chodra{}, \chsb{}, \chsc{},
\chodrc{}, \chsd{}, \chodrg{}, \chdac{}, \chodrac{},  \chodracz{},
\chodraczz{}, \chdad{}, \chodrad{}, \chdadz{a}, \chdbc{}, \chodrbc{},
\chdbcz{a}, \chdbd{}, become character formulae for $sl(4/N)$
for the same values of the representation parameters by just
discarding the counter-terms $\car$, $\car_{\rm long}\,$, resp.

Finally, let us mention that we explicate our results for $N=1,2$ in
Appendix A. There we display explicitly all decompositions \decog{},
\decggg{}, and when these do not hold, all quasi-decompositions
(like \dectaz{}) that replace them.  We leave similar detailed
discussion for\ $N=4$\ for the follow-up paper.

\vskip 5mm

\nt {\bf Acknowledgments.}
~The author would like to thank V.B. Petkova and E. Sokatchev
for useful discussions. The author would like to thank for hospitality
the Abdus Salam International Center for Theoretical Physics,
where work on the revision was done. This work was supported in part by
the Bulgarian National Council for Scientific Research, grant
F-1205/02,  the Alexander von Humboldt
Foundation in the framework of the
Clausthal-Leipzig-Sofia Cooperation,
the TMR Network EUCLID, contract HPRN-CT-2002-00325,
 and the European RTN  'Forces-Universe',
contract MRTN-CT-2004-005104.

\np

\appendix{A}{Explicit character formulae for ~N=1,2}

\subsec{{\bf N=1}}

\nt
For $N=1$ the displayed results are almost explicit,
so we can allow telegram style.

\nt\bu{\bf ~~ Long superfields}

\nt If ~$j_1j_2> 0$~ then ~$\hL_\L$~ has the maximum
possible number of states: ~$16$. The character formula is \chll{}.

If ~$j_1=0,j_2> 0$~ then the generator ~$X^+_{15}$~ can appear
only together with the generator ~$X^+_{25}\,$, and ~$\hL_\L$~
has $12$ states = 3(chiral)$\times${}4(anti-chiral)
states.\foot{In statements like this each sector
includes the vacuum.}
The character formula is \chlll{} with:
\eqn\chnoa{
\car ~=~ e(\a_{15}) (1 + e(\a_{35})) (1 + e(\a_{45}))}

The next case is conjugate.
If ~$j_1> 0,j_2= 0$~ then the generator ~$X^+_{35}$~ can appear
only together with the generator ~$X^+_{45}\,$, and ~$\hL_\L$~
has $12$ states. The character
formula is \chlll{} with:
\eqn\chnob{
\car ~=~ e(\a_{35}) (1 + e(\a_{15})) (1 + e(\a_{25}))}

The next case combines the previous two.
If ~$j_1=j_2= 0$~ then the generator ~$X^+_{15}$~ can appear
only together with the generator ~$X^+_{25}\,$,
the generator ~$X^+_{35}$~ can appear
only together with the generator ~$X^+_{45}\,$, and ~$\hL_\L$~
has $9$ states = 3(chiral)$\times${}3(anti-chiral) states.
The character formula is \chlll{} with:
\eqn\chnoc{
\car ~=~ e(\a_{15}) (1 + e(\a_{35})) (1 + e(\a_{45})) ~+~
e(\a_{35}) (1 + e(\a_{15})) (1 + e(\a_{25}))
~-~ e(\a_{15}) e(\a_{35}) \ ,}
i.e., we combine the counter-terms of the previous two cases,
but need to subtract a counter-term that is counted twice.

\nt\bu ~~~{\bf SRC cases}

\nt\bu{\bf a} ~~~$d ~=~ d_{\rm max} ~=~ d^1_{11} ~=~
2 +2j_2 +z ~>~ d^3_{11}$ \ .

\nt\bu ~~$j_1> 0$. The generator $X^+_{35}$ is eliminated
(though for different reasons for $j_2> 0$ and $j_2 =0$,
cf. \nulla, resp. \nulnraz)
and there are only $8$ states.\foot{For brevity, here and often
below we shall say "there are M states" meaning
"there are M states in ~$\hL_\L$~".} Then the character formula
is \chs{} (or equivalently \chodr{})
without counter-terms:
\eqn\chsnoa{\eqalign{
ch~\hL_\L ~=&~
 \prod_{\a \in\D^+_\I \atop \a\neq \a_{35}}
\ (1 +e(\a)) ~, \cr
&~ d ~=~ d_{\rm max} ~=~ d^1_{11} > d^3_{11} \ ,
\quad j_1> 0 \ .}}

For ~$j_2> 0$~ the decomposition \deca{} is fulfilled with
~$\hLL$~ having $16$ states as the maximal long superfield with
~$j_1j_2> 0$, while $\hL_{\L+\a_{35}}$ has 8 states (being the
same type as $\hL$).

For ~$j_2= 0$~ the decomposition \decaa{} is fulfilled with
~$\hLL$~ having $12$ states as the long superfield with
~$j_1> 0,j_2=0$, ~$\b ~=~ \a_{35}+\a_{45}$~ and
~$\hL_{\L+\a_{35}+\a_{45}}$~ having 4 states
- it actually belongs to case {\bf b} below (for
$j_1> 0,j_2=0$).

\nt\bu ~~$j_1=0$. The generator $X^+_{35}$ is eliminated,
the generator ~$X^+_{15}$~ can appear
only together with the generator ~$X^+_{25}$~
and there are only $6$ states.
Then the character formula is \chs{} (equivalently \chodr{}):
\eqn\chsnoaa{\eqalign{
ch~\hL_\L ~=&~
 \prod_{\a \in\D^+_\I
\atop {
\atop \a\neq \a_{35} }}
\ (1 +e(\a)) ~-~ \car ~, \cr
&\car ~=~ e(\a_{15}) (1 + e(\a_{45})) \cr
&~ d ~=~ d_{\rm max} ~=~ d^1_{11} > d^3_{11} \ ,
\quad j_1=0 \ .}}
This formula is equivalent also to \chct{}, noting:
\eqn\chctt{
\car ~=~ e(\a_{15}) (1 + e(\a_{45})) ~=~
(1-\hs_{\a_{35}})\ \cdot \car_{\rm long} \ ,}
taking ~$\car_{\rm long}$~ from \chnoa{}.

For ~$j_2> 0$~ the decomposition \deca{} is fulfilled with
~$\hLL$~ having $12$ states as the long superfield with
~$j_1=0,j_2> 0$, while $\hL_{\L+\a_{35}}$ has 6 states (being the
same type as $\hL$).

For ~$j_2= 0$~ the decomposition \decaa{} is fulfilled with
~$\hLL$~ having $9$ states as the long superfield with
~$j_1=j_2=0$, while ~$\hL_{\L+\a_{35}+\a_{45}}$~ has 3 states
- it actually belongs to the next case {\bf b}, cf. below
(for $j_1=j_2=0$).

\nt{\bu{\bf b}}~~~~~
$d ~=~ d^2_{11} ~=~ z ~>~ d^3_{11}\ , ~~ j_2=0$\ .

\nt\bu ~~$j_1> 0$. The generators $X^+_{35}$ and
$X^+_{45}$ are eliminated and there are only $4$ states.
Then the character formula is \chsb{} (for $i_0=0$)
without counter-terms:
\eqn\chsnob{\eqalign{ch~\hL_\L ~=&~
 (1 +e(\a_{15})) \ (1 +e(\a_{25})) ~, \cr
&d ~=~ d^2_{11} > d^3_{11}\ , ~~ j_1> 0, j_2=0\ .}}
These UIRs and the next subcase enter formula \deca{} together with
UIRs of case {\bf a} as we have shown above.

\nt\bu ~~$j_1= 0$. The generators $X^+_{35}$ and
$X^+_{45}$ are eliminated,
the generator ~$X^+_{15}$~ can appear
only together with the generator ~$X^+_{25}\,$,
and there are only $3$ states.
Then the character formula is \chsb{} (for $i_0=0$) with counter-term
~$\car = e(\a_{15})$~:
\eqn\chsnobz{\eqalign{
ch~\hL_\L ~=&~
~ 1 +e(\a_{25}) + e(\a_{15}) e(\a_{25})
 ~, \cr
&d ~=~ d^2_{11} > d^3_{11}\ , ~~ j_1=j_2=0\ .}}
Here holds also an analog of \chct{} with ~$\hW_\b$~ replaced by
~$\hW^b_0$~ and ~$\car_{\rm long}$~ from \chnoa{}.

\nt\bu{\bf c} ~~~$d ~=~ d_{\rm max} ~=~ d^3_{11}
~=~ 2 +2j_1 -z ~>~ d^1_{11}$ \ .

\nt\bu ~~$j_2> 0$. The generator $X^+_{15}$ is eliminated
(though for different reasons for $j_1> 0$ and $j_1 =0$,
cf. \nullc, resp. \nulnray)
and there are only $8$ states.
Then the character formula is \chsc{} without counter-terms:
\eqn\chsnoc{\eqalign{
ch~\hL_\L ~=&~
\prod_{\a \in\D^+_\I \atop \a\neq \a_{15}}
\ (1 +e(\a)) ~, \cr
&~ d ~=~ d_{\rm max} ~=~ d^3_{11} > d^1_{11} \ ,
\quad j_2> 0 \ .}}

For ~$j_1> 0$~ the decomposition \decc{} is fulfilled with
~$\hLL$~ having $16$ states as the maximal long superfield with
~$j_1j_2> 0$.
For ~$j_1= 0$~ the decomposition \deccc{} is fulfilled with
~$\hLL$~ having $12$ states as the long superfield with
~$j_1= 0,j_2> 0$, and ~$L_{\L+\a_{15}+\a_{25}}$~ having 4 states
- it actually belongs to the next case {\bf d}, cf. below (for
$j_1= 0,j_2>0$).

\nt\bu ~~$j_2=0$. The generator $X^+_{15}$ is eliminated,
the generator ~$X^+_{35}$~ can appear
only together with the generator ~$X^+_{45}$~
and there are only $6$ states.
Then the character formula is \chsc{}:
\eqn\chsnocc{\eqalign{
ch~\hL_\L ~=&~
\prod_{\a \in\D^+_\I
\atop {
\atop \a\neq \a_{15} }}
\ (1 +e(\a)) ~-~ \car ~, \cr
&\car ~=~ e(\a_{35}) (1 + e(\a_{25})) \cr
&~ d ~=~ d_{\rm max} ~=~ d^3_{11} > d^1_{11} \ ,
\quad j_2=0 \ .}}
This formula is equivalent also to \chct{}
with ~$\car_{\rm long}$~ from \chnob{}.

For ~$j_1> 0$~ the decomposition \decc{} is fulfilled with
~$\hLL$~ having $12$ states as the long superfield with
~$j_1> 0,j_2= 0$.
For ~$j_1= 0$~ the decomposition \deccc{} is fulfilled with
~$\hLL$~ having $9$ states as the long superfield with
~$j_1=j_2=0$,  and ~$L_{\L+\a_{15}+\a_{25}}$~ having 3 states
- it actually belongs to the next case {\bf d}, cf. below,
(for $j_1= j_2=0$).

\nt{\bu{\bf d}}~~~~~
$d ~=~ d^4_{11} ~=~ -z ~>~ d^1_{11}\ , ~~ j_1=0$\ .

\nt\bu ~~$j_2> 0$. The generators $X^+_{15}$ and
$X^+_{25}$ are eliminated and there are only $4$ states.
Then the character formula is \chsd{} (for $i'_0=0$)
without counter-terms:
\eqn\chsnod{\eqalign{
ch~\hL_\L ~=&~
 (1 +e(\a_{35})) \ (1 +e(\a_{45})) ~, \cr
&d ~=~ d^4_{11} > d^1_{11}\ , ~~ j_1=0, j_2> 0\ .}}
These UIRs and the next subcase enter formula \decc{} together with
UIRs of case {\bf c} as we have shown above.

\nt\bu ~~$j_2= 0$. The generators $X^+_{15}$ and
$X^+_{25}$ are eliminated,
the generator ~$X^+_{35}$~ can appear
only together with the generator ~$X^+_{45}\,$,
and there are only $3$ states.
Then the character formula is \chsd{} (for $i'_0=0$)
with counter-term
~$\car = e(\a_{35})$~:
\eqn\chsnodd{\eqalign{
ch~\hL_\L ~=&~
 1 +e(\a_{45}) + e(\a_{35}) e(\a_{45})
 ~, \cr
&d ~=~ d^4_{11} > d^1_{11}\ , ~~ j_1=j_2=0\ .}}
Here holds also an analog of \chct{} with ~$\hW_\b$~ replaced by
~$\hW^d_0$~ and ~$\car_{\rm long}$~ from \chnob{}.

\nt\bu ~~~{\bf DRC cases}

\nt{\bu{\bf ac}}~~~
$d ~=~ d_{\rm max} ~=~ d^1_{11} = d^3_{11} ~=~ d^{ac}
~=~ 2 + j_1 + j_2$\ ,
~~~$z ~=~ j_1-j_2$\ .

\nt
The generators $X^+_{15}$ and $X^+_{35}$ are eliminated
(though for different reasons for $j_1> 0$ and $j_1 =0$,
resp., for $j_2> 0$ and $j_2 =0$). There are only $4$ states and
the character formula is \chdac{} (for $i_0=i'_0=0$)
without counterterms:
\eqn\chdnoac{\eqalign{
ch~\hL_\L ~=&~ (1 +e(\a_{25})) \ (1 +e(\a_{45})) ~=\cr
=&~ ch~\hV^\L ~-~ {1\over 1 +e(\a_{15})}\ ch~\hV^{\L+\a_{15}}
~-~ {1\over 1 +e(\a_{35})}\ ch~\hV^{\L+\a_{35}}
~+\cr
&+~ {1\over (1 +e(\a_{15})) (1 +e(\a_{35}))}\ ch~\hV^{\L+\a_{15}+\a_{35}}
\ ,}}
where the terms with minus may be interpreted as taking out states,
while the last term indicates adding back what was taken two
times. This may written also in the form of the following
pseudo-decomposition:
\eqn\decss{ \(\hL_{\rm long}\)_{\vert_{d=d^{ac}}}
~=~ \hL_{\L} ~\oplus~ \hL_{\L +\a_{15}} ~\oplus~
\hL_{\L + \a_{35}} ~\ominus~ \hL_{\L +\a_{15}+ \a_{35}} }
where ~$\hL_{\L +\a_{15}}\,$, ~$\hL_{\L + \a_{35}}\,$,
are SRC UIRs with 8 states each described above in cases {\bf c},
{\bf a}, resp. They are embedded in ~$\hV^\L$~ via the generators
\ $X^+_{15}\,$, $X^+_{35}\,$, resp.
Together with ~$\hL_{\L}$~ this brings in terms which has to be
taken out with the last term in which the
 representation denoted ~$\hL_{\L +\a_{15}+ \a_{35}}$~ is
supposed to have the same 4 states as $\hL_{\L}$ and these excessive
states are de-embedded via the composition of the other two maps,
i.e., via the product of generators
\ $X^+_{15}\, X^+_{35}\,$.

\nt{\bu{\bf ad}}~~~
$d ~=~ d^1_{11} = d^4_{11} ~=~ 1 + j_2 = -z\ , ~~ j_1=0$.

\nt
The generators $X^+_{15}$, $X^+_{25}$ and $X^+_{35}$ are eliminated
(for the latter for different reasons for $j_2> 0$ and $j_2 =0$).
These are the first series of massless UIRs, and everything is
already explicit in the general formulae.
There are only $2$ states and the character formula is
\chdadz{} for $N=1$~:
\eqn\chdnoac{ch~\hL_\L ~=~ 1 +e(\a_{45}) \ . }

\nt{\bu{\bf bc}}~~~
$d ~=~ d^2_{11} = d^3_{11} ~=~ 1 + j_1 = z\ , ~~ j_2=0$.

\nt
The generators $X^+_{15}$, $X^+_{35}$ and $X^+_{45}$ are eliminated
(for the first for different reasons for $j_1> 0$ and $j_1 =0$).
These are the second series of massless UIRs.
There are only $2$ states and the character formula is
\chdbcz{} for $N=1$~:
\eqn\chdnoac{
ch~\hL_\L ~=~ 1 +e(\a_{25}) \ .}

\vskip 3mm\nt{\bu{\bf bd}}~~~
$d ~=~ d^2_{11} = d^4_{11} ~=~ j_1=j_2=z ~=0$

\nt
As we explained in the detail this is the trivial
1-dimensional irrep consisting of the vacuum.

\vskip 5mm

\subsec{N=2}

\nt\bu{\bf ~~ Long superfields}

\nt
We first write down conditions \conz{} explicitly for $N=2$:
\eqna\onznt
$$\eqalignno{
&\ve_{15} + \ve_{16} ~\leq~ \ve_{25} + \ve_{26} + 2j_1
&\onznt a\cr
&\ve_{35} + \ve_{36} ~\leq~ \ve_{45} + \ve_{46} + 2j_2
&\onznt b\cr
& \ve_{16} + \ve_{26} + \ve_{35} + \ve_{45}
~\leq~ \ve_{15} + \ve_{25} + \ve_{36} + \ve_{46} + r_1
&\onznt c\cr}$$
To simplify the exposition we classify the generators by their
contribution to \onznt{}. Namely, the chiral, anti-chiral,
operators:
\eqn\chiro{\eqalign{
&\Phi^c ~=~ (X^+_{16})^{\ve_{16}}\, (X^+_{15})^{\ve_{15}} \,
 (X^+_{26})^{\ve_{26}}\, (X^+_{25})^{\ve_{25}} \ ,\cr
&\Phi^a ~=~ (X^+_{35})^{\ve_{35}} \,
(X^+_{36})^{\ve_{36}} \,
     (X^+_{45})^{\ve_{45}} \,
     (X^+_{46})^{\ve_{46}} \ , }}
will be distinguished by the values (cf. also \indd{}):
\eqn\inde{\eqalign{
& \ve^c_j ~=~ \ve_{25} + \ve_{26} - \ve_{15} - \ve_{16} \ ,\cr
& \ve^a_j ~=~ \ve_{45} + \ve_{46} - \ve_{35} - \ve_{36} \ ,\cr
& \ve^c_r ~\equiv~ \ve_{15} + \ve_{25} - \ve_{16} - \ve_{26} \ ,\cr
& \ve^a_r ~\equiv~ \ve_{36} + \ve_{46} - \ve_{35} - \ve_{45} \ ,\cr
& \ve_r ~\equiv~ \ve_r^1 ~=~ \ve^c_r+\ve^a_r \ .
}}
Explicitly, the chiral operators are arranged as follows:
\eqna\incaa
$$\eqalignno{
&X^+_{15} \, X^+_{25} \ ,
\qquad \ve^c_r =2\ , ~\ve^c_j =0\ , \cr
&X^+_{25} \ , \quad
X^+_{26} \, X^+_{15} \, X^+_{25} \ ,
\qquad \ve^c_r =1\ , ~\ve^c_j =1\ , \cr
&X^+_{15} \ , \quad
X^+_{16} \, X^+_{15} \, X^+_{25} \ ,
\qquad \ve^c_r =1\ , ~\ve^c_j =-1\ , \cr
&X^+_{26} \, X^+_{25} \ ,
\qquad \ve^c_r =0\ , ~\ve^c_j = 2\ , \cr
&{\bf 1}\ ,\quad
X^+_{16}\, X^+_{25} \ ,\quad
~X^+_{26} \, X^+_{15} \ , \quad
~X^+_{16} \,X^+_{26} \, X^+_{15} \, X^+_{25} \ ,
\qquad \ve^c_r =0\ , ~\ve^c_j = 0\ , \cr
&X^+_{16} \, X^+_{15} \ ,
\qquad \ve^c_r =0\ , ~\ve^c_j = -2\ , \cr
&X^+_{26} \ , \quad
X^+_{26} \, X^+_{16} \, X^+_{25} \ ,
\qquad \ve^c_r =-1\ , ~\ve^c_j = 1\ , \cr
&X^+_{16} \ , \quad
X^+_{16} \, X^+_{15} \, X^+_{26} \ ,
\qquad \ve^c_r =-1\ , ~\ve^c_j = -1\ , \cr
&X^+_{16} \, X^+_{26} \ ,
\qquad \ve^c_r =-2\ , ~\ve^c_j = 0\ , &\incaa{}
}$$
while the anti-chiral operators are arranged as follows:
\eqna\incla
$$\eqalignno{
&X^+_{36} \, X^+_{46} \ ,
\qquad \ve^a_r =2\ , ~\ve^a_j =0\ , \cr
&X^+_{46} \ , \quad
X^+_{45} \, X^+_{36} \, X^+_{46} \ ,
\qquad \ve^a_r =1\ , ~\ve^a_j =1\ , \cr
&X^+_{36} \ , \quad
X^+_{35} \, X^+_{36} \, X^+_{46} \ ,
\qquad \ve^a_r =1\ , ~\ve^a_j =-1\ , \cr
&X^+_{45} \, X^+_{46} \ ,
\qquad \ve^a_r =0\ , ~\ve^a_j = 2\ , \cr
&{\bf 1}\ ,\quad
X^+_{35}\, X^+_{46} \ ,\quad
~X^+_{45} \, X^+_{36} \ , \quad
~X^+_{35} \,X^+_{45} \, X^+_{36} \, X^+_{46} \ ,
\qquad \ve^a_r =0\ , ~\ve^a_j = 0\ , \cr
&X^+_{35} \, X^+_{36} \ ,
\qquad \ve^a_r =0\ , ~\ve^a_j = -2\ , \cr
&X^+_{45} \ , \quad
X^+_{45} \, X^+_{35} \, X^+_{46} \ ,
\qquad \ve^a_r =-1\ , ~\ve^a_j = 1\ , \cr
&X^+_{35} \ , \quad
X^+_{35} \, X^+_{36} \, X^+_{45} \ ,
\qquad \ve^a_r =-1\ , ~\ve^a_j = -1\ , \cr
&X^+_{35} \, X^+_{45} \ ,
\qquad \ve^a_r =-2\ , ~\ve^a_j = 0\ . &\incla{}
}$$
The same arrangement applies to the states obtained by
applying the operators on the vacuum (for which all these indices
naturally have zero value). We have added also the identity
operator ~{\bf 1}~ in order to be able to take into account
the vacuum automatically.

The allowed states satisfy: ~$\ve^c_j + 2j_1 \geq 0$,
~$\ve^a_j + 2j_2 \geq 0$, ~$\ve_r + r_1 \geq 0$, cf. \conz{}.
Now we are ready to classify the allowed states depending on the
values of ~$j_1,j_2,r_1\,$.
Actually what we do below amounts to giving explicitly formula
\chlll{}.

\nt\bu ~~First we give the possible states when ~$j_1,j_2\geq 1$~:
\eqna\clasf 
$$\eqalignno{
\Phi^c\,\Phi^a\,|\L\rg \ , \quad &j_1,j_2\geq 1, ~r_1\geq
4,\quad 256
~{\rm states} \ ; &\clasf a\cr
\Phi^c\,\Phi^a\,|\L\rg \ , \quad &j_1,j_2\geq 1, ~r_1= 3,
\quad 255
~{\rm states} \ , \cr
&{\rm excluding\ the\ state} ~~X^+_{16} \, X^+_{26}\, X^+_{35} \,
X^+_{45}\ , ~{\rm with} ~\ve_r =-4\ ; &\clasf b\cr
\Phi^c\,\Phi^a\,|\L\rg \ , \quad &j_1,j_2\geq 1, ~r_1= 2,
\quad 247
~{\rm states} \ , \cr
&{\rm excluding\ the\ 9\ states\ with}~ ~\ve_r \leq -3
 \ ; &\clasf c\cr
    \Phi^c\,\Phi^a\,|\L\rg \ , \quad &j_1,j_2\geq 1, ~r_1= 1,
\quad 219
~{\rm states} \ , \cr
&{\rm excluding\ the\ 37\ states\ with}~ ~\ve_r \leq -2
 \ ; &\clasf d\cr
\Phi^c\,\Phi^a\,|\L\rg \ , \quad &j_1,j_2\geq 1, ~r_1= 0,
\quad 163
~{\rm states} \ , \cr
&{\rm excluding\ the\ 93\ states\ with}~ ~\ve_r \leq -1
 \ . &\clasf e\cr
}$$

Further we classify the states when ~$j_1,j_2\geq 1$ is not
fulfilled using the five cases in \clasf{} as reference point.

\nt\bu ~~$j_1\geq 1$, $j_2 =\half\,$.\nl
With respect to \clasf{} we exclude 16 states with $\ve^a_j=-2$,
(so $\ve^a_j + 2j_2 =-1$)~:
\eqn\exa{ X^+_{35} \, X^+_{36} \, \Phi^c\, |\L\rg \ .}
However, for \clasf{d,e} the case when ~$\Phi^c = X^+_{16} \,
X^+_{26}$, (with $\ve^c_r=-2$) is already taken out, and for
\clasf{e} the four cases of $\Phi^c$
 with $\ve^c_r=-1$ are already taken out.
Thus, altogether, in the five cases corresponding to
\clasf{a,b,c,d,e} we take out ~$16,16,16,15,11$~ states and so
there remain now ~$240,239,231,204,152$~ states.

\nt\bu ~~$j_1=\half\,$, $j_2\geq 1$.\nl
This is the case conjugate to the previous.
With respect to \clasf{} we exclude 16 states with $\ve^c_j=-2$,
(so $\ve^c_j + 2j_1 =-1$)~:
\eqn\exb{ X^+_{16} \, X^+_{15} \, \Phi^a\, |\L\rg \ .}
Noting the double-counting for the five cases $\Phi^a$
with $\ve^a_r=-2,-1$, in the cases corresponding to
\clasf{a,b,c,d,e} now have ~$240,239,231,204,152$~ states.

\nt\bu ~~$j_1=j_2=\half\,$.\nl
This is a combination of the previous two cases.
With respect to \clasf{} we exclude the states we excluded
in both, which would double the numbers (to $32,32,32,30,22$),
however, we have to take into account that the state
~$X^+_{16} \, X^+_{15} \, X^+_{35} \, X^+_{36} \, |\L\rg$~ is
counted two times.
Thus, altogether, in the five cases corresponding to
\clasf{a,b,c,d,e} we take out ~$31,31,31,29,21$~ states and so
there remain now ~$225,224,216,190,142$~ states.

\nt\bu ~~$j_1\geq 1$, $j_2 =0$.\nl
In addition to the states excluded in the case
~$j_1\geq 1$, $j_2 =\half\,$, we exclude 64 states with $\ve^a_j=-1$,
(so $\ve^a_j + 2j_2 =-1$)~:
\eqna\exc
$$\eqalignno{
&X^+_{36}\, \Phi^c\,|\L\rg \ , \quad
X^+_{35} \, X^+_{36} \, X^+_{46}\, \Phi^c\,|\L\rg \ , &\exc a\cr
&X^+_{35}\, \Phi^c\,|\L\rg \ , \quad
X^+_{35} \, X^+_{36} \, X^+_{45}\, \Phi^c\,|\L\rg \ .&\exc b\cr}$$
We have to take into account that certain states were
already taken out, namely, the following:\nl
- for \clasf{c,d,e} the two cases \exc{b} with
~$\Phi^c = X^+_{16} \,
X^+_{26}$, (so that $\ve_r=-3$);\nl
- for \clasf{d,e} the eight cases obtained by
combining \exc{b} with $\Phi^c$
 with $\ve^c_r=-1$ (so that $\ve_r=-2$);\nl
- for \clasf{e} the 12 cases obtained by
combining \exc{b} with $\Phi^c$
 with $\ve^c_r=0$ (so that $\ve_r=-1$);\nl
- for \clasf{e} the two cases \exc{a} with
~$\Phi^c = X^+_{16} \,
X^+_{26}$, (so that $\ve_r=-1$).\nl
Altogether, for \clasf{c,d,e} the overcounting is by ~$2, 10,24$~
states. Thus, the states we actually take out
w.r.t. the case ~$j_1\geq 1$, $j_2 =\half$~ are
~$64,64,62,54,40$. Finally, for \clasf{e} we have to take out the
impossible state ~$X^+_{36}\,X^+_{45}\,|\L\rg$, cf. \impy.
Altogether the states remaining in the cases
corresponding to \clasf{a,b,c,d,e} are
 ~$176,175,169,150,111$, resp.

\nt\bu ~~$j_1= 0$, $j_2 \geq 1$.\nl
This is the case conjugate to the previous.
In addition to the states excluded in the case
~$j_1=\half\,$, $j_2\geq 1$, we exclude 64 states with $\ve^c_j=-1$,
(so $\ve^c_j + 2j_1 =-1$)~:
\eqna\exd
$$\eqalignno{
&X^+_{15}\, \Phi^a\,|\L\rg \ , \quad
X^+_{16} \, X^+_{15} \, X^+_{25}\, \Phi^a\,|\L\rg \ , &\exd a\cr
&X^+_{16}\, \Phi^a\,|\L\rg \ , \quad
X^+_{16} \, X^+_{15} \, X^+_{26}\, \Phi^a\,|\L\rg \ .&\exd b\cr}$$
We have to take into account that certain states were
already taken out, namely, the following:\nl
- for \clasf{c,d,e} the two cases \exd{b} with
~$\Phi^a = X^+_{35} \,
X^+_{45}$, (so that $\ve_r=-3$);\nl
- for \clasf{d,e} the eight cases obtained by
combining \exd{b} with $\Phi^a$
 with $\ve^a_r=-1$ (so that $\ve_r=-2$);\nl
- for \clasf{e} the 12 cases obtained by
combining \exd{b} with $\Phi^a$
 with $\ve^a_r=0$ (so that $\ve_r=-1$);\nl
- for \clasf{e} the two cases \exd{a} with
~$\Phi^a = X^+_{35} \,X^+_{45}$, (so that
$\ve_r=-1$).\nl
Altogether, excluding also the impossible state
 ~$X^+_{15}\,X^+_{26}\,|\L\rg$ (when $r_1=0$, cf. \impz),
in the five cases corresponding to
\clasf{a,b,c,d,e} we now have ~$176,175,169,150,111$~ states.

\nt\bu ~~$j_1=\half\,$, $j_2 =0$.\nl
This is a combination of previous cases so w.r.t. \clasf{} we exclude
the states in \exa{}, \exb{}, \exc{}. Due to overlaps there are
five states which
are counted two times - those in \exa,\exc{} when ~$\Phi^c=
X^+_{16} \, X^+_{15}$. Thus, w.r.t. the case ~$j_1\geq 1$, $j_2
=\half$ we would take out 11 states. However, from those
the state \exb{} with ~$\Phi^a=X^+_{35} \, X^+_{45}$~
was taken out in \clasf{d,e} and the
states \exb{} with ~$\Phi^a= X^+_{45}$~
~$\Phi^a=X^+_{45} \, X^+_{35} \, X^+_{46}$~
were taken out in \clasf{e}.
Thus, w.r.t. the case ~$j_1\geq 1$, $j_2
=\half$~ we take out ~$11,11,11,10,8$~ states. Thus,
in the cases corresponding to \clasf{a,b,c,d,e}
there are ~$165,164,158,140,103$~ states.

\nt\bu ~~$j_1=0$, $j_2 =\half\,$.\nl
This case is conjugate to the previous one and
so w.r.t. \clasf{} we exclude the states in \exa{}, \exb{},
\exd{}. W.r.t. the case ~$j_1=\half\,$, $j_2
\geq 1$ we take out ~$11,11,11,10,8$~ states. Thus,
in the cases corresponding to \clasf{a,b,c,d,e}
there are ~$165,164,158,140,103$.

\nt\bu ~~$j_1=j_2 =0$.\nl
This is a combination of previous cases so we exclude
the states in \exa{}, \exb{}, \exc{}, \exd{}.
Due to overlaps of \exd{} with \exa{} and \exb{}
w.r.t. to the case $j_1=\half\,$, $j_2 =0$ we would take out 44
states (instead of 64). However, from those
the two states \exd{b} with ~$\Phi^a=X^+_{35} \, X^+_{45}$~
were taken out in \clasf{c,d,e},
the four states obtained from \exd{b} with ~$\Phi^a= X^+_{45}$~
~$\Phi^a=X^+_{45} \, X^+_{35} \, X^+_{46}$~
were taken out in \clasf{d,e},
the two states \exd{b} with ~$\Phi^a=X^+_{45} \, X^+_{46}$~
were taken out in \clasf{e},
the eight states obtained from \exd{b} with
~$\Phi^a=${\bf 1},
~$\Phi^a=X^+_{35} \, X^+_{46}$,
~$\Phi^a=X^+_{45} \, X^+_{36}$,
~$\Phi^a=X^+_{35} \, X^+_{45}\, X^+_{36} \, X^+_{46}$,
were taken out in \clasf{e},
the two states \exd{a} with ~$\Phi^a=X^+_{35} \, X^+_{45}$~
were taken out in \clasf{e}.
Thus, the states we actually take out
w.r.t. the case $j_1=\half\,$, $j_2 =0$ are
~$44,44,42,38,26$. For \clasf{e} we have also to take out
two impossible states: \impz{} and its combination with \impy{}:
\eqn\impt{X^+_{15}\, X^+_{26} \,X^+_{36}\, X^+_{45} \, |\L\rg\ .}
Altogether the states remaining in the cases
corresponding to \clasf{a,b,c,d,e} are
 ~$121,120,116,102,75$~ states.

Thus, the smallest $N=2$ long superfield has ~$75$~
states in ~$\hL_\L$. Since above the states we described
by exclusion we would like to list these 75 states.
First there are 6 chiral states:
\eqna\chir
$$\eqalignno{
&X^+_{25} \, |\L\rg \ , \quad
X^+_{15} \, X^+_{25} \, |\L\rg \ \quad
X^+_{26} \, X^+_{15} \, X^+_{25} \, |\L\rg \ , &\chir a\cr
&X^+_{26} \, X^+_{25} \, |\L\rg \ , \quad
X^+_{16} \, X^+_{25} \, |\L\rg \ , \quad
X^+_{16} \,X^+_{26} \, X^+_{15} \, X^+_{25} \, |\L\rg \ . &\chir b}$$
and 6 anti-chiral states:
\eqna\achir
$$\eqalignno{
&X^+_{46} \, |\L\rg \ , \quad
X^+_{36} \, X^+_{46} \, |\L\rg \ , \quad
X^+_{45} \, X^+_{36} \, X^+_{46} \, |\L\rg \ ,
&\achir a\cr
&X^+_{45} \, X^+_{46} \, |\L\rg \ , \quad
X^+_{35} \, X^+_{46} \, |\L\rg \ , \quad
X^+_{35} \,X^+_{45} \, X^+_{36} \, X^+_{46} \, |\L\rg \ ,
&\achir b}$$
Now let ~$\Phi_c \, |\L\rg$, ~$\Phi_a \, |\L\rg$, denote any of the
six states in \chir{}, \achir{}, resp.,
~$\Phi'_c \, |\L\rg$, ~$\Phi'_a \, |\L\rg$, denote any of the
three states in \chir{a}, \achir{a}, resp.
Then, there are the following states:
\eqna\comb
$$\eqalignno{
&|\L\rg \ , \quad
 \Phi_c \, \Phi_a \, |\L\rg \ ,
&\comb a\cr
&X^+_{15}\, X^+_{26} \, \Phi_a \, |\L\rg\ , &\comb b\cr
&X^+_{36}\, X^+_{45} \, \Phi_c \, |\L\rg\ , &\comb c\cr
&X^+_{26} \, \Phi'_a \, |\L\rg\ , \quad
X^+_{16} \,X^+_{26}\, X^+_{25} \, \Phi'_a \, |\L\rg\ , &\comb d\cr
&X^+_{45} \, \Phi'_c \, |\L\rg\ , \quad
X^+_{35} \,X^+_{45}\, X^+_{46} \, \Phi'_c \, |\L\rg\ , &\comb e\cr
&X^+_{35}\, X^+_{45} \, X^+_{15}\, X^+_{25} \, |\L\rg\ , \quad
X^+_{16}\, X^+_{26} \, X^+_{36}\, X^+_{46} \, |\L\rg\ . &\comb f\cr
}$$
Obviously, there are $63$ states in \comb{} ($37+6+6+6+6+2$),
and altogether $75$ states in \achir, \chir{} and
\comb{}.

\vskip 5mm

\nt\bu ~~~{\bf SRC cases}

\nt
Here we consider the SRC cases similarly to the long superfields
taking again the five cases in \clasf{} as reference point.

\nt\bu{\bf a} ~~~$d ~=~ d_{\rm max} ~=~ d^1_{21}
~=~ 2 +2j_2 +z+r_1 ~>~ d^3_{22}$ \ .\nl
The maximal number of states is $128 = 16$(chiral)$\times
8$(anti-chiral), achieved for $j_1\geq 1,r_1\geq 4$.

\nt\bu ~~$j_2> 0\,$.\nl
Here hold character formulae
\chs{}, or equivalently \chodr{} or \chct{} when ~$r_1> 0$,
while for ~$r_1=0$~ the character formula is \chsa{} (for
$i_0=1$). We give more detailed description.

The generator $X^+_{36}$ is eliminated.
The eight states in the anti-chiral sector are obtained by applying
to the vacuum the following operators:
\eqna\incz
$$\eqalignno{
&X^+_{46} \ , \qquad \ve^a_r =1\ , ~\ve^a_j =1\ , \cr
&X^+_{45} \, X^+_{46} \ ,
\qquad \ve^a_r =0\ , ~\ve^a_j = 2\ , \cr
&{\bf 1}\ ,\quad
X^+_{35}\, X^+_{46} \ ,
\qquad \ve^a_r =0\ , ~\ve^a_j = 0\ , &\incz{a}\cr
&X^+_{45} \ , \quad
X^+_{45} \, X^+_{35} \, X^+_{46} \ ,
\qquad \ve^a_r =-1\ , ~\ve^a_j = 1\ , \cr
&X^+_{35} \ ,
\qquad \ve^a_r =-1\ , ~\ve^a_j = -1\ , \cr
&X^+_{35} \, X^+_{45} \ ,
\qquad \ve^a_r =-2\ , ~\ve^a_j = 0\ . \cr
}$$
The above is equivalent to the anti-chiral part of character formula
\chs{}~:
$$\eqalignno{&
(1+e(\a_{46}))\ (1 +e(\a_{35}))\ (1 +e(\a_{45})) &\incz{b}} $$
however, the more detailed description in \incz{a} is necessary
to obtain the results on the counter-terms.
In particular, for ~$r_1=1$~ the last operator does not
contribute to the anti-chiral sector, while for ~$r_1=0$~
only the first three operators contribute to the anti-chiral sector,
and the generator ~$X^+_{35}$~ is also eliminated
from the whole basis.

In summary, the results are.
When ~$j_1\geq 1$~ correspondingly to the cases in
\clasf{a,b,c,d,e} we have now ~$128,127,120,99,42$~ states.
When ~$j_1=\half$~ correspondingly to the cases in
\clasf{a,b,c,d,e} we have now ~$120,119,112,92,39$~ states.
When ~$j_1=0$~ correspondingly to the cases in
\clasf{a,b,c,d,e} we have now ~$88,87,82,68,28$~ states.

When ~$r_1> 0$~ holds formula \deca{} with ~$\b = \a_{36}\,$,
where ~$\hL_{\rm long}$~ is a long superfield with the same
values of ~$j_1$~ and ~$r_i$~ as $\L$,  and with
~$j_2\geq 1$. Note that when the
weight ~$\L$~ corresponds to cases \clasf{a,b,c,d,e} then
the weight ~$\L+\a_{36}$~ corresponds to cases \clasf{a,a,b,c,d}
(since the value of $r_1$ is increased by 1). Thus,
when ~$j_1\geq 1$~ the UIR ~$\hL_{\L+\a_{36}}$~ has
~$128,128,127,120,99$ states, when ~$j_1=\half$~ it has
~$120,120,119,112,92$ states, when ~$j_1=0$~ it has
~$88,88,87,82,68$ states. Summed together with the numbers for
the UIR ~$\hL_{\L}$~ from above we obtain the following
contributions to ~$\hL_{\rm long}$~:
when ~$j_1\geq 1$~ there are
~$256,255,247,219,141$ states, when ~$j_1=\half$~ there are
~$240,239,231,204,131$ states, when ~$j_1=0$~ there are
~$176,175,169,150,96$ states. Except the last cases (in which
$r_1=0$) these cases match exactly (not only by numbers)
the cases of long superfields
for the corresponding values of ~$j_1=1,\half,0$~ and
~$j_2\geq 1$.

When $r_1=0$ the long superfields have ~$163,152,111$ states
i.e., a mis-match of ~$22,21,15$ states. All these extra states
contain the generator ~$X^+_{35}$~ and do not contain the
generator ~$X^+_{36}$. Explicitly, when ~$j_1\geq1$~ the $22$
states are:
\eqn\pseu{ \eqalign{&X^+_{35}\, \Phi^c_1\,\wt \cr
&X^+_{35}\,X^+_{45} \,X^+_{46}\, \Phi^c_1\,\wt \cr
&X^+_{35}\,X^+_{46}\, \Phi^c_2\,\wt \cr
&X^+_{35}\,X^+_{45} \,X^+_{15}\, X^+_{25}\,
\wt \ ,}}
where ~$\Phi^c_1$~ denotes the 5 chiral operators of the first
three rows of \incaa{}, ~$\Phi^c_2$~ denotes the 11 chiral
operators of the first six rows of \incaa{}. When ~$j_1=\half$~ the $21$
states are as in \pseu{} except the state
~$X^+_{35}\,X^+_{46}\, X^+_{16}\,X^+_{15}\, \wt$ which is not in
the long superfield (since ~$\ve^c_j + 2j_1 = -1$).
When ~$j_1=0$~ the $16$ states are as in \pseu{} except the state
excluded for ~$j_1=\half$~ and six states which are obtained
for ~$\Phi^c_1 ~=~ \Phi^c_2 ~=~ X^+_{16}\,,
~X^+_{16}X^+_{15}\,X^+_{25}$ (i.e., excluding the third row of
\incaa{}, since for them ~$\ve^c_j + 2j_1 = \ve^c_j = -1$).
Altogether, instead of the decomposition \deca{} we have
the quasi-decomposition \dectaz{}:
\eqn\decta{ \(\hL_{\rm long}\)_{\vert_{d=d^a}} ~=~ \hL_\L\ \oplus\ \hL_{\L+\a_{36}}
\oplus\ \hL'_{\L+\a_{35}} \ ,\qquad r_1=0\ .}

The 28 states of the minimal case are given as follows. There are
two anti-chiral states:
\eqn\achirr{
X^+_{46} \, |\L\rg \ , \quad
X^+_{45} \, X^+_{46} \, |\L\rg \ ,}
and six chiral states (just as in \chir{}):
\eqna\chirr
$$\eqalignno{
&X^+_{25} \, |\L\rg \ , \quad
X^+_{15} \, X^+_{25} \, |\L\rg \ \quad
X^+_{26} \, X^+_{15} \, X^+_{25} \, |\L\rg \ , &\chirr a\cr
&X^+_{26} \, X^+_{25} \, |\L\rg \ , \quad
X^+_{16} \, X^+_{25} \, |\L\rg \ , \quad
X^+_{16} \,X^+_{26} \, X^+_{15} \, X^+_{25} \, |\L\rg \ . &\chirr b}$$
Combining the chiral and chiral states
 would give further $12$ states. The rest of the states
are obtained by combining these states with impossible states
from the opposite chirality, yet obtaining allowed states.
Explicitly, the list looks like this.
Let ~$\Phi_a \, |\L\rg$, ~$\Phi_c \, |\L\rg$, denote any of the
 states in \achirr{}, \chirr{}, resp.,
~$\Phi'_c \, |\L\rg$, denote any of the
three states in \chirr{a}, resp.
Thus, there are the following states:
\eqna\combb
$$\eqalignno{
&|\L\rg \ , \quad \Phi_c \, \Phi_a \, |\L\rg \ ,
&\combb a\cr
&X^+_{15}\, X^+_{26} \, \Phi_a \, |\L\rg\ , &\combb b\cr
&X^+_{26} \, X^+_{46} \, |\L\rg\ ,
X^+_{26} \, X^+_{16} \, X^+_{25} \, X^+_{46} \, |\L\rg\ ,
&\combb c\cr
&X^+_{45} \, \Phi'_c \, |\L\rg\ . &\combb d\cr
}$$
Obviously, there are $20$ states in \combb{} ($13+2+2+3$).
Altogether, there are $28$ states in \achirr, \chirr{} and
\combb{}. This list amounts to giving explicitly character formula
\chsa{} (for $N=2$, $i_0=N-1=1$) without counter-terms.
This superfield and its conjugate (considered below)
are the shortest semi-short SRC $N=2$ superfields.

\nt\bu ~~$j_2=0$.\nl
Here holds character formula \chodra{} and
for ~$r_1=0$ holds also character formula \chsaaz{}.
A more detailed description follows.

~The state ~$X^+_{36}\,X^+_{46}
\, |\L\rg$~ and its descendants are eliminated (due to \nulnraz{}).
This elimination is described by the second term in character
formula \chodra{a}.
The eight states in the anti-chiral sector here come from:
\eqn\inczz{\eqalign{
&X^+_{46} \ , \qquad \ve^a_r =1\ , ~\ve^a_j =1\ , \cr
&X^+_{45} \, X^+_{46} \ ,
\qquad \ve^a_r =0\ , ~\ve^a_j = 2\ , \cr
&{\bf 1}\ ,\quad
X^+_{35}\, X^+_{46} \ ,\quad
~X^+_{45} \, X^+_{36} \ ,
\qquad \ve^a_r =0\ , ~\ve^a_j = 0\ , \cr
&X^+_{45} \ , \quad
X^+_{45} \, X^+_{35} \, X^+_{46} \ ,
\qquad \ve^a_r =-1\ , ~\ve^a_j = 1\ , \cr
&X^+_{35} \, X^+_{45} \ ,
\qquad \ve^a_r =-2\ , ~\ve^a_j = 0\ . \cr
}}
The above eight differ from \incz{} by one operator:
~$X^+_{35}$~ is replaced here by ~$X^+_{45} \, X^+_{36}\,$.
For ~$r_1=1$~ the last operator does not
contribute to the anti-chiral sector.
Whenever ~$r_1=0$~ the generators $X^+_{35}$ and
$X^+_{36}$ are eliminated from the anti-chiral part of the basis,
which is further restricted due to \conz{c}
and there are only two anti-chiral states as given in \achirr.

In summary, when ~$j_1\geq 1$~ correspondingly to the cases in
\clasf{a,b,c,d,e} we have now ~$128,127,121,103,68$~ states.
When ~$j_1=\half$~ correspondingly to the cases in
\clasf{a,b,c,d,e} we have now ~$120,119,113,96,63$~ states.
When ~$j_1=0$~ correspondingly to the cases in
\clasf{a,b,c,d,e} we have now ~$88,87,83,70,45$~ states.

We know that when ~$r_1> 0$~ holds formula \decaa{} for ~$\hL_{\rm long}$~
with the same values of ~$j_1,j_2(=0),r_1$~ as  ~$\L$~ and
with ~$\b = \b_{12} ~=~ \a_{36}+\a_{46}\,$.
In more detail, when the
weight ~$\L$~ corresponds to cases \clasf{a,b,c,d,e} then
the weight ~$\L+\b_{12}$~ corresponds to cases \clasf{a,a,a,b,c}
(since the value of $r_1$ is increased by 2) and furthermore
~$\hL_{\L+\b_{12}}$~ is actually a SRC of type {\bf b},
see below from where we take the numbers:
When ~$j_1\geq 1$~ the UIR ~$\hL_{\L+\b_{12}}$~ has
~$48,48,48,47,42$ states, when ~$j_1=\half$~ it has
~$45,45,45,44,39$ states, when ~$j_1=0$~ it has
~$33,33,33,32,29$ states. Summed together with the numbers for
the UIR ~$\hL_{\L}$~ from above we obtain the following
contributions to ~$\hL_{\rm long}$~:
when ~$j_1\geq 1$~ there are
~$176,175,169,150,110$ states, when ~$j_1=\half$~ there are
~$165,164,158,140,102$ states, when ~$j_1=0$~ there are
~$121,120,116,102,74$ states. Except the last cases (when
$r_1=0$) these cases match exactly the cases of long superfields
for the corresponding values of ~$j_1=1,\half,0$~ and ~$j_2=0$.
For completeness one may check that the states of
~$\hL_{\L+\b_{12}}$~ appear in ~$\hL_{\rm long}$~
being multiplied by ~$X^+_{36}\,X^+_{46}\,$.
In the cases when $r_1=0$ there is a mis-match of one state
and that extra state  is
~$X^+_{35}\,X^+_{46}\,\wt$~ which is excluded from ~$\hL_\L$~
as explained in general, cf. \nulnrb{}. (It is also excluded in
case {\bf b} below.) Thus, instead of \decaa{} we have
the quasi-decomposition:
\eqn\dectaa{  \(\hL_{\rm long}\)_{\vert_{d=d^a}}
~=~ \hL_\L ~\oplus \hL_{\L+\b_{12}}
~\oplus \hL'_{\L+\a_{35} +\a_{46}}
\ , \qquad r_1=0 \ ,}
where as in \decta{} we have put a prime on the last term indicating
that this is not a genuine irrep.

\nt\bu{\bf b}~~~~~
$d ~=~ d^2_{21} ~=~ z+r_1 ~>~ d^3_{22}\ , ~~ j_2=0$\ .

\nt
The character formula is \chsb{}.
The generators $X^+_{36}$ and $X^+_{46}$ are eliminated
due to \nullb{} and \nullbb{}. Due to \conz{b}
there are at most two anti-chiral states:
\eqn\achirrz{
X^+_{45} \, |\L\rg \ , \quad
X^+_{35} \, X^+_{45} \, |\L\rg \ .}
Thus, the maximal number of states is $48 (16\times3)$ achieved for
$r_1\geq 4$, $j_1\geq 1$. These states are given explicitly as:
\eqn\basa{ \eqalign{
\Psi_{\bar \ve} ~=&~
 (X^+_{16})^{\ve_{16}}\, (X^+_{15})^{\ve_{15}} \,
 (X^+_{26})^{\ve_{26}}\, (X^+_{25})^{\ve_{25}} \,
 (X^+_{35})^{\ve_{35}} \, (X^+_{45})^{\ve_{45}} \,
 |\L\rg \ , \cr
& \ve_{aj} = 0,1;\ \ve_{35} \leq \ve_{45} \,;
~r_1\geq 4,\ j_1\geq 1\ .}}
In summary, when ~$j_1\geq 1$~ we have correspondingly to the
cases in \clasf{a,b,c,d,e} ~$48,47,42,31,10$~ states.
When ~$j_1=\half$~ we have correspondingly to the cases in
\clasf{a,b,c,d,e} ~$45,44,39,29,9$~ states.
When ~$j_1=0$~ we have correspondingly to the cases in
\clasf{a,b,c,d,e} ~$33,32,29,23,7$~ states.
The cases when ~$r_1> 2$~ were included in decompositions \decaa{}
in the previous case {\bf a}.
(The cases when ~$r_1= 2$~ were included in
quasi-decompositions \dectaa{}
in the previous case {\bf a}.)

The minimal number when $r_1> 0$ is
$23$ achieved for $r_1=1$, $j_1=0$. Besides the obvious states
which include ~$X^+_{45} \, |\L\rg$, nine chiral states,
their combinations and the vacuum, there are the following states:
\eqn\addd{\eqalign{ &X^+_{35} \, X^+_{45} \,\Phi'\, |\L\rg \ ,\cr
&\Phi' ~=~ X^+_{25} \ , ~~~
X^+_{15} \, X^+_{25} \ , ~~~
X^+_{26}\, X^+_{15} \, X^+_{25} \ . }}

Whenever ~$r_1=0$~ the generators ~$X^+_{35}$~ and
~$X^+_{45}$~ are also eliminated from the basis due to
\nullbbb{}. Thus, these UIRs are chiral. Due to \conz{c}
and excluding the state \impz{}
there are $10,9,7$ states for $j_1\geq 1,\half,0$, resp.
(as stated above). These states explicitly are:
\eqna\semm
$$\eqalignno{
&|\L\rg\ , ~~~X^+_{25}\, |\L\rg\ , ~~~
X^+_{15}\, X^+_{25}\, |\L\rg\ ,~~~
X^+_{16}\, X^+_{25}\, |\L\rg\ ,~~~
X^+_{26}\, X^+_{25}\, |\L\rg\ , \cr
&X^+_{26}\, X^+_{15}\, X^+_{25}\, |\L\rg\ , ~~~
X^+_{16}\, X^+_{26}\, X^+_{15}\, X^+_{25}\, |\L\rg\ ,
\qquad j_1\geq0 \ , &\semm a\cr
&X^+_{15}\, |\L\rg\ , ~~~
X^+_{16}\, X^+_{15}\, X^+_{25}\, |\L\rg\ ,
\qquad j_1\geq\half \ , &\semm b\cr
& X^+_{16}\, X^+_{15}\, |\L\rg\ ,
\qquad j_1\geq 1 \ . &\semm c}$$
For $j_1=0$ the superfield in \semm{a} and its conjugate
(considered below) are the shortest short SRC $N=2$ superfields.

\nt\bu{\bf c} ~~~$d ~=~ d_{\rm max} ~=~ d^3_{22}
~=~ 2 +2j_1 -z + r_1 ~>~ d^1_{21}$ \ .\nl
This case is the conjugate to \bu{\bf a}\
and the maximal number of states is $128 = 8$(chiral)$\times
16$(anti-chiral) achieved for $j_2\geq 1,r_1\geq 4$.

\nt\bu ~~$j_1> 0\,$.\nl
The generator $X^+_{15}$ is eliminated.
The eight states in the chiral sector
are obtained from the following operators:
\eqn\aincz{\eqalign{
&X^+_{25} \ , \qquad \ve^c_r =1\ , ~\ve^c_j =1\ , \cr
&X^+_{26} \, X^+_{25} \ ,
\qquad \ve^c_r =0\ , ~\ve^c_j = 2\ , \cr
&{\bf 1}\ ,\quad
X^+_{16}\, X^+_{25} \ ,
\qquad \ve^c_r =0\ , ~\ve^c_j = 0\ , \cr
&X^+_{26} \ , \quad
X^+_{26} \, X^+_{16} \, X^+_{25} \ ,
\qquad \ve^c_r =-1\ , ~\ve^c_j = 1\ , \cr
&X^+_{16} \ ,
\qquad \ve^c_r =-1\ , ~\ve^c_j = -1\ , \cr
&X^+_{16} \, X^+_{26} \ ,
\qquad \ve^c_r =-2\ , ~\ve^c_j = 0\ . \cr
}}
In summary, when ~$j_2\geq 1$~ correspondingly to the cases in
\clasf{a,b,c,d,e} we have now ~$128,127,120,99,42$~ states.
When ~$j_2=\half$~ correspondingly to the cases in
\clasf{a,b,c,d,e} we have now ~$120,119,112,92,39$~ states.
When ~$j_2=0$~ correspondingly to the cases in
\clasf{a,b,c,d,e} we have now ~$88,87,82,68,28$~ states.
Whenever ~$r_1=0$~ the generator ~$X^+_{16}$~ is also eliminated
from the basis.

When ~$r_1> 0$~ holds decomposition \decc{} with ~$\b = \a_{15}\,$.
When ~$r_1 = 0$~ holds the quasi-decomposition:
\eqn\dectc{  \(\hL_{\rm long}\)_{\vert_{d=d^c}}
~=~ \hL_\L\ \oplus\ \hL_{\L+\a_{15}}
\oplus\ \hL'_{\L+\a_{16}} \ ,\qquad r_1=0\ ,}
cf. \decta{}. We omit most details since all results and formulae are
by conjugation from case \bu{\bf a} (when $j_2\neq0$).

We still give the 28 states of the minimal case. There are two
chiral states:
\eqn\aachirr{
X^+_{25} \, |\L\rg \ , \quad
X^+_{26} \, X^+_{25} \, |\L\rg \ ,}
and six anti-chiral states (just as in \achir{}):
\eqna\achirrrr
$$\eqalignno{
&X^+_{46} \, |\L\rg \ , \quad
X^+_{36} \, X^+_{46} \, |\L\rg \ \quad
X^+_{45} \, X^+_{36} \, X^+_{46} \, |\L\rg \ , &\achirrrr a\cr
&X^+_{45} \, X^+_{46} \, |\L\rg \ , \quad
X^+_{35} \, X^+_{46} \, |\L\rg \ , \quad
X^+_{35} \,X^+_{45} \, X^+_{36} \, X^+_{46} \, |\L\rg \ . &\achirrrr b}$$
 The rest of the states are obtained as follows.
Let ~${\hat\Phi}_a \, |\L\rg$, ~${\hat\Phi}_c \, |\L\rg$, denote any of the
 states in \aachirr{}, \achirrrr{}, resp.,
~${\hat\Phi}'_c \, |\L\rg$, denote any of the
three states in \achirrrr{a}, resp.
Thus, there are the following states:
\eqna\combbz
$$\eqalignno{
&|\L\rg \ , \quad {\hat\Phi}_c \,{\hat\Phi}_a \, |\L\rg \ ,
&\combbz a\cr
&X^+_{36}\, X^+_{45} \,{\hat\Phi}_c \, |\L\rg\ , &\combbz b\cr
&X^+_{45} \, X^+_{25} \, |\L\rg\ ,
X^+_{45} \, X^+_{35} \, X^+_{46} \, X^+_{25} \, |\L\rg\ ,
&\combbz c\cr
&X^+_{26} \,{\hat\Phi}'_c \, |\L\rg\ . &\combbz d\cr
}$$
This superfield and its conjugate (considered in \bu{\bf a})
are the shortest semi-short SRC $N=2$ superfields.

\nt\bu ~~$j_1=0$.\nl
~The state ~$X^+_{15}\,X^+_{25}
\, |\L\rg$~ and its descendants are eliminated (due to \nulnraz{}).
The eight states in the chiral sector here come from:
\eqn\ainczz{\eqalign{
&X^+_{25} \ , \qquad \ve^c_r =1\ , ~\ve^c_j =1\ , \cr
&X^+_{26} \, X^+_{25} \ ,
\qquad \ve^c_r =0\ , ~\ve^c_j = 2\ , \cr
&{\bf 1}\ ,\quad
X^+_{16}\, X^+_{25} \ ,\quad
~X^+_{26} \, X^+_{15} \ ,
\qquad \ve^c_r =0\ , ~\ve^c_j = 0\ , \cr
&X^+_{26} \ , \quad
X^+_{26} \, X^+_{16} \, X^+_{25} \ ,
\qquad \ve^c_r =-1\ , ~\ve^c_j = 1\ , \cr
&X^+_{16} \, X^+_{26} \ ,
\qquad \ve^c_r =-2\ , ~\ve^c_j = 0\ . \cr
}}
The above eight differ from \aincz{} by one operator:
~$X^+_{16}$ is replaced here by
~$X^+_{26} \, X^+_{15}\,$. In summary, when ~$j_2\geq 1$~
correspondingly to the cases in
\clasf{a,b,c,d,e} we have now ~$128,127,121,103,68$~ cases.
When ~$j_2=\half$~ correspondingly to the cases in
\clasf{a,b,c,d,e} we have now ~$120,119,113,96,63$~ cases.
When ~$j_2=0$~ correspondingly to the cases in
\clasf{a,b,c,d,e} we have now ~$88,87,83,70,45$~ cases.
Whenever ~$r_1=0$~ the generators $X^+_{16}$ and
$X^+_{15}$ are eliminated from the chiral part of the basis,
which is further restricted due to \conz{c}
and there are only two chiral states as in \aachirr.

When ~$r_1> 0$~ holds formula \deccc{} for ~$\hL_{\rm long}$~
with the same values of ~$j_1(=0),j_2,r_1$~ as for ~$\L$~ and
with ~$\b = \b_{34} ~=~ \a_{15}+\a_{25}\,$.
When ~$r_1= 0$~ this decomposition is spoiled by one state
~$X^+_{16}\,X^+_{25}\,\wt$~ which is excluded from ~$\hL_\L$~
as explained in general, cf. \nulnrbb{}, and instead of
\deccc{} we have the quasi-decomposition:
\eqn\dectcc{ \(\hL_{\rm long}\)_{\vert_{d=d^c}}
 ~=~ \hL_\L ~\oplus \hL_{\L+\b_{34}}
~\oplus \hL'_{\L+\a_{16}+\a_{25}}
\ , \qquad r_1=0 \ .}

\nt{\bu{\bf d}}~~~~~
$d ~=~ d^4_{22} ~=~ -z + r_1 ~>~ d^1_{21}\ , ~~ j_1=0$\ .\nl
This case is the conjugate to \bu{\bf b}.

\nt
The generators $X^+_{15}$ and $X^+_{25}$ are eliminated
due to \nulld{} and \nulldd{}. Due to \conz{b}
there are at most two chiral states depending on the value of $r_1$~:
\eqn\chirrz{\eqalign{
&X^+_{26} \, |\L\rg \ , \quad r_1\geq 1 \ , \cr
&X^+_{16} \, X^+_{26} \, |\L\rg \ , \quad r_1\geq 2 \ . }}
The maximal number of states is $48 (3\times16)$ achieved for
$r_1\geq 4$, $j_2\geq 1$. These states are given explicitly as:
\eqn\basa{ \eqalign{
\Psi_{\bar \ve} ~=&~
 (X^+_{35})^{\ve_{35}}\, (X^+_{36})^{\ve_{36}} \,
 (X^+_{45})^{\ve_{45}}\, (X^+_{46})^{\ve_{46}} \,
(X^+_{16})^{\ve_{16}} \, (X^+_{26})^{\ve_{26}} \,
 |\L\rg \ , \cr
& \ve_{aj} = 0,1;\ \ve_{35} \leq \ve_{45} \,;
~r_1\geq 2,\ j_2\geq 1
.}}
In summary, when ~$j_2\geq 1$~ correspondingly to the cases in
\clasf{a,b,c,d,e} we have now ~$48,47,42,31,10$~ states.
When ~$j_2=\half$~ correspondingly to the cases in
\clasf{a,b,c,d,e} we have now ~$45,44,39,29,9$~ states.
When ~$j_2=0$~ correspondingly to the cases in
\clasf{a,b,c,d,e} we have now ~$33,32,29,23,7$~ states.
The cases when ~$r_1> 2$~ were included in decompositions \deccc{}
in the previous case {\bf c}.
(The cases when ~$r_1= 2$~ were included in decompositions \dectcc{}
in the previous case {\bf c}.)

The minimal number when $r_1> 0$ is
$23$ achieved for $r_1=1$, $j_2=0$. Besides the obvious states
which include ~$X^+_{26} \, |\L\rg$, nine anti-chiral states,
their combinations and the vacuum, there the following states:
\eqn\addd{\eqalign{ &X^+_{16} \, X^+_{26} \,\Phi'\, |\L\rg \ ,\cr
&\Phi' ~=~ X^+_{46} \ , ~~~
X^+_{36} \, X^+_{46} \ , ~~~
X^+_{45}\, X^+_{36} \, X^+_{46} \ . }}

Whenever ~$r_1=0$~ the generators ~$X^+_{16}$~ and
~$X^+_{26}$~ are also eliminated from the basis due to
\nullbbb{}. Thus, these UIRs are anti-chiral. Due to \conz{c}
and excluding the state \impy{}
there are $10,9,7$ states for $j_2\geq 1,\half,0$, resp.
These states explicitly are:
\eqna\semm
$$\eqalignno{
&|\L\rg\ , ~~~X^+_{46}\, |\L\rg\ , ~~~
X^+_{36}\, X^+_{46}\, |\L\rg\ ,~~~
X^+_{35}\, X^+_{46}\, |\L\rg\ ,~~~
X^+_{45}\, X^+_{46}\, |\L\rg\ , \cr
&X^+_{45}\, X^+_{36}\, X^+_{46}\, |\L\rg\ , ~~~
X^+_{35}\, X^+_{45}\, X^+_{36}\, X^+_{46}\, |\L\rg\ ,
\qquad j_2\geq0 \ , &\semm a\cr
&X^+_{36}\, |\L\rg\ , ~~~
X^+_{35}\, X^+_{36}\, X^+_{46}\, |\L\rg\ ,
\qquad j_2> 0 \ , &\semm b\cr
& X^+_{35}\, X^+_{36}\, |\L\rg\ ,
\qquad j_2\geq 1 \ . &\semm c}$$
For $j_2=0$ the superfield in \semm{a} and its conjugate
(considered above) are the shortest short SRC $N=2$ superfields.

\vskip 5mm

\nt\bu ~~~{\bf DRC cases}

\vskip 3mm

\nt
Here we consider the DRC cases taking again the five cases of
long superfields in \clasf{} as reference point.

\nt{\bu{\bf ac}}~~~
$d ~=~ d_{\rm max}~=~ d^1_{21} = d^3_{22} ~=~ 2 + j_1 + j_2 +
r_1$\ , ~~~$z ~=~ j_1-j_2$\ .\nl
~The maximal number of states is
$64 = 8$(chiral)$\times 8$(anti-chiral), achieved for
$r_1\geq 4$. The 8 anti-chiral, chiral,
states are as described in \bu{\bf a},\bu{\bf c}, resp.,
(differing for $j_2> 0$ and $j_2=0$, $j_1> 0$ and $j_1=0$,
resp.).

\nt\bu ~~$j_1j_2> 0\,$.\nl
Here hold character formulae \chdac{} (without counterterms for
$r_1\geq 4$). The states ~$X^+_{15}\, |\L\rg$,
 ~$X^+_{36}\,|\L\rg$~ and their descendants are
eliminated. Correspondingly to the cases in
\clasf{a,b,c,d,e} we have now ~$64,63,57,42,11$~ states.
In the last case, (where $r_1=0$), we eliminate also the
generators ~$X^+_{35}$~ and ~$X^+_{16}$.

For ~$r_1> 0$ holds decomposition \decgg{}
with $\b=\a_{15}$, $\b'=\a_{36}$ as stated in the
general exposition. We would like demonstrate this and also see
how it breaks down for ~$r_1=0$, thus, we include for the moment
the case $r_1=0$. Referring to \decgg{} we note when the
weight ~$\L$~ corresponds to cases \clasf{a,b,c,d,e} then
the weights ~$\L+\a_{15}\,$, $\L+\a_{36}\,$, correspond to cases
\clasf{a,a,b,c,d} (since the value of $r_1$ is increased by 1),
i.e., the corresponding UIRs have ~$64,64,63,57,42$~ states each.
The weight ~$\L+\a_{15}+\a_{36}\,$, corresponds to cases
\clasf{a,a,a,b,c} (since the value of $r_1$ is increased by 2),
i.e., the corresponding UIRs have ~$64,64,64,63,57$~ states.
Summed together with the numbers for
the UIR ~$\hL_{\L}$~ from above we obtain the following
contributions to ~$\hL_{\rm long}$~:
~$256,255,247,219,152$. Except the last case (in which
$r_1=0$) these cases match exactly
the cases of long superfields for the case ~$j_1,j_2\geq 1$.

When $r_1=0$ the long superfields for the cases ~$j_1,j_2\geq 1$~
have ~$163$ states, i.e., a mis-match of ~$11$
states.\foot{The reader may wonder whether the long
superfield with ~$j_1=\half,j_2\geq 1,r_1=0$~ may not be used since it has
152 states, however, this is only a coincidence of the total
number.} These extra states
contain either the generator ~$X^+_{16}$~ or ~$X^+_{35}$~ or both,
and they do not contain either ~$X^+_{15}$~ or ~$X^+_{36}\,$.
Explicitly, these extra states are:
\eqn\dtecd{\eqalign{
&X^+_{16} \, X^+_{46} \, |\L\rg \ , ~~
X^+_{16} \, X^+_{25} \, |\L\rg \ , ~~
X^+_{16} \, X^+_{25} \, X^+_{46} \, |\L\rg \ , \cr
&X^+_{16} \, X^+_{26} \, X^+_{25} \, X^+_{46} \, |\L\rg \ , ~~
X^+_{16} \, X^+_{25} \, X^+_{45} \, X^+_{46} \, |\L\rg \ , \cr
&X^+_{35} \, X^+_{46} \, |\L\rg \ ,
~~X^+_{35} \, X^+_{25} \,  |\L\rg \ ,
~~X^+_{35} \, X^+_{25} \, X^+_{46} \, |\L\rg \ , \cr
&X^+_{35} \, X^+_{26} \, X^+_{25} \, X^+_{46} \, |\L\rg \ ,
~~X^+_{35} \, X^+_{25} \, X^+_{45} \, X^+_{46} \, |\L\rg \ , \cr
&X^+_{16} \, X^+_{35} \, X^+_{25} \, X^+_{46} \, |\L\rg \ .
}}
Altogether, instead of \decgg{} we may write:
\eqn\decact{\eqalign{
\(\hL_{\rm long}\)_{\vert_{d=d^{ac}}}
~=&~ \hL_{\L} ~\oplus~ \hL_{\L +\a_{15}} ~\oplus~
\hL_{\L + \a_{36}} ~\oplus~ \hL_{\L +\a_{15}+ \a_{36}} ~\oplus\cr
&~ \oplus~ \hL'_{\L+\a_{16}} ~ \oplus ~ \hL'_{\L+\a_{35}}
\oplus\ \hL'_{\L+\a_{16}+\a_{35}}\ , \quad r_1=0\ ,}}
where we have represented the extra states by the last three
terms (corresponding to first and second line of \dtecd{},
third and fourth line of \dtecd{}, fifth line of \dtecd{},
resp.), and we have put primes on these since they are
not genuine irreps.

Finally, we give the 11 states of the UIR at ~$r_1=0$~:
\eqn\dtacd{\eqalign{
&|\L\rg \ , \quad
X^+_{25} \, X^+_{46} \, |\L\rg \ ,
\quad
X^+_{26} \, X^+_{45} \,
X^+_{25} \, X^+_{46} \, |\L\rg \ ,
\cr
&X^+_{46} \, |\L\rg \ , \quad
X^+_{45} \, X^+_{46} \, |\L\rg \ ,\cr
&X^+_{25} \, |\L\rg \ , \quad
X^+_{26} \, X^+_{25} \, |\L\rg \ ,\cr
&\Phi_c^0\, X^+_{46} \, |\L\rg \ , \quad
\Phi_c^0 ~=~ X^+_{26} \, , \quad
X^+_{26} \, X^+_{25} \, ,
\cr
&\Phi_a^0\, X^+_{25} \, |\L\rg \ , \quad
\Phi_a^0 ~=~ X^+_{45} \ , \quad
X^+_{45} \, X^+_{46} \
.}}
This superfield is the shortest semi-short $N=2$ superfield.

\nt\bu ~~$j_1> 0\,,j_2=0\,$. ~~
Here hold character formulae \chodrac{} (without counterterms for
$r_1\geq 4$). The states ~$X^+_{36}\,X^+_{46}\, |\L\rg$,
 ~$X^+_{15}\,|\L\rg$~ and their descendants are
eliminated. Correspondingly to the cases in
\clasf{a,b,c,d,e} we have now ~$64,63,58,45,16$~ states.
In the last case, where $r_1=0$, we eliminate the generator ~$X^+_{16}$~
and exclude the generators ~$X^+_{3,4+k}$~
from the anti-chiral sector.

For ~$r_1> 0$ holds decomposition \decacp{}. Note that when the
weight ~$\L$~ corresponds to cases \clasf{a,b,c,d,e} then
the weight ~$\L+\a_{15}$~ corresponds to cases
\clasf{a,a,b,c,d} (since the value of $r_1$ is increased by 1),
i.e., the corresponding UIRs have ~$64,64,63,58,45$~ states.
The weight ~$\L+\b_{12}$~ corresponds to cases
\clasf{a,a,a,b,c} (since the value of $r_1$ is increased by 2),
but from type {\bf bc} considered below,
i.e., the corresponding UIRs have ~$24,24,24,23,19$~ states.
The weight ~$\L+\a_{15}+\b_{12}$~ corresponds to cases
\clasf{a,a,a,a,b} (since the value of $r_1$ is increased by 3),
also from type {\bf bc},
i.e., the corresponding UIRs have ~$24,24,24,24,23$~ states.
Summed together with the numbers for
the UIR ~$\hL_{\L}$~ from above we obtain the following
contributions to ~$\hL_{\rm long}$~:
~$176,175,169,150,103$. Except the last case (in which
$r_1=0$) these cases match exactly the cases of long superfields
for the cases when ~$j_1\geq 1,j_2=0$.

When $r_1=0$ the corresponding long superfields
have ~$111$ states, i.e., there is a mis-match of ~$8$
states.\foot{Again the long superfield with correct number of
states 103 (with $j_1=\half,j_2=r_1=0$) does not fit.}
These extra states
contain either the generator ~$X^+_{16}$~ or ~$X^+_{35}$~ or both,
and they do not contain ~$X^+_{15}\,$.
Explicitly, they are:
\eqn\dtecdz{\eqalign{
&X^+_{16} \, X^+_{46} \, |\L\rg \ , ~~
X^+_{16} \, X^+_{25} \, |\L\rg \ , ~~
X^+_{16} \, X^+_{25} \, X^+_{46} \, |\L\rg \ , ~~
X^+_{16} \, X^+_{26} \, X^+_{25} \, X^+_{46} \, |\L\rg \ , \cr
&X^+_{16} \, X^+_{25} \, X^+_{45} \, X^+_{46} \, |\L\rg \ , ~~
X^+_{16} \, X^+_{25} \, X^+_{36} \, X^+_{45} \, |\L\rg \ ,\cr
&X^+_{35} \, X^+_{46} \, |\L\rg \ , \cr
&X^+_{16} \, X^+_{35} \, X^+_{25} \, X^+_{46} \, |\L\rg \ .
}}
Altogether, instead of \decacp{} we may write:
\eqn\decactz{\eqalign{  \(\hL_{\rm long}\)_{\vert_{d=d^{ac}}}
~=&~ \hL_{\L} ~\oplus~ \hL_{\L +\a_{15}} ~\oplus~
\hL_{\L + \b_{12}} ~\oplus~ \hL_{\L +\a_{15}+ \b_{12}} ~\oplus\cr
&~ \oplus~ \hL'_{\L+\a_{16}} ~ \oplus ~ \hL'_{\L+\a_{35}}
\oplus\ \hL'_{\L+\a_{16}+\a_{35}}\ ,\quad r_1=0\ , }}
where (as in \decact{})
we have represented the extra states by the last three
terms (corresponding to first and second line of \dtecdz{},
third line of \dtecdz{}, fourth line of \dtecdz{},
resp.), and we have put primes on these since they are
not genuine irreps.

Finally, we give the 16 states of the UIR at ~$r_1=0$~:
\eqn\dtaca{\eqalign{
&|\L\rg \ , \quad
X^+_{25} \, X^+_{46} \, |\L\rg \ , \cr
&X^+_{46} \, |\L\rg \ , \quad
X^+_{45} \, X^+_{46} \, |\L\rg \ ,\cr
&X^+_{25} \, |\L\rg \ , \quad
X^+_{26} \, X^+_{25} \, |\L\rg \ ,\cr
&\Phi_a^0\, X^+_{25} \, |\L\rg \ , \quad
\Phi_a^0 ~=~ X^+_{45} \, X^+_{46} \, ,
~X^+_{35} \, X^+_{46} \, ,
~X^+_{36} \, X^+_{45} \, , \cr&
\Phi_a^-\, X^+_{25} \, |\L\rg \ , \quad
\Phi_a^- ~=~ X^+_{45} \, ,
~X^+_{35} \, X^+_{45} \, X^+_{46} \, , \cr
&{\hat\Phi}_c\, X^+_{46} \, |\L\rg \ , \quad
{\hat\Phi}_c ~=~ X^+_{26} \, ,
X^+_{26} \, X^+_{25} \, , \cr
&\Phi_a^0\,X^+_{45} \, X^+_{25} \, |\L\rg \ . }}
The states of \dtacd{} are a subset of \dtaca{}.

The next case is conjugate to the preceding.

\nt\bu ~~$j_1=0,j_2>0\,$. ~~
Here hold character formulae \chodracz{} (without counterterms for
$r_1\geq 4$).
The states ~$X^+_{15}\,X^+_{25}\, |\L\rg$,
 ~$X^+_{36}\,|\L\rg$~ and their descendants are
eliminated. Correspondingly to the cases in
\clasf{a,b,c,d,e} we have now ~$64,63,58,45,16$~ states.
In the last case, when $r_1=0$, we eliminate the generator ~$X^+_{35}$~
and exclude the generators ~$X^+_{1,4+k}$~
from the chiral sector.

For ~$r_1> 0$ holds decomposition \decacq{}.
When $r_1=0$ the corresponding long superfields
have ~$111$ states, i.e., there is a mis-match of ~$8$
states. These extra states
contain either the generator ~$X^+_{16}$~ or ~$X^+_{35}$~ or both,
and they do not contain ~$X^+_{36}\,$.
Explicitly, these extra states are:
\eqn\dtecdy{\eqalign{
&X^+_{35} \, X^+_{25} \, |\L\rg \ , ~~
X^+_{35} \, X^+_{46} \, |\L\rg \ , ~~
X^+_{35} \, X^+_{46} \, X^+_{25} \, |\L\rg \ , ~~
X^+_{35} \, X^+_{45} \, X^+_{46} \, X^+_{25} \, |\L\rg \ , \cr
&X^+_{35} \, X^+_{46} \, X^+_{26} \, X^+_{25} \, |\L\rg \ , ~~
X^+_{35} \, X^+_{46} \, X^+_{15} \, X^+_{26} \, |\L\rg \ ,\cr
&X^+_{16} \, X^+_{25} \, |\L\rg \ , \cr
&X^+_{35} \, X^+_{16} \, X^+_{46} \, X^+_{25} \, |\L\rg \ .
}}
Altogether, instead of \decacq{} we may write:
\eqn\decactzz{\eqalign{
\(\hL_{\rm long}\)_{\vert_{d=d^{ac}}}
~=&~ \hL_{\L} ~\oplus~ \hL_{\L +\a_{36}} ~\oplus~
\hL_{\L + \b_{34}} ~\oplus~ \hL_{\L +\a_{36}+ \b_{34}} ~\oplus\cr
&~ \oplus~ \hL'_{\L+\a_{16}} ~ \oplus ~ \hL'_{\L+\a_{35}}
\oplus\ \hL'_{\L+\a_{16}+\a_{35}}\ ,\quad r_1=0\ . }}

Finally, we give the 16 states of the UIR at ~$r_1=0$~:
\eqn\dtacb{\eqalign{
&|\L\rg \ , \quad
X^+_{25} \, X^+_{46} \, |\L\rg \ , \cr
&X^+_{46} \, |\L\rg \ , \quad
X^+_{45} \, X^+_{46} \, |\L\rg \ ,\cr
&X^+_{25} \, |\L\rg \ , \quad
X^+_{26} \, X^+_{25} \, |\L\rg \ ,\cr
&\Phi_c^0\, X^+_{46} \, |\L\rg \ , \quad
\Phi_c^0 ~=~ X^+_{26} \, X^+_{25} \, ,
~X^+_{16} \, X^+_{25} \, ,
~X^+_{15} \, X^+_{26} \, , \cr&
\Phi_c^-\, X^+_{46} \, |\L\rg \ , \quad
\Phi_c^- ~=~ X^+_{26} \, ,
~X^+_{16} \, X^+_{26} \, X^+_{25} \, , \cr&
{\hat\Phi}_a\, X^+_{25} \, |\L\rg \ , \quad
{\hat\Phi}_a ~=~ X^+_{45} \, ,
X^+_{45} \, X^+_{46} \, , \cr
&\Phi_c^0\,X^+_{45} \, X^+_{46} \, |\L\rg \ . }}
The states of \dtacd{} are a subset of \dtacb{}.

\nt\bu ~~$j_1=j_2=0$. ~~
Here hold character formulae \chodraczz{} (without counterterms for
$r_1\geq 4$). The states ~$X^+_{15}\,X^+_{25}\, |\L\rg$,
 ~$X^+_{36}\,X^+_{46}\, |\L\rg$~ and their descendants are
eliminated. Correspondingly to the cases in
\clasf{a,b,c,d,e} we have now ~$64,63,59,47,24$~ states.
In the last case, when $r_1=0$, we exclude the generators ~$X^+_{3,4+k}$~
from the anti-chiral sector and the generators ~$X^+_{1,4+k}$~
from the chiral sector and also the combination of impossible
states \impt{} as explained in the general exposition.

For ~$r_1> 0$ holds decomposition \decacr{}. Note that when the
weight ~$\L$~ corresponds to cases \clasf{a,b,c,d,e} then
the weights ~$\L+\b_{12}\,$, ~$\L+\b_{34}$~ corresponds to cases
\clasf{a,a,a,b,c} (since the value of $r_1$ is increased by 2),
but from types {\bf bc}, {\bf ad}, resp., considered below,
i.e., the corresponding UIRs have ~$24,24,24,23,20$~ states each.
The weight ~$\L+\b_{12}+\b_{34}$~ corresponds to cases
\clasf{a,a,a,a,a} (since the value of $r_1$ is increased by 4),
but from type {\bf bd},
i.e., the corresponding UIRs have ~$9,9,9,9,9$~ states.
Summed together with the numbers for
the UIR ~$\hL_{\L}$~ from above we obtain the following
contributions to ~$\hL_{\rm long}$~:
~$121,120,116,102,73$. Except the last case (in which
$r_1=0$) these cases match exactly the cases of long superfields
for the cases when ~$j_1=j_2=0$.

When $r_1=0$ the corresponding long superfields
have ~$75$ states, i.e., there is a mis-match of ~$2$
states. These extra states are:
\eqn\dtecdzz{X^+_{16} \, X^+_{25} \, |\L\rg \ , \quad
X^+_{35} \, X^+_{46} \, |\L\rg \ , }

Altogether, instead of \decacr{} we may write:
\eqn\decacty{\eqalign{
\(\hL_{\rm long}\)_{\vert_{d=d^{ac}}}
~=&~ \hL_{\L} ~\oplus~ \hL_{\L +\b_{12}} ~\oplus~
\hL_{\L + \b_{34}} ~\oplus~ \hL_{\L +\b_{12}+ \b_{34}} ~\oplus\cr
&~ \oplus~ \hL'_{\L+\a_{16}} ~ \oplus ~ \hL'_{\L+\a_{35}}
\ ,\quad r_1=0\ , }}

Finally, we give the 24 states of the UIR at ~$r_1=0$~:
\eqn\dtac{\eqalign{
&|\L\rg \ , \quad
X^+_{25} \, X^+_{46} \, |\L\rg \ , \cr
&X^+_{46} \, |\L\rg \ , \quad
X^+_{45} \, X^+_{46} \, |\L\rg \ ,\cr
&X^+_{25} \, |\L\rg \ , \quad
X^+_{26} \, X^+_{25} \, |\L\rg \ ,\cr
&\Phi_c^0\, X^+_{46} \, |\L\rg \ , \quad
\Phi_c^0 ~=~ X^+_{26} \, X^+_{25} \, ,
~X^+_{16} \, X^+_{25} \, ,
~X^+_{15} \, X^+_{26} \, , \cr&
\Phi_c^-\, X^+_{46} \, |\L\rg \ , \quad
\Phi_c^- ~=~ X^+_{26} \, ,
~X^+_{16} \, X^+_{26} \, X^+_{25} \, , \cr
&\Phi_a^0\, X^+_{25} \, |\L\rg \ , \quad
\Phi_a^0 ~=~ X^+_{45} \, X^+_{46} \, ,
~X^+_{35} \, X^+_{46} \, ,
~X^+_{36} \, X^+_{45} \, , \cr&
\Phi_a^-\, X^+_{25} \, |\L\rg \ , \quad
\Phi_a^- ~=~ X^+_{45} \, ,
~X^+_{35} \, X^+_{45} \, X^+_{46} \, , \cr
&\Phi_c^0\, \Phi_a^0\, |\L\rg \ {\rm excluding\
the\ state:}\ X^+_{15}\, X^+_{26} \,X^+_{36}\, X^+_{45} \, |\L\rg\ .}}
The states of \dtaca{},\dtacb{} are subsets of \dtac{}.

\vskip 3mm

\nt{\bu{\bf ad}}~~~
$d ~=~ d^1_{21} = d^4_{22} ~=~ 1 + j_2 + r_1\ , ~~ j_1=0$,
~~~$z ~=~ -1-j_2$\ .\nl
Here hold character formulae \chdad{} when $j_2r_1> 0$,
\chodrad{} when $j_2=0,r_1> 0$, (both these cases without
counterterms for $r_1\geq4$), and finally when $r_1=0$ holds \chdadz{}
independently of the value of $j_2$ - these are the anti-chiral
massless UIRs.

The generators $X^+_{15}$,
$X^+_{25}$, and in addition $X^+_{36}$ for $j_2> 0$
(resp. the state ~$X^+_{36}\,X^+_{46}\, |\L\rg$,
 and its descendants for $j_2=0$) are eliminated.
The maximal number of states is $24 = 3$(chiral)$\times
8$(anti-chiral), achieved for $r_1\geq 4$.
The chiral sector for $r_1>0$
 consists of the two states in \chirrz{}
and the vacuum,
while the anti-chiral sector is given by \incz{} for $j_2>0$ and by
\inczz{} for $j_2=0$.

The 24 states for $j_2>0$ are given explicitly as:
\eqna\tada
$$\eqalignno{
&|\L\rg\ ,~
X^+_{46} \,|\L\rg\ , ~ X^+_{45} \, X^+_{46} \,|\L\rg\ ,
\qquad r_1\geq0\ ,\cr
&X^+_{35}\, X^+_{46} \,|\L\rg\ ,~
~ X^+_{26}\, X^+_{46} \,|\L\rg\ , \qquad r_1\geq1\ ,\cr
&X^+_{45} \,|\L\rg\ ,
~X^+_{45} \, X^+_{35} \, X^+_{46} \,|\L\rg\ ,
~X^+_{26}\, X^+_{45} \, X^+_{46} \,|\L\rg\ ,
~X^+_{35} \,|\L\rg\ ,
\qquad r_1\geq1\ ,\cr
&X^+_{26}\,|\L\rg\ ,
~X^+_{26}\, X^+_{35}\, X^+_{46} \,|\L\rg\ ,
~X^+_{16} \, X^+_{26}\, X^+_{46} \,|\L\rg\ , \qquad r_1\geq1\ ,\cr
&X^+_{26}\, X^+_{45} \,|\L\rg\ ,
~X^+_{26}\, X^+_{45} \, X^+_{35} \, X^+_{46} \,|\L\rg\ ,
~X^+_{35} \, X^+_{45} \,|\L\rg\ ,
~X^+_{16} \, X^+_{26}\, X^+_{45} \, X^+_{46} \,|\L\rg\ ,
\qquad r_1\geq2\ ,\cr
&X^+_{16} \, X^+_{26}\,|\L\rg\ ,
~X^+_{16} \, X^+_{26}\, X^+_{35}\, X^+_{46} \,|\L\rg\ ,
~X^+_{26}\, X^+_{35} \,|\L\rg\ ,
\qquad r_1\geq2\ ,\cr
&X^+_{26}\, X^+_{35} \, X^+_{45} \,|\L\rg\ ,
~X^+_{16} \, X^+_{26}\, X^+_{45} \,|\L\rg\ ,
~X^+_{16} \, X^+_{26}\, X^+_{45} \, X^+_{35} \, X^+_{46} \,|\L\rg\ ,
~X^+_{16} \, X^+_{26}\, X^+_{35} \,|\L\rg\ ,
\quad r_1\geq3\ ,\cr
&X^+_{16} \, X^+_{26}\, X^+_{35} \, X^+_{45} \,|\L\rg\ ,
\qquad r_1\geq4\ . &\tada{}}$$
Thus, correspondingly to the cases in
\clasf{a,b,c,d,e} we have now ~$24,23,19,12,3$~ states.

The irreps with ~$r_1> 2$~  appear (two times if $r_1> 3$)  in decomposition
\decacq{} as explained in detail in the main text for type {\bf ad}.
(The irreps with ~$r_1= 2$~ have appeared in quasi-decomposition
\decactzz{}.)

The 24 states for $j_2=0$ are given explicitly as:
\eqna\tad
$$\eqalignno{
&|\L\rg\ ,~
X^+_{46} \,|\L\rg\ , ~ X^+_{45} \, X^+_{46} \,|\L\rg\ ,
\qquad r_1\geq0\ ,\cr
&X^+_{35}\, X^+_{46} \,|\L\rg\ ,
~ X^+_{26}\, X^+_{46} \,|\L\rg\ ,
~X^+_{45} \, X^+_{36} \,|\L\rg\ ,
\qquad r_1\geq1\ ,\cr
&X^+_{45} \,|\L\rg\ ,
~X^+_{45} \, X^+_{35} \, X^+_{46} \,|\L\rg\ ,
~X^+_{26}\, X^+_{45} \, X^+_{46} \,|\L\rg\ ,
\qquad r_1\geq1\ ,\cr
&X^+_{26}\,|\L\rg\ ,
~X^+_{26}\, X^+_{35}\, X^+_{46} \,|\L\rg\ ,
~X^+_{16} \, X^+_{26}\, X^+_{46} \,|\L\rg\ ,
~X^+_{26}\, X^+_{45} \, X^+_{36} \,|\L\rg\ ,
\qquad r_1\geq1\ ,\cr
&X^+_{26}\, X^+_{45} \,|\L\rg\ ,
~X^+_{26}\, X^+_{45} \, X^+_{35} \, X^+_{46} \,|\L\rg\ ,
~X^+_{35} \, X^+_{45} \,|\L\rg\ ,
~X^+_{16} \, X^+_{26}\, X^+_{45} \, X^+_{46} \,|\L\rg\ ,
\qquad r_1\geq2\ ,\cr
&X^+_{16} \, X^+_{26}\,|\L\rg\ ,
~X^+_{16} \, X^+_{26}\, X^+_{35}\, X^+_{46} \,|\L\rg\ ,
~X^+_{16} \, X^+_{26}\, X^+_{45} \, X^+_{36} \,|\L\rg\ ,
\qquad r_1\geq2\ ,\cr
&X^+_{26}\, X^+_{35} \, X^+_{45} \,|\L\rg\ ,
~X^+_{16} \, X^+_{26}\, X^+_{45} \,|\L\rg\ ,
~X^+_{16} \, X^+_{26}\, X^+_{45} \, X^+_{35} \, X^+_{46} \,|\L\rg\ ,
\qquad r_1\geq3\ ,\cr
&X^+_{16} \, X^+_{26}\, X^+_{35} \, X^+_{45} \,|\L\rg\ ,
\qquad r_1\geq4\ . &\tad{}}$$
Thus, correspondingly to the cases in
\clasf{a,b,c,d,e} we have now ~$24,23,20,13,3$~ states.

The irreps with ~$r_1> 2$~    appear as the term
~$\hL_{\L + \b_{34}}$~ of \decacr{}, while those with
~$r_1> 3$~ appear also as the term
~$\hL_{\L + \a_{3,4+N} + \b_{34}}$~ of \decacq{} but only when
~$j_2=\half$~ in ~$\L$~ there.
(The irreps with ~$r_1= 2$~ have appeared in quasi-decompositions
\decacty{}.)

The cases \tada{} and \tad{} share 21 states (for $r_1\geq4$).
The 3 states by which they differ are the last states on the
3rd,6th,7th lines of \tada{} and 2nd,4th,6th lines of \tad{}.

\vskip 3mm

\nt{\bu{\bf bc}}~~~
$d ~=~ d^2_{21} = d^3_{22} ~=~ 1 + j_1 + r_1\ , ~~ j_2=0$,
~~~$z ~=~ 1+j_1$\ .\nl
Here hold character formulae \chdbc{} when $j_1r_1> 0$,
\chodrbc{} when $j_1=0,r_1> 0$, (both these cases without
counterterms for $r_1\geq4$), and finally when $r_1=0$ holds \chdbcz{}
independently of the value of $j_1$ - these are the chiral
massless UIRs.

The generators $X^+_{36}$,
$X^+_{46}$, and in addition $X^+_{15}$ for $j_1> 0$
(resp. the state ~$X^+_{15}\,X^+_{25}\, |\L\rg$,
 and its descendants for $j_1=0$) are eliminated.
The maximal number of states is $24 = 8$(chiral)$\times
3$(anti-chiral), achieved for $r_1\geq 4$.
The anti-chiral sector for $r_1>$
 consists of the two states in \achirrz{}
and the vacuum,
while the chiral sector is given by \aincz{} for $j_1> 0$ and by
\ainczz{} for $j_1=0$.

The 24 states for $j_1> 0$ are given explicitly as:
\eqna\tbc
$$\eqalignno{
&|\L\rg\ ,~
X^+_{25} \,|\L\rg\ , ~ X^+_{26} \, X^+_{25} \,|\L\rg\ ,
\qquad r_1\geq0\ ,\cr
&X^+_{16}\, X^+_{25} \,|\L\rg\ ,~
~ X^+_{45}\, X^+_{25} \,|\L\rg\ , \qquad r_1\geq1\ ,\cr
&X^+_{26} \,|\L\rg\ ,
~X^+_{26} \, X^+_{16} \, X^+_{25} \,|\L\rg\ ,
~X^+_{45}\, X^+_{26} \, X^+_{25} \,|\L\rg\ , ~X^+_{16} \,|\L\rg\ ,
\qquad r_1\geq1\ ,\cr
&X^+_{45}\,|\L\rg\ ,
~X^+_{45}\, X^+_{16}\, X^+_{25} \,|\L\rg\ ,
~X^+_{35} \, X^+_{45}\, X^+_{25} \,|\L\rg\ , \qquad r_1\geq1\ ,\cr
&X^+_{45}\, X^+_{26} \,|\L\rg\ ,
~X^+_{45}\, X^+_{26} \, X^+_{16} \, X^+_{25} \,|\L\rg\ ,
~X^+_{16} \, X^+_{26} \,|\L\rg\ ,
~X^+_{35} \, X^+_{45}\, X^+_{26} \, X^+_{25} \,|\L\rg\ ,
\qquad r_1\geq2\ ,\cr
&X^+_{35} \, X^+_{45}\,|\L\rg\ ,
~X^+_{35} \, X^+_{45}\, X^+_{16}\, X^+_{25} \,|\L\rg\ ,
~X^+_{45}\, X^+_{16} \,|\L\rg\ ,
\qquad r_1\geq2\ ,\cr
&X^+_{45}\, X^+_{16} \, X^+_{26} \,|\L\rg\ ,
~X^+_{35} \, X^+_{45}\, X^+_{26} \,|\L\rg\ ,
~X^+_{35} \, X^+_{45}\, X^+_{26} \, X^+_{16} \, X^+_{25} \,|\L\rg\ ,
X^+_{35} \, X^+_{45}\, X^+_{16} \,|\L\rg\ ,
\quad r_1\geq3\ ,\cr
&X^+_{35} \, X^+_{45}\, X^+_{16} \, X^+_{26} \,|\L\rg\ ,
\qquad r_1\geq4\ . &\tbc{}}$$

Thus, correspondingly to the cases in
\clasf{a,b,c,d,e} we have now ~$24,23,19,12,3$~ states.

The irreps with ~$r_1> 2$~  appear (up to two times)  in decomposition
\decacp{} as explained in detail in the main text for type {\bf bc}.
(The irreps with ~$r_1= 2$~ have appeared in quasi-decomposition
 \decactz{}.)

The 24 states for $j_1=0$ are given explicitly as:
\eqna\tbca
$$\eqalignno{
&|\L\rg\ ,~
X^+_{25} \,|\L\rg\ , ~ X^+_{26} \, X^+_{25} \,|\L\rg\ ,
\qquad r_1\geq0\ ,\cr
&X^+_{16}\, X^+_{25} \,|\L\rg\ ,~
~X^+_{45}\, X^+_{25} \,|\L\rg\ ,
~X^+_{26} \, X^+_{15} \,|\L\rg\ ,
\qquad r_1\geq1\ ,\cr
&X^+_{26} \,|\L\rg\ ,
~X^+_{26} \, X^+_{16} \, X^+_{25} \,|\L\rg\ ,
~X^+_{45}\, X^+_{26} \, X^+_{25} \,|\L\rg\ ,
\qquad r_1\geq1\ ,\cr
&X^+_{45}\,|\L\rg\ ,
~X^+_{45}\, X^+_{16}\, X^+_{25} \,|\L\rg\ ,
~X^+_{35} \, X^+_{45}\, X^+_{25} \,|\L\rg\ ,
~X^+_{45}\, X^+_{26} \, X^+_{15} \,|\L\rg\ ,
\qquad r_1\geq1\ ,\cr
&X^+_{45}\, X^+_{26} \,|\L\rg\ ,
~X^+_{45}\, X^+_{26} \, X^+_{16} \, X^+_{25} \,|\L\rg\ ,
~X^+_{16} \, X^+_{26} \,|\L\rg\ ,
~X^+_{35} \, X^+_{45}\, X^+_{26} \, X^+_{25} \,|\L\rg\ ,
\qquad r_1\geq2\ ,\cr
&X^+_{35} \, X^+_{45}\,|\L\rg\ ,
~X^+_{35} \, X^+_{45}\, X^+_{16}\, X^+_{25} \,|\L\rg\ ,
~X^+_{35} \, X^+_{45}\, X^+_{26} \, X^+_{15} \,|\L\rg\ ,
\qquad r_1\geq2\ ,\cr
&X^+_{45}\, X^+_{16} \, X^+_{26} \,|\L\rg\ ,
~X^+_{35} \, X^+_{45}\, X^+_{26} \,|\L\rg\ ,
~X^+_{35} \, X^+_{45}\, X^+_{26} \, X^+_{16} \, X^+_{25} \,|\L\rg\ ,
\qquad r_1\geq3\ ,\cr
&X^+_{35} \, X^+_{45}\, X^+_{16} \, X^+_{26} \,|\L\rg\ ,
\qquad r_1\geq4\ .&\tbca{}}$$
Thus, correspondingly to the cases in
\clasf{a,b,c,d,e} we have now ~$24,23,20,13,3$~ states.

The irreps with ~$r_1> 2$~    appear as the term
~$\hL_{\L + \b_{12}}$~ of \decacr{}, while those with
~$r_1> 3$~ appear also as the term
~$\hL_{\L + \a_{15} + \b_{12}}$~ of \decacp{} but only when
~$j_1=\half$~ in ~$\L$~ there.
(The irreps with ~$r_1= 2$~ have appeared in quasi-decomposition
\decacty{}.)

\vskip 3mm

\nt{\bu{\bf bd}}~~~
$d ~=~ d^2_{21} = d^4_{22} ~=~ r_1\ , ~~ j_1=j_2=0=z$\ .\nl
~~The generators
~$X^+_{15}$, ~$X^+_{25}$, ~$X^+_{36}$,
~$X^+_{46}$ are eliminated.
For ~$r_1=1$~ also the generators
~$X^+_{16}$, ~$X^+_{35}$ are eliminated.
For ~$r_1=0$~ the remaining two generators
~$X^+_{26}$, ~$X^+_{45}$ are eliminated and we have the trivial
irrep as explained in general.

For ~$r_1> 0$~ the character formula is \chdbd{} with
~$i_0=i'_0=0$. The maximal number of states is
nine and the list of states together with the conditions when
they exist are:
\eqn\tbd{\eqalign{
&|\L\rg \ , \qquad r_1\geq 0 \ , \cr
&X^+_{26} \, |\L\rg \ , \quad
X^+_{45} \, |\L\rg\ , \qquad r_1\geq 1 \ , \cr
&X^+_{16} \, X^+_{26} \, |\L\rg\ , \quad
X^+_{35} \, X^+_{45} \, |\L\rg\ , \quad
X^+_{26} \, X^+_{45} \, |\L\rg\ , \qquad r_1\geq 2 \ , \cr
&X^+_{16} \, X^+_{26} \, X^+_{45} \, |\L\rg\ , \quad
X^+_{26} \, X^+_{35} \, X^+_{45} \, |\L\rg\ ,
\qquad r_1\geq 3 \ , \cr
&X^+_{16} \, X^+_{26} \, X^+_{35} \,
X^+_{45} \, |\L\rg\ , \qquad r_1\geq 4 \
. }}
Thus, correspondingly to the cases in
\clasf{a,b,c,d,e} we have now ~$9,8,6,3,1$~ states.
The mixed massless irrep is obtained for ~$d=r_1=1$~ and consists
of the first three states above - as was shown in general.

The irreps with ~$r_1> 4$~ have appeared in decomposition
\decacr{}, cf. type {\bf ac} above.
(The irreps with ~$r_1= 4$~ have appeared in quasi-decomposition
 \decacty{}.)

\np

\appendix{B}{Odd Reflections}

\vskip 3mm

We consider two ways to extend the Weyl group $W$ of the even subalgebra $\goo$
to a larger group by   ~{\it odd reflections} \DPf,\Serg{}. One way we
introduced in \oddr{}. The other way is as follows.
For ~$\b\in \D_\I\,$, ~$\L\in \chi^*$, we define:
\eqna\suwe
$$\eqalignno{
&s_\b\, \Lambda = \Lambda -2 {(\b,\Lambda)\over
(\b,\b)}\,\b , \qquad (\b,\b)\neq
0 \cr
&s_\b\, \Lambda = \Lambda +\b , \qquad (\b,\b)= 0 ,
\quad (\b,\Lambda)\neq 0 \cr
&s_\b\, \Lambda = \Lambda , \qquad (\b,\b)= 0 , \quad
(\b,\Lambda)= 0, \quad
\b\neq \Lambda \cr
&s_\b\, \b = -\b ,
&\suwe{}}$$
where $(\cdot,\cdot)$ is the standard bilinear product in $\chi^*$.
As in the even case one has: $s^{-1}_{\b} = s_{-\b}$,
and ~$s_\b^2 = $id$_{\chi^*}$.

The two versions of the odd reflections have different uses. The definition \suwe{}
is useful for the fact that
each such odd reflection transforms the root system $\D$ into a root system $\D'$
so that the corresponding Lie superalgebras $\cg(\D)$ and $\cg(\D')$ are isomorphic
\Serg,\Yamn.

The generalized odd reflections ~$\hs_\b$~ introduced in \DPf\ and \oddr{} are useful
for the identification of the generalized Weyl group with the multiplets
of reducible Verma modules. 
For this we used the fact that each ~$\hs_\b$~ generates an
 infinite discrete abelian group given in \oddwb{}.
In fact, the generalized odd reflections act as  translations and
  do not preserve $\Delta$.

Finally, note that if $\alpha,\beta,\alpha +\beta \in\Delta$ and
$(\alpha,\alpha)(\beta,\beta)(\alpha +\beta,\alpha +\beta)=0$
then $\hs_{\alpha +\beta}$ can not be expressed in terms
of $\hs_{\alpha}$, $\hs_{\beta}$.

\np

\appendix{C}{Characters of the even subalgebra}

For the characters of the even subalgebra we first recall its structure:
~$\cg^\bac_0 ~=~ sl(4) \oplus gl(1) \oplus sl(N)$~ of
~$\cg^\bac$. We choose a basis in which the Cartan subalgebra
~$\ch$~ of ~$\cg^\bac$~ is also a Cartan subalgebra of
~$\cg^\bac_0\,$.
Since the subalgebra ~$\cg^\bac_0$~ is reductive the
corresponding character formulae will be given by the
products of the character formulae of the two simple factors
~$sl(4)$~ and ~$sl(N)$.

We start with the ~$sl(4)$~ case. We denoted the six positive
roots of $sl(4)$ by $\a_{ij}\,$, $1\leq i<j\leq 4$.
For the simplification of
the character formulae we use notation for the formal
exponents corresponding to the $sl(4)$ simple roots:
~$t_j ~\equiv~ e(\a_{j,j+1})$, $j=1,2,3$; then for the three
non-simple roots we have: $e(\a_{13}) = t_1t_2$, $e(\a_{24}) = t_2t_3$,
$e(\a_{14}) = t_1t_2t_3\,$. In terms of these the
character formula for a Verma module over $sl(4)$ is:
\eqn\chas{ch_0~V^{\L^s} ~~=~~ {e(\L^s)\over (1-t_1) (1-t_2) (1-t_3)
(1-t_1t_2) (1-t_2t_3) (1-t_1t_2t_3)} }
where by ~$\L^s$~ we denote the $sl(4)$ lowest weight.

The representations of $sl(4)$ which we consider are
infinite-dimensional. When ~$d>d_{\rm max}$~ then all the numbers:
~$n_2,n_{13},n_{24},n_{14}\,$~ from \redd{} can not be positive
integers. Then
the only reducibilities of the $sl(4)$ Verma module are related
to the complexification of the Lorentz subalgebra of $su(2,2)$,
i.e., with $sl(2) \oplus sl(2)$,
and the character formula is given by the product of the two
character formulae for finite-dimensional $sl(2)$ irreps.
In short, the $sl(4)$ character formula is:
\eqn\slgen{\eqalign{
ch_0~L_{\L^s} ~=&~ ch_0~V^{\L^s} ~-~
ch_0~V^{\L^s + n_1\a_{12}} ~-~ ch_0~V^{\L^s + n_3\a_{34}}
~+~ ch_0~V^{\L^s + n_1\a_{12} + n_3\a_{34}} ~=\cr
=&~ {e(\L^s)\ (1-t_1^{n_1})\ (1-t_3^{n_3})
 \over (1-t_1) (1-t_2) (1-t_3)
(1-t_1t_2) (1-t_2t_3) (1-t_1t_2t_3)} ~=\cr
=&~ e(\L^s)\ \cq^s_{n_1,n_2} \ , \cr
&n_1 = 2j_1+1,\ n_3 = 2j_2+1,\ ~d > d_{\rm max} \ ,
}}
and we have introduced for later use notation $\cq^s_{n_1,n_2}$
for the character factorized by $e(\L^s)$.
The above formula obviously has the form $\chm$ replacing ~$W
~\mt~ W_2 \times W_2\,$, where ~$W_2$~ is the two-element Weyl
group of $sl(2)$.

When ~$d\leq d_{\rm max}$~ there are additional even reducibilities,
cf. \disda{}, \embde{}, \paory{}, and the discussion in-between.

Thus, we need additional formulae for $ch_0~L_{\L^s}$~:
\eqna\chaory
$$\eqalignno{ch_0~L_{\L^s}
 ~=&~\cr ~=&~
e(\L^s)\ \cq^s_{n_1,n_2}\ \cdot\ (1-t_1t_2t_3) = {e(\L^s)\ (1-t_1^{n_1})\ (1-t_3^{n_3})
 \over (1-t_1) (1-t_2) (1-t_3)
(1-t_1t_2) (1-t_2t_3)}\  , &  \cr & {\rm for}~ \paory{a},
~~~d ~=~ d^1_{N1} = d^3_{NN} ~=~ 2 + j_1 + j_2, ~~j_1j_2> 0 \ ;
&\chaory{a} \cr
~=&~
e(\L^s)\ \cq^s_{1,2}\ \cdot\ (1-t_2t_3) = {e(\L^s)\   (1+t_3)
 \over   (1-t_2) (1-t_1t_2) (1-t_1t_2t_3)}\  , &  \cr & {\rm for}~ \paory{b},
~~~d ~=~ d^1_{N1} = d^4_{NN} ~=~ 3/2, ~~j_1=0, j_2=\ha \ ;
&\chaory{b} \cr
~=&~
e(\L^s)\ \cq^s_{2,1}\ \cdot\ (1-t_1t_2) = {e(\L^s)\ (1+t_1)
 \over  (1-t_2)  (1-t_2t_3) (1-t_1t_2t_3)}\  , &  \cr & {\rm for}~ \paory{d},
~~~d ~=~ d^2_{N1} = d^3_{NN} ~=~ 3/2, ~~j_1=\ha, j_2=0 \ ;
&\chaory{c} \cr
~=&~
e(\L^s)\ \cq^s_{1,1}\ \cdot\ (1-t_1t^2_2t_3) = {e(\L^s)\  (1-t_1t^2_2t_3)
 \over  (1-t_2) (1-t_1t_2) (1-t_2t_3)(1-t_1t_2t_3)}\  , &  \cr & {\rm for}~ \paory{c,e,f},
~~~d ~=~   1 , ~~j_1=j_2=0  \ .
&\chaory{d} \cr
}$$

In the case of ~$sl(N)$~ the representations are
finite-dimensional since we induce from UIRs of $su(N)$.
The character formula is \chm, which we repeat in order to
introduce the corresponding notation:
\eqn\chu{
ch_0~L_{\L^u} (r_1,\ldots,r_{N-1})
~~=~~ \sum_{w\in W_u}(-1)^{\ell(w)} ~ch_0~V^{w\cdot\L^u} \,, \quad
\L^u\in -\G^u_+}
The index $u$ is to distinguish the quantities pertinent to the case.

\vskip 5mm

\np

\parskip=0pt \baselineskip=10pt
\listrefs
\np\end